%% file: main.tex
\documentclass[english,autodetect-engine,11pt]{article} 
\input{preamble}

\title{The Walras-Bowley Lecture: Fragmentation of Matching Markets and How Economics Can Help Integrate Them\thanks{\tiny{Kamada: Haas School of Business, University of California, Berkeley, and Department of Economics, University of Tokyo, Tokyo, Japan, y.cam.24@gmail.com. Kojima: Department of Economics, University of Tokyo, Tokyo, Japan,
fuhitokojima1979@gmail.com. Matsushita: Graduate School of Informatics, Kyoto University, Kyoto, Japan, amatsushita@i.kyoto-u.ac.jp. This paper was presented as 
the Walras-Bowley Lecture delivered by Fuhito Kojima at the 2023 North American Summer Meeting of the Econometric Society. 
 }
We thank Itai Ashlagi, David Cantala, Julien Combe, Damian Gibaja, Josue Ortega,
Anand Siththaranjan, Shengwu Li, Marek Pycia, Bernard Salanie, and participants at various seminars and conferences for their helpful comments. We thank CyberAgent AI Lab, especially Yoshihiro Takenami, for invaluable assistance in obtaining data. 
We thank officials in Bunkyo Ward and Koriyama City for providing data used in the present paper. Nanami Aoi, Masato Eguchi, Shinji Koiso, Kazuki Sekiya, Meina Takahashi, Rikuto Ueda, Kazuma Takakura, and Jiarui Xie provided excellent research assistance. Fuhito Kojima is supported by the JSPS KAKENHI Grant-In-Aid 21H04979 and JST ERATO Grant Number JPMJER2301, Japan. }
}
\author{Yuichiro Kamada, Fuhito Kojima, and Akira Matsushita}
\date{\today}

\begin{document}
\maketitle

\input{sections/abst}
\input{sections/intro}
\input{sections/fragmentation_examples}
\input{sections/background}

\input{sections/theoretical_model}

\input{sections/data}
\input{sections/est_results}

\input{sections/counterfactual}
\input{sections/conclusion}

\bibliographystyle{aer}
\bibliography{ref.bib}

\end{document}

%% file: preamble.tex
\usepackage[left=1.0in,right=1.0in,top=1.0in,bottom=1.0in]{geometry}
\usepackage[T1]{fontenc}
\usepackage[dvipsnames]{xcolor}
\usepackage{xurl}
\usepackage[colorlinks=true,urlcolor=blue,citecolor=blue,linkcolor=blue,bookmarks=true,breaklinks=true]{hyperref}
\usepackage{graphicx,lscape,afterpage,bbm,mathtools,amscd,amsmath,amsthm,amssymb,setspace,titlesec,enumerate,enumitem,multirow,multicol,algorithm,algpseudocode,lscape,tabularray,verbatim,chngcntr,array,soul,makecell}
\usepackage[flushleft]{threeparttable}
\usepackage[export]{adjustbox}
\usepackage[at]{easylist}
\usepackage[normalem]{ulem}
\usepackage[capitalise]{cleveref}
\usepackage[labelsep=colon]{caption}

\usepackage[authoryear,round,comma]{natbib}
\input{natbib_setup}

\UseTblrLibrary{booktabs,siunitx,notes}
\def\sym#1{\ifmmode^{#1}\else\(^{#1}\)\fi}

\numberwithin{equation}{section}

\definecolor{color}{RGB}{0,0,0}
\usepackage{subcaption}

\newtheorem{assumption}{Assumption}
\newtheorem{definition}{Definition}

\newcolumntype{C}{>{\centering\arraybackslash}X}

\theoremstyle{definition}

\newcolumntype{f}[1]{D{.}{.}{#1}}
\newcolumntype{d}{S[input-symbols=()]}
\NewColumnType{d}{S[input-symbols=()]}

\newcommand{\mulc}[1]{\multicolumn{1}{@{}c@{}}{#1}}
\newcommand{\mulcb}[1]{\multicolumn{1}{@{}c@{}}{\textbf{#1}}}

\let\oldfootnote\footnote
\renewcommand{\footnote}[1]{\unskip\oldfootnote{#1}}

\newlist{enumerate-mechanism}{enumerate}{1}
\setlist[enumerate-mechanism]{
  label=Mechanism~\arabic*.,
  ref=\arabic*,
  wide=1em,
  leftmargin=1em,
  font=\bfseries
}
\crefname{enumerate-mechanismi}{Mechanism}{Mechanisms}   
\Crefname{enumerate-mechanismi}{Mechanism}{Mechanisms}   

\newcommand{\prob}{\text{Pr}}

\newcommand{\size}[1]{{\left\vert{#1}\right\vert}}

\newcommand{\indep}{\mathop{\perp\!\!\!\!\perp}}
\newcommand{\ind}[1]{{\mathbf{1}_{ \left\{#1\right\} }}}

\renewcommand{\st}{i}
\newcommand{\chil}{i}
\newcommand{\sch}{s}
\newcommand{\St}{I}
\newcommand{\Sch}{S}
\newcommand{\R}{\mathbb{R}}
\newcommand{\n}{\emptyset}
\newcommand{\nset}{\{ \emptyset \}}

\newcommand{\nppt}[1]{\textbf{\small +#1pt}}
\newcommand{\nppts}[1]{\textbf{\small +#1pts}}

%% file: natbib_setup.tex
\usepackage{xparse}
\usepackage{expl3}

\ExplSyntaxOn

\cs_new_protected:Nn \klp_print_author_year:n {%
  \citeauthor{#1}\ (\citeyear{#1})
}

\NewDocumentCommand \citeptlist { m }
 {%
  \seq_set_split:Nnn \l_tmpa_seq { , } { #1 }%
  \int_set:Nn \l_tmpb_int { \seq_count:N \l_tmpa_seq }%
  \int_zero:N \l_tmpa_int%
  \seq_map_inline:Nn \l_tmpa_seq%
   {%
    \int_incr:N \l_tmpa_int%
    \klp_print_author_year:n { ##1 }%
    \int_compare:nNnT { \l_tmpa_int } < { \l_tmpb_int }%
     {%
       \int_compare:nNnTF { \l_tmpb_int } = { 2 }%
        { ~and~ }%
        {%
         \int_compare:nNnTF { \l_tmpa_int } = { \l_tmpb_int - 1 }%
          { ,~and~ }%
          { ,~ }%
        }%
     }%
   }%
 }

\ExplSyntaxOff

\RenewDocumentCommand{\citet}{m}{\citeptlist{#1}}

%% file: sections/abst.tex
\begin{abstract}  Fragmentation of matching markets is a ubiquitous problem across countries and across applications.  In order to study the implications of fragmentation and possibilities for integration, we first document and discuss a variety of fragmentation cases in practice such as school choice, medical residency matching, and so forth.
Using the real-life dataset of daycare matching markets in Japan, we then empirically evaluate the impact of interregional transfer of students by estimating student utility functions under a variety of specifications and then using them for counterfactual simulation. 
Our simulation compares a fully integrated market and a partially integrated one with a ``balancedness'' constraint---for each region, the inflow of students from the other regions must be equal to the outflow to the other areas. We find that partial integration achieves 39.2 to 59.6\% of the increase in the child welfare that can be attained under full integration, which is equivalent to a 3.3 to 4.9\% reduction of travel time. The percentage decrease in the unmatch rate is 40.0 to 52.8\% under partial integration compared to the case of full integration. The results suggest that even in environments where full integration is not a realistic option,  partial integration, i.e., integration that respects the balancedness constraint, has a potential to recover a nontrivial portion of the loss from fragmentation. 
\end{abstract}

%% file: sections/intro.tex
\label{sec:intro}

\section{Introduction}
Many of the most consequential markets in our societies---school admissions, medical resident matching, daycare placements, kidney exchanges---are matching markets, where centralized mechanisms are often employed to improve efficiency and fairness. Over the past several decades, the field of market design has made substantial progress in developing and implementing such mechanisms (e.g., \citet{abdulkadirouglu2003school} for school choice, \citet{roth1984evolution} for medical residency matching, \citet{kamakoji-fair} for daycare placements, and \citet{roth2004kidney} for kidney exchanges). 
In doing so, it has led economists to assume a dual role as both analysts and engineers \citep{roth:02}.

How are those sophisticated mechanisms implemented in practice? 
Typically, these mechanisms are run by individual cities, districts or institutions, and the implementation is usually confined to narrowly defined administrative or political boundaries. These boundaries can reflect long-standing institutional arrangements, localized funding responsibilities, or jurisdictional autonomy. Regardless of their origins, the consequence is that agents on different sides of a boundary are matched as if they participated in entirely separate markets---even when they live mere blocks apart.

The aim of this paper is to study the implications of fragmentation of matching markets and possibilities for integration. To do so, we begin by offering a detailed descriptive account of fragmented matching markets in practice. We observe that fragmentation is prevalent across a variety of settings globally, from public school systems and childcare allocation to medical residency assignments, foster care placements, and public housing markets, among others. We highlight how institutional boundaries and localized governance create fragmented, parallel markets. Each case underscores the potential inefficiencies due to constrained choices caused by fragmentation, motivating our inquiry into mechanisms that can integrate markets effectively.

Against this background, we then investigate public daycare assignment in Japan in detail. Daycare services in Japan are administered by municipal governments, and children---if they wish to attend a daycare---are typically required to attend a daycare within their home municipality. This rule applies even when a facility just across the border is closer, more desirable, or more convenient for the parents' commute than any facilities in the home municipality. Although a 2018 amendment to the Children and Childcare Act encourages intermunicipal admissions---known in Japanese as \emph{ekkyo} (literally, ``crossing boundaries'') matching---, the provision of this option by municipalities  remains extremely limited. This appears to be an especially problematic feature in large urban areas such as Tokyo, as it features a highly fragmented municipal structure in which many children live near administrative borders  and most parents commute across municipal borders.

The potential benefits of interregional coordination are easy to imagine: reduced commuting times, better accommodation of families' needs and preferences,  and fewer unassigned children. However, integration is not simply a matter of merging databases. Municipal governments face real constraints---political, legal, and fiscal. In particular, daycare services are financed from municipal budgets to a large extent, creating strong incentives for municipalities to serve only their own residents. This fiscal structure would generate resistance to intermunicipal admission, even if such admission is desirable from the perspective of children or aggregate social welfare.
Resistance to interdistrict transfer is neither unique to the Japanese context nor specific to daycare allocation. In the context of school choice in the U.S., for instance, schools receive funding based on enrollment, so school districts are concerned about sending their residents to schools outside. In Hebei province of China, an intercity high school admission was abolished in 2024 because the system led some cities to lose high-achieving students to schools in other cities.\footnote{Interestingly, the authorities' concerns about losing students in those two examples are in the opposite direction to those expressed in the context of  Japanese daycare allocation where the concerns are about serving outside students. However, the balancedness constraint we adopt addresses those concerns just as well as it addresses the concern in the Japanese daycare context.}

To reconcile the potential for integration with these political realities, \citet{kamakoji-ekkyo} propose a new class of mechanisms that satisfy a 
``balancedness constraint'' in a model where students and schools are partitioned into ``regions'': for each region, the number of incoming students from other regions must equal the number of outgoing students to other regions. In the Japanese daycare context, for instance, this constraint  implies that each municipality's burden incurred by children coming from other municipalities would be compensated by the same number of its own residents taken care of by daycares outside of the municipality.   This paper builds on that theoretical framework and estimates the empirical effects of both unconstrained, or ``full,'' integration as well as ``partial'' integration by balanced matching using real-world data from two distinct geographical settings, namely Tokyo City and Koriyama City in Japan.

We structurally estimate the preferences of children and use them for counterfactual simulations of alternative matching mechanisms.\footnote{We acknowledge that the preferences are an aggregation of those of the family members in reality, but we use the term ``child'' throughout to simplify the exposition.} For Tokyo City, we use rank-ordered lists (ROLs) submitted to   Bunkyo Ward---one of the 23  municipalities in Tokyo City---in  2022 and estimate the children's preferences over daycare  attributes, particularly focusing on the disutility of  traveling, under various behavioral assumptions.\footnote{By ``Tokyo City,'' we refer to the collection of 23 special wards of Tokyo (called ``Tokyo-to Kubu'' or ``Tokyo 23 ku'') which comprises the central urban part of Tokyo, not the Tokyo Prefecture that contains more regions than the 23 wards. Tokyo City was once a single municipality administering the area now covered by the special wards of Tokyo, but  it was divided into the current 23 wards in 1947.} 
Then, in our counterfactual analysis, 
we simulate the outcomes of partial integration by balanced matching. 
Moreover, we also simulate the outcomes under \emph{full integration}, i.e., in case a centralized clearinghouse can match children to daycares in the 23 wards of Tokyo without the balancedness constraint. 

To complement this analysis, we conduct a parallel exercise using data from Koriyama City, which is administered as a single municipality despite covering a geographic area roughly as large as Tokyo's 23 wards combined. Because Koriyama's current matching process is city-wide, we are able to observe preferences over all available daycares. This allows us to estimate utility functions with arguably better precision and to simulate the effects of hypothetical fragmentation---dividing the city into districts corresponding to municipalities before they merged to form the present Koriyama City---with separate matching processes. The analyses of  Tokyo and Koriyama thus enable a two-sided investigation: what happens when we integrate a fragmented market, and what happens when we fragment an integrated one? 

Our findings demonstrate that market integration improves outcomes, and a nontrivial portion of the improvement under full integration can be achieved  through partial integration. In Tokyo, under intermunicipal admissions between wards, partial integration achieves 41.5\% of the increase in the child welfare that can be achieved under full integration. This is equivalent to a 3.5\% reduction in travel time in welfare terms.\footnote{We test two specifications for partial integration, which we detail later, and here we report the numbers for one of such specifications.}  Furthermore, the decrease in the unmatch rate  under partial integration is 41.9\% of the one under full integration (5.5\% compared to 13.1\%). For Koriyama, we run an analogous analysis, considering full and partial integration from the hypothetical status quo of the fragmented market. We find that partial integration achieves 59.6\% of the increase in the child welfare that can be achieved under full integration, and this is equivalent to 4.9\% reduction in travel time in welfare terms. 
The decrease in the unmatch rate  under partial integration is 52.8\% of the one under full integration (3.7\% compared to 7.0\%). 
We also quantify the extent to which the benefit of integration is concentrated for children living near regional borders. We find that the welfare gain is not only due to the improvement of children who live close to the boundary, but it is spread out more widely through the seats freed up by the interregional transfer.

From a broader perspective, this paper contributes to the growing literature on matching with constraints \citep{kamakoji-basic}. Recent work has explored how to incorporate distributional objectives, diversity goals, and policy constraints into matching mechanisms. Our contribution lies in developing and empirically evaluating a method for achieving  interregional integration in a politically feasible way. We view the balancedness constraint not only as a technical device but as a formalization of real-world political acceptability. It allows us to design mechanisms that are not just theoretically optimal under unrealistic assumptions but implementable in environments where governance is decentralized and local autonomy is paramount.

Our work also contributes to the expanding empirical market design literature that estimates participants' underlying preferences from strategically submitted ROLs.
In the daycare matching markets of both Bunkyo Ward and Koriyama City, the number of daycares each child can rank is capped at five  and ten,  respectively, as in many school choice applications. Under such constraints, the employed mechanisms fail  to be strategy-proof: children can sometimes improve their chances of obtaining a more preferred daycare by strategically reporting their preferences. To recover underlying utility functions despite this strategic behavior, we utilize the framework of \citet{fack2019beyond} that imposes several behavioral assumptions, such as stability  of the matching, to generate moment conditions that identify preferences without requiring a full model of belief formation. 
In addition to strict truth-telling, weak truth-telling, and undominated strategies and stability assumptions they used, we additionally introduce what we call an ``optimistic expectation truth-telling'' assumption, under which children optimistically predict the cutoff scores of daycares and exclude those they believe they cannot possibly gain admission to. 
Overall, we estimate sixteen candidate models---each corresponding to a different modeling assumption---to identify children's utility parameters.\footnote{We have four behavioral assumptions and four utility specifications, resulting in sixteen models.} For our counterfactual exercises, we select the best-fitting specification, which turns out to be one under optimistic expectation truth-telling.

The remainder of this paper is organized as follows. Section~\ref{sec:fragmentation_examples} gives a descriptive account of fragmentation in a variety of matching markets. Section~\ref{sec:background} describes the institutional background of daycare matching in Japan. Section~\ref{sec:background-ekkyo} introduces the theoretical model. Section~\ref{sec:data} presents the data and estimation methods. Section~\ref{sec:est_results} discusses empirical results. Section~\ref{sec:simulation} simulates the effects of various integration scenarios. Section~\ref{sec:conclusion} concludes with broader implications and directions for future research.

%% file: sections/fragmentation_examples.tex
 \section{Fragmented Matching Markets in Practice}
\label{sec:fragmentation_examples} 
Matching markets in the real world are often fragmented across multiple jurisdictions or institutions, which means that separate matching mechanisms operate in parallel rather than through a single  unified  clearinghouse. Such fragmentation is prevalent in many domains, leading to constrained choices for participants and potential inefficiencies. We illustrate this phenomenon with examples from school admissions, childcare allocation, medical residencies, foster care, and public housing, among others, drawing on documented cases in various countries. Each example underscores how limited integration can hinder the efficiency of outcomes and highlights recent developments or policies in response to these challenges.
We also point out that policymakers trying to integrate matching markets often face resistance which limits the extent of integration.

\subsection{School Admissions} Public school choice processes are frequently organized at a local level, with each city or district running its own admissions system rather than a larger regional or national system \citep{kamakoji-ekkyo,kamakoji-integration}. For instance, in the United States,  children often must apply separately to different districts and charter schools due to uncoordinated enrollment systems. 
In China, high school admissions are typically administered at the city level with virtually no intercity transfers allowed; one province (Hebei) even abolished a policy allowing intercity high school choice in 2024 due to concerns that elite students disproportionately left certain cities, thereby shutting down the gains from a broader market. 
These cases illustrate how the scope of school choice is artificially narrowed by jurisdictional boundaries and also how a fully integrated school choice system might be difficult in practice. Still, integration has been implemented: New Orleans, for instance, runs a centralized enrollment system called OneApp that allows children to apply to multiple schools, including not only  traditional public schools but also charter schools \citep{abdulkadiroglu2017minimizing}. Evidence of efficiency gain from integration has been provided by \citet{klein2024school}, for instance, who study integration of school districts in Hungary and find significant welfare improvement for students.

To partially address fragmentation, many U.S. states have introduced interdistrict open enrollment policies, allowing students to attend public schools outside their residential districts. As of the early 2020s, 43 states had some form of interdistrict choice in place.\footnote{\url{https://www.ecs.org/open-enrollment-policies}.} These programs expand  children's choice sets beyond their residential district, partially integrating otherwise isolated local matching markets. However, their effectiveness is limited, e.g.,  in states where open enrollment is voluntary, many districts choose not to accept non-resident students or selectively admit high-achieving applicants, exacerbating inequalities.\footnote{
\url{https://reason.org/commentary/why-open-enrollment-laws-that-let-public-schools-reject-transfer-students-arent-good-enough/} and \url{https://files.eric.ed.gov/fulltext/ED628507.pdf}.
}

From a market-design perspective, these interdistrict enrollment systems relate closely to formal matching theory. \citet{hafalir2018interdistrict} provide a theoretical framework for interdistrict school choice, proposing mechanisms that yield stable  student assignments across district borders. They find that conditions under which such mechanisms exist are quite restrictive. Consistent with their analysis, interdistrict choice policies currently serve as important but only imperfect solutions to the broader fragmentation problem in school admissions.

\subsection{Childcare and Daycare Allocation} 

Fragmentation is also pronounced in early childhood education and childcare markets \citep{kamakoji-ekkyo}. In Japan, daycare slots are allocated separately by each municipality: for example, Tokyo City is divided into 23 wards, each conducting its own matching process for daycare placements. Under this local residency rule, a child is typically eligible only for daycares in their home ward.  However, many children live near ward boundaries or their parents commute across them, so their most preferred daycare options may lie outside their residential ward. Although attending a daycare in a neighboring municipality is legally possible, it is exceedingly rare in practice under the current system. This means children often cannot access a nearby daycare if it falls under a different local authority's domain, leading to inefficient outcomes such as unfilled spots in one ward and unmet demand in a neighboring ward.   The fragmentation of  childcare markets is not unique to Japan. Major Chinese cities run kindergarten admissions at the district level, with some cities outright prohibiting interdistrict enrollment and others allowing it only in limited circumstances. In Beijing, for example, each of the 16 districts admits children independently; this uncoordinated approach can result in the same child receiving multiple offers from different districts while other children get none, a clear inefficiency of parallel matching processes. A few policymakers have experimented with partial integration: in one notable case, the adjacent cities of Yokohama and Kawasaki in Japan opened certain daycares to each other's residents by reserving a quota of spots for non-locals. While this arrangement made some intercity matching possible, the fixed quotas created their own imbalances (analogous to the Chinese college admissions quotas discussed below) and did not fully eliminate inefficiency. Overall, the childcare domain illustrates how fragmenting a market by locality limits children's  options and can leave capacity underutilized. Consolidating or coordinating these matching markets---even if only partially---has the potential to improve child welfare by widening children's choice sets.

\subsection{Medical Residency Matches} 
The assignment of medical residents to hospitals is another context where multiple matching markets operate in parallel. In the United States, the National Resident Matching Program (NRMP) provides a centralized match for most residency positions nationwide. However, several specialties historically have  run matching mechanisms separately, outside of the NRMP. Ophthalmology and urology are prime examples: 
rather than  participating in the main NRMP system, each uses its own matching process (the SF Match for ophthalmology and a specialty match for urology) that runs earlier than the NRMP matching each academic year \citep{massenzio2023navigating}. As a result, medical graduates interested in those fields and the fields participating in the NRMP must navigate distinct matching processes with different timelines and rules. 
Until recently, there was also a separate match for osteopathic medical residencies in the U.S., run by the American Osteopathic Association (AOA). This parallel AOA match operated alongside the NRMP match for decades, meaning osteopathic students and programs matched in a separate pool. In 2020, the two systems were unified under a single accreditation and matching system, finally ending the split process \citep{almarzooq2021single}. The existence of multiple specialty-specific matches  can lead to inefficient outcomes. For example, applicants who  succeeded  to match in an early specialty-specific process might have missed opportunities in the main match because  assignment  across the two were not coordinated. Similarly, residency allocation in other countries are also  fragmented: for example, until the early 2010s,  the United Kingdom's postgraduate medical training placements were handled through regional hiring rather than a unified national match, a practice that was reformed (amid much controversy) to a centralized nationwide system.\footnote{See \url{https://www.medschools.ac.uk/media/2578/aspiring-to-excellence-report.pdf} for a failed attempt at establishing a nationwide matching market in the 2000s.  Since the 2010s, the U.K. has been using a national matching system called Oriel: \url{https://www.nwpgmd.nhs.uk/oriel} and \url{https://www.oriel.nhs.uk/Web}.} These examples show that even in highly organized markets like physician labor, fragmentation along specialty or institutional lines can pose challenges for participants. Recent moves toward integration suggest recognition of the benefits of a unified matching market in medicine.

\subsection{Foster Care Placements} 
Foster care placement systems often exhibit fragmentation at the level of small administrative units. In Los Angeles County, for instance, placements are managed by nineteen regional offices within the Department of Children and Family Services. This sub-county fragmentation means that a child may go unmatched in one region despite the availability of suitable foster homes in neighboring regions, even within the same county. Such localized matching limits the scope of choice for both children and prospective foster parents, leading to significant inefficiencies. \citet{robinson2019gets} analyzes this structure and finds that assigning placements at the county level, rather than within individual regions,  would successfully  decrease the average number of placements per child by 8\% and reduce the average distance between foster homes and children's schools by 54\%. These findings underscore the potential benefits of integrating placement processes across regional boundaries to improve outcomes in foster care systems.
Recognizing these challenges, some child welfare agencies have initiated efforts to enhance intercounty cooperation and develop centralized databases of foster families. 
For example, the Inter County Transfer (ICT) Protocol in California provides guidelines to facilitate the transfer of case records for dependents between counties, aiming to streamline processes and improve placement outcomes.\footnote{\url{https://www.calsaws.org/wp-content/uploads/2022/01/CIT-0291-21-ICT-Communication-Protocol-Baseline.pdf}} However, such initiatives are not yet widespread. For instance, the Interstate Compact on the Placement of Children (ICPC) provides a legal framework for interjurisdictional placements, but its implementation varies across states and often involves complex procedures.\footnote{\url{https://www.myflfamilies.com/sites/default/files/2025-04/ICPC\%20-\%20A\%20Manual\%20and\%20Instructional\%20Guide\%20for\%20Juvenile\%20and\%20Family\%20Court\%20Judges.pdf}} 

\subsection{Public Housing Allocation} The allocation of public housing and housing vouchers offers another clear example of fragmented matching in practice. In the United States, rather than a single national or even statewide system, public housing assistance is administered through hundreds of local housing authorities, each with its own waiting lists and selection processes: a low-income family seeking housing assistance typically must apply separately to a waiting lists of each public housing agency (PHA) in each city or county of interest. These lists are not integrated: a family might be on the waitlist in a city but that has no bearing on their status in another city. Indeed, even within the same locality, PHAs often maintain distinct waiting lists for different programs (e.g., one list for public housing units and another for Housing Choice Vouchers).\footnote{ \url{https://www.hud.gov/sites/dfiles/PIH/documents/PHOG_Waiting_List_Chapter.pdf}}
This fragmented approach can lead to misallocation in several ways. Families might sit idle on a long waitlist in one city while public housing units remain vacant in another city for which the family did not or could not submit an application. Applicants bear greater burdens in managing applications across multiple PHAs or programs, which favors those with more resources and information, thereby raising concerns about equity. 
Some regions have attempted reforms: a few metropolitan areas have created centralized or joint waiting lists that cover multiple housing authorities, aiming to give applicants a one-stop application for a broader area. For example, a regional waiting list in the Boston area allows applicants to be considered for public housing across several neighboring towns.\footnote{\url{https://www.section8listmass.org}} However, such initiatives are rather  rare.  By and large, public housing allocation markets remain highly  fragmented. Official guidelines acknowledge that a PHA operating across multiple municipalities may keep separate waiting lists for each locality, though this can raise issues of fragmentation in housing allocations.\footnote{\url{https://www.law.cornell.edu/cfr/text/24/982.204}} \citet{bloch2020matching} propose mechanisms for integrated public housing assignment across multiple housing authorities.

\subsection{Other Contexts} Many other matching markets exhibit similar fragmentation. A notable example is kidney exchange for transplant patients with incompatible donors. In principle, a single   exchange pool yields the best chance to find compatible kidney donor-recipient matches. In practice, the kidney exchange market has long been split among multiple programs and hospitals. In the U.S., several independent kidney exchange clearinghouses operate concurrently, and numerous hospitals conduct internal exchanges only among their own patients \citep{ashlagi2014free,agarwal2019market}. This lack of integration means a patient-donor pair at one hospital might go unmatched even though a compatible donor exists at another hospital's registry, simply because the two pools are not merged. The fragmentation of kidney exchange has been identified as a source of inefficiency and fewer transplants, and efforts are underway to encourage hospitals to participate in larger nationwide programs. 
Kidney exchange markets could be integrated at an international level as well, leading to the so-called global kidney exchange. The proponents of global kidney exchange such as the Alliance for Paired Kidney Donation in the U.S. and the European Network for Collaboration on Kidney Exchange Programmes in Europe point out that it could achieve efficiency gain from kidney exchanges across national borders, but opponents raise concerns such as the possibility of detracting from the transplantation community's ethical norm of self-sufficiency \citep{ambagtsheer2020global}.

Exchange markets exist for other organs beyond kidneys such as livers and lungs \citep{ergin2020efficient,ergin2017dual}. 
Although theoretical advancement has been made for the possibility of integrating the markets for different organs \citep{dickerson2017multi,watanabe2022multiorgan,anand2025Multi} and there are real examples in which a multi-organ exchange took place, these markets have mostly operated independently and remain fragmented.\footnote{See  \url{https://www.cmu.edu/news/stories/archives/2019/may/kidney-liver-swap.html} for a real case of a multi-organ exchange.}

Similar issues arise in college admissions as well: in China's university placement system, each province runs its own admission quota for local and non-local students, effectively segmenting the national higher education market by province \citep{kamakoji-ekkyo}. This quota system, while ensuring regional representation, has been noted to cause inefficiencies---some colleges cannot fill seats with qualified students from one region even as high-scoring candidates from other regions are rejected due to quota limits.  

%% file: sections/background.tex
\section{Daycare Matching in Japan}
\label{sec:background}

Daycare services in Japan are administered at the municipal level. Children are eligible to attend daycares from two months old until they enter elementary school at the age of six.
Parents who wish to enroll their children in a licensed daycare---whether public or private---must submit an application to the centralized municipal matching clearinghouse in their municipality of residence.\footnote{By ``licensed daycare'' we refer to \emph{Ninka Hoiku En}, a childcare facility licensed by the prefectural governor under the Children and Childcare Act. In the formal analysis, we simply use the term ``daycare'' to mean ``licensed daycare.''} Although a 2018 amendment to the Children and Childcare Act encourages enrolling children in daycares outside their home municipality (a practice known as \textit{ekkyo} matching), most placements still occur locally. For example, for Bunkyo Ward and Koriyama City, the two municipalities that we analyze,  6.7\% and 2.8\% of their applicants live outside the respective municipalities.

Each municipality conducts a matching process every month, allocating available daycare seats to the children who made the applications.\footnote{If a child is not matched to any daycare, they remain on a waiting list until the next monthly matching cycle or until a position opens up. For our analysis, we focus on a single admission process and therefore treat all such children simply as ``unmatched.''} 
Yet, the majority of children are placed in one session corresponding to the start date in April, which is the beginning of the academic year in Japan.\footnote{
Among children enrolling in daycare for the first time at their current facility, 59.2\% began in April, with the next highest month accounting for only 4.6\%. See  \url{https://www.mhlw.go.jp/content/11900000/000756549.pdf}.
}

In most municipalities including Bunkyo Ward and Koriyama City, a matching is determined by  \textbf{Serial Dictatorship}. In  Serial Dictatorship, there is a single priority order over the children  in the municipality, which is strict and based on a score that reflects the family's need for daycare services. The child with the highest priority is matched to a daycare (or the outside option) which she prefers most, the child with the second-highest priority is matched to a daycare (or the outside option) which she prefers most among those that still have capacity, and so forth. 
The capacity of each daycare is set for each age, and the matching process is conducted separately for children at each age.\footnote{In what follows, when we refer to the age of a child, we consider the age as of the beginning of April.}

\begin{table}[ht]
\centering
\caption{Example of priority scores and priority orders in Bunkyo Ward}
\label{tab:3_priority_scores}
\input{tables/section3/sample_priority_scores_Bunkyo_3}
\end{table}

Priority criteria differ by municipality, but common factors include the parents' employment status (defined by hours per day and days per week), the presence of siblings already enrolled in some facility, and the availability of grandparents who can provide childcare. 
Many children share the same priority scores.\footnote{In both municipalities,  more than 20\% of the scores are concentrated at a single point mass. See \Cref{fig:cutoff_socres_in_ROL} for the distribution of scores in each municipality.} 
When multiple children have the same score, the municipality uses additional child and family characteristics to specify a suborder.  These characteristics  include the number of days the  parents have lived in the municipality and the amount of income-based residence tax paid in the previous year, which in practice allows no ties. The suborder based on those characteristics is combined with the priority score to determine  a strict ordering of children. 
\Cref{tab:3_priority_scores} uses a hypothetical example to illustrate how the priority order is determined by the scores and suborders in Bunkyo Ward. For instance, a child with both parents working full-time and no grandparents living nearby would usually receive a higher score than a child with parents having more flexible working arrangements or grandparents available to help. 

\begin{figure}[tbhp]
\centering
\includegraphics[width=9cm
]{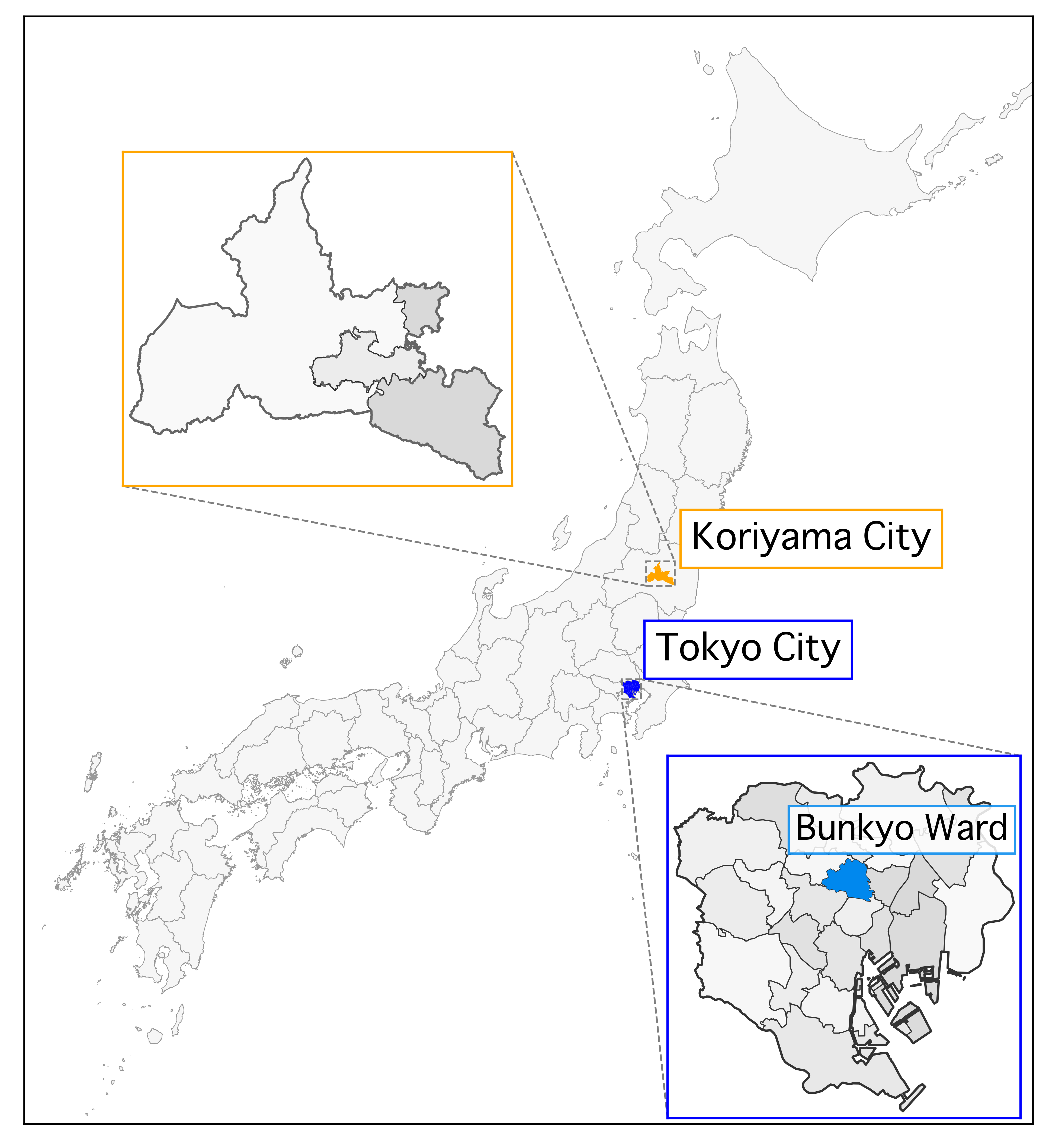}
\caption{Locations of Tokyo City and Koriyama City. The enlarged maps of Tokyo and Koriyama are shown at the same scale. Tokyo City is divided into 23 wards, with Bunkyo Ward located roughly at its center. Koriyama City is a single municipality now, while it used to be organized as three distinct municipalities; the boundaries illustrated on Koriyama's map represent the borders of those former municipalities.
} 
\label{fig:3_JP_map}
\end{figure}

Our analysis covers two regions: Tokyo City and Koriyama City, whose locations are illustrated in \Cref{fig:3_JP_map}. The two cities have similar sizes in terms of area (\Cref{tab:3_city_size}). However, Tokyo City is a dense urban center, whereas Koriyama City is a less-populated suburban municipality. In the following subsections, we outline each region's demographic and institutional characteristics.

\input{tables/section3/tab_background_city_comparison}

\subsection{Tokyo City}\label{sec:background-tokyo}

\begin{figure}[tbp]
\centering
\hspace{10em}
\includegraphics[width=12cm]{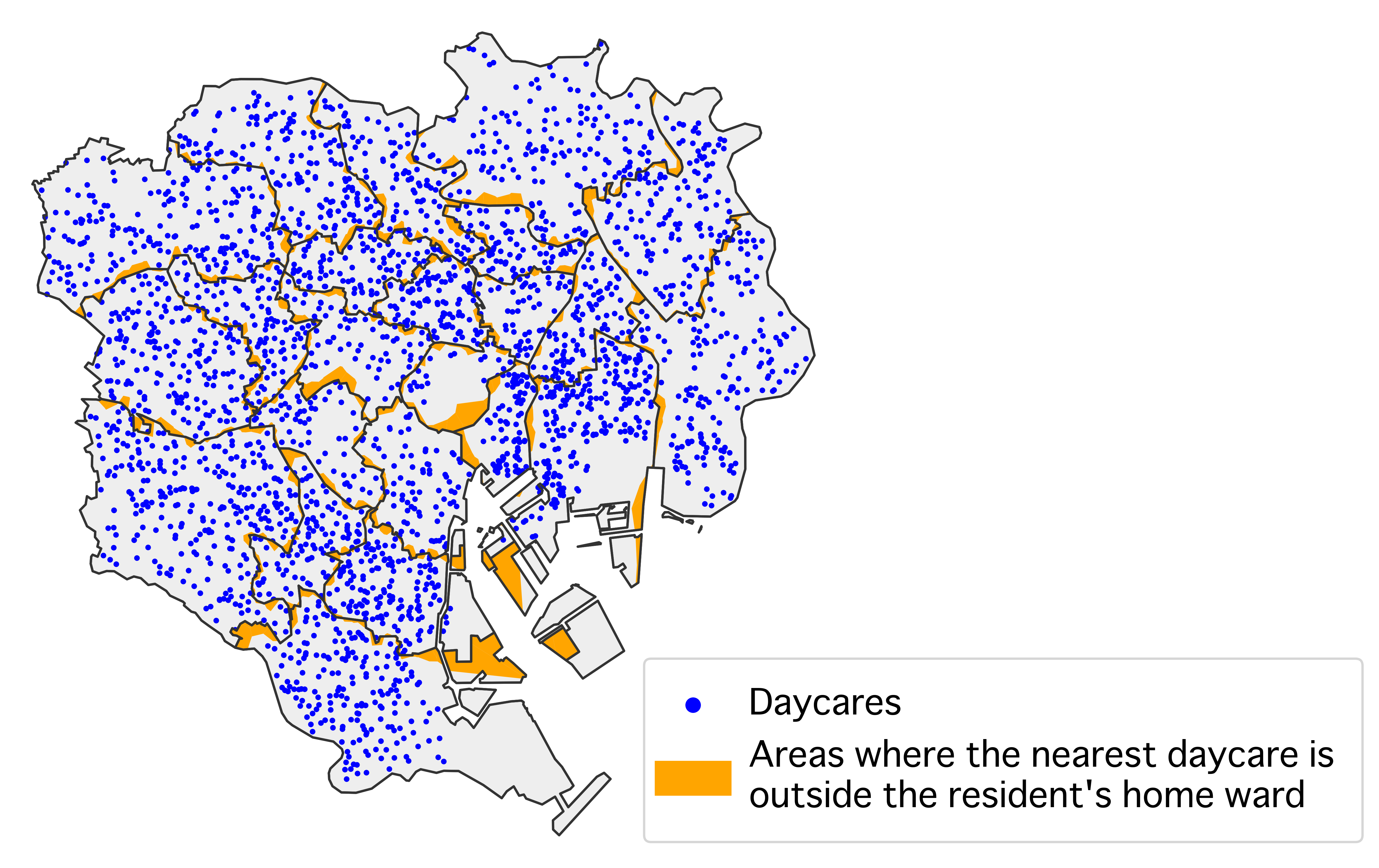}
\caption{Location of daycares in Tokyo City. Each blue dot represents a daycare. Orange-shaded areas indicate the places where the nearest daycare is located outside the resident's home ward, covering 4.29\% of the population at age 0-4. Population figures are drawn from 2020 Population Census in Japan (\citeyear{kokuseichousa2020}), in which population data are available only in five-year age bands (such as ages 0-4, ages 5-9, etc.).}
\label{fig:3_voronoi_tokyo}
\end{figure}

Tokyo City---the nation's capital---is divided into 23 special wards, each operating as an independent municipality. These wards are relatively small, 27.04 square kilometers on average, and each one operates its daycare matching system independently. Because each ward is very small, many children find that their closest daycare lies outside their home ward. The orange regions in \Cref{fig:3_voronoi_tokyo} highlight where the nearest daycare falls beyond ward limits. 
Additionally, since a large number of residents commute across ward boundaries, there may also be significant demand for intermunicipal matching by those living far from municipal borders as well.%
\footnote{For example, in Bunkyo Ward, 65.3\% of employed residents commute to jobs outside the ward, while 78.3\% of people working in Bunkyo Ward live elsewhere. \textit{Source}: 2020 Population Census in Japan (\citeyear{kokuseichousa2020}).}

In terms of scale, the 23 wards administer over 2,800 licensed daycares in total, collectively serving more than 220,000 children. These figures reflect the extensive demand for childcare services in Tokyo's urban environment and underscore the potential benefits of reducing administrative fragmentation to improve overall childcare accessibility.

We obtained the submitted rank-ordered lists (ROLs) from Bunkyo Ward, located at the center of Tokyo City. 
While our goal is to assess the impact of integrating multiple daycare markets, it is difficult to directly observe children's preferences over daycares outside their municipality of residence, since such preferences are rarely elicited under the current system. To address this challenge, we first estimate the underlying utility function using the ROL data obtained from Bunkyo Ward. We then conduct counterfactual simulations assuming that children are also allowed to apply to daycares outside their home ward. This enables us to evaluate the potential effects of market integration.

\subsection{Koriyama City}\label{sec:background-koriyama}

\begin{figure}[tbhp]
\centering
\hspace{6em}
\includegraphics[width=12cm]{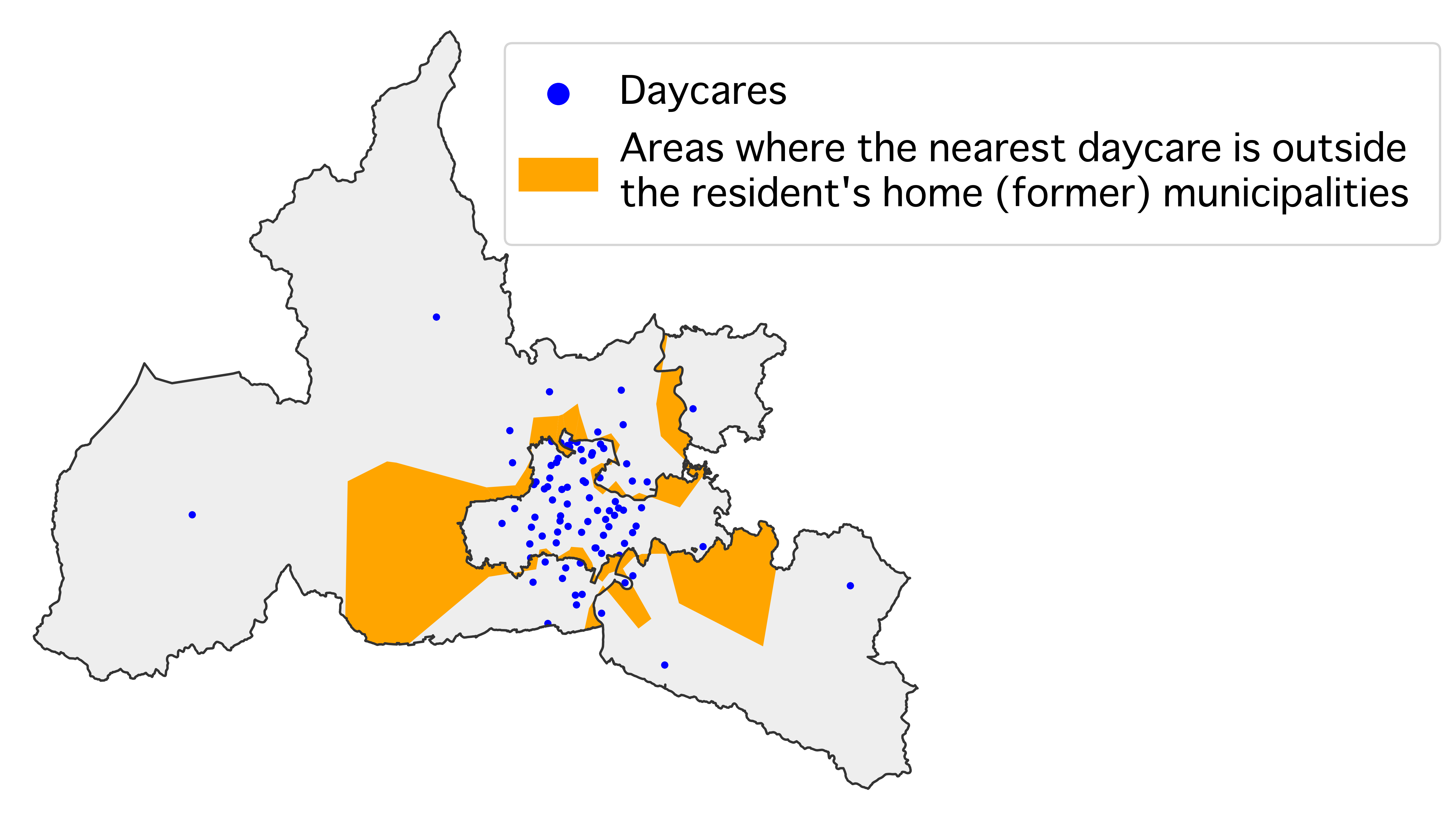}
\caption{Location of daycares in Koriyama City. Each blue dot represents a daycare. Orange-shaded areas indicate the places where the nearest daycare is located outside the resident's home (former) municipalities, covering 11.67\% of the population at age 0-4.}
\label{fig:3_voronoi_koriyama}
\end{figure}

Koriyama City is located in Fukushima Prefecture in  Northeastern Japan. Its area covers about 757 square kilometers, making it similar in size to Tokyo City, which encompasses about 623 square kilometers. Unlike Tokyo City, Koriyama City is administered as a single municipality. Thus, despite its large territory, the city operates a unified daycare matching system under a single local government. 
As of 2024, Koriyama City maintains 89 licensed daycares and provides daycare services to 5,533 children, exhibiting overall scale and distribution very different from those found in Tokyo City. 

To understand the impact of fragmentation on daycare allocation and efficiency, we artificially divide Koriyama City into three smaller administrative regions, each with its own daycare matching system, much like Tokyo's wards.
More specifically, we consider a counterfactual situation in which the city is partitioned into three regions that used to be municipalities on their own. 
\footnote{The former municipalities were Asaka County, Tamura County, and Koriyama City (which then occupied the central area shown in \Cref{fig:3_voronoi_koriyama}). These three areas used to be municipalities on their own, while the two counties were later divided into twelve municipalities. Then in 1965, Koriyama City absorbed those municipalities from Asaka and Tamura Counties,  forming the current Koriyama City.
} \Cref{fig:3_voronoi_koriyama} maps the distribution of daycares across Koriyama City. The orange regions illustrate areas where the nearest daycare lies beyond the former municipal boundaries. 

We obtained the submitted ROLs from Koriyama City. An advantage of this dataset over Tokyo's is that it captures how strongly children prefer daycares outside their former home municipalities because the city government currently runs a unified daycare matching system throughout the city. 
By comparing the results of the simulated fragmentation to those of the current unified system, it becomes possible to analyze how dividing a large municipality into smaller units affects daycare accessibility and efficiency.   This approach can help policymakers assess the potential benefits or drawbacks of restructuring daycare admissions and inform decisions on how best to manage childcare services in large city areas.

%% file: tables/section3/sample_priority_scores_Bunkyo_3.tex
\footnotesize
\begin{tblr}{
    width = \linewidth,
    colspec = {X[c,0.7] X[c,0.6] X[c,1.3] X[c,1.3] X[c,1] X[c,1.2] X[c,1.2] X[c,0.9]},
    row{1,2} = {font=\bfseries},
    hlines
}
\SetCell[r=2]{c} Priority Order & \SetCell[r=2]{c} Score & \SetCell[c=5]{c} Score Breakdown &&&&& \SetCell[r=2]{c} Suborder \\ 
& & Parent 1 Employment & Parent 2 Employment & {Twins/\\Siblings} & Grandparents Nearby & Residence & \\ 
1 & 31 & Full-time \nppts{10} & {Single parent\\ \nppts{13}} & Twins \nppts{2} & Not Nearby \nppt{1} & Resident \nppts{5} &  \\ 
2 & 28 & Full-time \nppts{10} & Full-time \nppts{10} & Siblings in daycares \nppts{2} & Not Nearby \nppt{1} & Resident \nppts{5} &  \\ 
3 & 21 & Full-time \nppts{10} & Part-time (4–6 hours/day, 3 days/week) \nppts{6} & & & Resident \nppts{5} & {{\scriptsize Resident is prioritized over commuters}\\1st\\(in 21pts)} \\ 
4 & 21 & Full-time \nppts{10} & Attending grad school \nppts{8} & & Not Nearby \nppt{1} & Commuter \nppts{2} & {2nd\\(in 21pts)} \\ 
5 & 20 & Full-time \nppts{10} & Unemployed \nppts{5} & & & Resident \nppts{5} &  \\ 
\end{tblr}

\begin{tablenotes}
\renewcommand{\baselinestretch}{1.0}
\footnotesize
\item \textit{Notes:} This table presents an example of how the priority  order is determined for five hypothetical children in Bunkyo Ward.
The priority order is determined first by arranging children in descending order of their scores; if two or more children have the same score, then a suborder is determined for those children.
The  suborder  is determined  lexicographically using 22 criteria. For example, the first criterion is being a Bunkyo Ward resident; the second is  having no delinquency in daycare fees; the third is receiving public assistance. Additional criteria include several detailed factors such as the length of the parents' residency in Bunkyo Ward and the amount of income-based resident tax paid. In practice, these criteria ensure a strict suborder among all children with the same score.
\end{tablenotes}

%% file: tables/section3/tab_background_city_comparison.tex
\begin{table}[tbhp]
\centering
\caption{Comparison of Tokyo City and Koriyama City}
\label{tab:3_city_size}
\small

\begin{talltblr}[
note{}={\textit{Notes:} The data for Tokyo City (and Bunkyo Ward) are as of 2022. Those in Koriyama City are as of 2024.},
label={none},
]{
  width={0.9\linewidth},
  colspec={l *{3}{X[r,1]}},
}
\toprule
& \SetCell[c=2]{c}{\textbf{Tokyo}} && \SetCell[r=2]{c}{\textbf{Koriyama}} \\
\cmidrule{2-3}
& \SetCell[c=1]{c}{\textbf{All}} & \SetCell[c=1]{c}{\textbf{Bunkyo}} & \\
\midrule
Area (km$^2$)             & 622.7   & 11.3   & 757.2  \\
Population at ages 0-5     & 413,261 & 11,839 & 13,052 \\
Number of daycare users   & 227,163 & 6,253  & 5,533  \\
Number of daycares &   2,817 &   127  &    89  \\
\bottomrule
\end{talltblr}

\end{table}

%% file: sections/theoretical_model.tex
\section{Model of Daycare Matching}
\label{sec:background-ekkyo}

This section outlines the model and the ``FIG cycles algorithm'' developed by \cite{kamakoji-ekkyo}, and relates them to our daycare matching market. Following the standard terminology in the school choice literature, we refer to children (applicants) as ``students'' and daycares as ``schools'' in this section.

\input{sections/model}
\input{sections/fig_cycles}
\input{sections/model_daycare_market}

%% file: sections/model.tex
\subsection{Model}
\label{sec:model}

Let there be a finite set of students $I$ and a finite set of schools $S$. 
Each student has a strict preference relation $\succ_\st$ over the set of schools and the outside option (which she receives if she is unmatched), denoted $\n$. 
For any $\sch, \sch' \in S \cup \{\n\}$, we write $\sch \succeq_\st 
\sch'$ if and only if $\sch \succ_\st \sch'$ or $\sch = \sch'$. Each school $\sch$
has a strict priority order $\succ_\sch$ over the set of students and leaving a position vacant (which is denoted by $\n$). 
For any $\st, \st' \in I \cup \{\n\}$, we write $\st \succeq_\sch 
\st'$ if and only if $\st \succ_\sch \st'$ or $\st = \st'$. 
Each school $\sch\in S$ is endowed with a  \textbf{capacity} $q_{\sch}$, which is
a nonnegative integer. 

School $\sch$ is said to be \textbf{acceptable} to student $\st$ if $\sch \succ_\st \n$. Similarly, student $\st$ is acceptable to school $\sch$ if $\st \succ_\sch \n$. 
We often write only acceptable partners to
denote preferences and priorities. For example,
$\succ_{\st}: \sch,\sch'$
means that school $\sch$ is the most preferred, $\sch'$ is the second most preferred, and $\sch$ and $\sch'$ are the only acceptable schools under preferences $\succ_\st$ of student $\st$. 

A \textbf{matching} $\mu$ is a mapping that satisfies (i) $\mu_\st\in S\cup\{\n\}$ for all $\st\in I$, (ii) $\mu_{\sch} \subseteq I$ and $|\mu_{\sch}|\leq q_{\sch}$ for all
$\sch \in S$, and (iii) for any $\st \in I$ and $\sch \in S$, $\mu_\st=\sch$  if and only if $\st \in \mu_{\sch}$.
 That is, a matching simply specifies which student is assigned to which school (if any).

A matching is \textbf{individually rational} if no student or school is matched with an unacceptable partner. 
Given  a matching $\mu$, we say that a student $\st$ has \textbf{justified envy}  to another student $j$ if there is a school $\sch \in S$ such that
(i) $\mu_j=\sch$, 
(ii) $\sch\succ_\st \mu_\st$, and 
(iii) $\st \succ_\sch j$. We say that a matching $\mu$ is \textbf{fair} if there is no pair of students $(i,j) \in I^2$ such that $i$ has justified envy to $j$. 
Finally, 
a matching $\mu$ \textbf{weakly Pareto dominates} a matching $\mu'$ if 
$\mu_\st \succeq_\st \mu_\st'$ for every $\st \in I$, and  $\mu$ \textbf{Pareto dominates}  $\mu'$ if
$\mu$ weakly Pareto dominates $\mu'$ and 
there exists $\st \in I$ such that 
$\mu_\st \succ_\st \mu_\st'$. 

\subsubsection{Regions, Balancedness, and Efficient iBF}\label{efficient-ibf}
Each student and each school belong to exactly one region. Formally, a set of regions $R$ is a partition of $I\cup S$ that satisfies the following conditions:
\begin{enumerate}
\item Each region $r\in R$ is a nonempty subset of $I\cup S$.
\item Regions are disjoint: $r\cap r'=\emptyset$ for any $r,r'\in R$ such that $r\neq r'$.
\item All students and schools belong to some region: $\bigcup_{r\in R}r=I \cup S$.
\end{enumerate}

\begin{definition}\rm A matching $\mu$ is \textbf{balanced} if for each $r\in R$,
\begin{equation*}\label{eq:feasible}\underbrace{\sum_{\sch\in r}|\{\st|\st\in\mu_\sch, i\not\in r\}|}_{\text{inflow to }r}=\underbrace{\sum_{\sch\not \in  r}|\{\st|\st\in\mu_\sch, i\in  r\}|}_{\text{outflow from }r}.\end{equation*}
\end{definition}

As its defining equality shows, balancedness means that, for each region $r$, the number of students attending a school in region $r$ while not living in region $r$  equals the number of students attending a school outside region $r$ while living in region $r$.
Note that balancedness is not a ``pairwise'' concept; that is, it does not require that for every pair of regions $r$ and $r'$, the number of students who live in $r$ and are matched to a school in $r'$ is the same as the number of students who live in $r'$ and are matched to a school in $r$. Instead, it allows for more complex exchanges involving more than two  regions.  For instance, balancedness permits three-way exchanges, such as the case where two students from region $r_A$ attend schools in region $r_B$,  two students from $r_B$ attend schools in region $r_C$, and two students from $r_C$ attend schools in $r_A$.

We say that a matching is an \textbf{iBF} if it is individually rational, balanced, and fair.

\begin{definition}\rm
A matching $\mu$ is said to be an 
\textbf{efficient iBF} 
if $\mu$ is  an iBF and there is no iBF $\mu'$ that Pareto dominates $\mu$. 
\end{definition}

The standard environment without the balancedness constraint can be subsumed by our model as a special case in which there is only one region. In that case, there is a unique efficient iBF, which corresponds to a ``student-optimal stable matching.''  In our setting, there may be multiple efficient iBFs.\footnote{See Example 2 in \cite{kamakoji-ekkyo}.} 

%% file: sections/fig_cycles.tex
\subsection{FIG (Fair Improvement Graph) Cycles }
\label{sec:FIGcycles}

We define a fair improvement graph to  define our algorithm. A bipartite directed graph on $I$ and $S$, $\mathcal G \subseteq (I\times S) \cup (S\times I)$, is a set of ordered pairs of agents in $I\cup S$. An interpretation is that if $(\st,\sch)\in \mathcal G$, then there is an arrow pointing from $\st$ to $\sch$. In this case, we say ``$\st$ points to $\sch$.'' The case of $(\sch,\st)\in \mathcal G$ is analogous. Given a graph $\mathcal G$, a \textbf{cycle in $\mathcal G$} is any sequence of the form
$(\st_1,\sch_1,\st_2,\sch_2, \dots, \st_m,\sch_m)$ such that for each $k \in \{1,\dots,m\}$, 
\begin{enumerate}
\item
$\st_k$ points to $\sch_k$, i.e., 
$(\st_k,\sch_k)\in \mathcal G$, 
\item
$\sch_k$ points to $\st_{k+1}$,  i.e., 
$(\sch_k,\st_{k+1})\in \mathcal G$, where we set $\st_{m+1}:=\st_1$,
\item $\st_k \neq \st_{k'}$ for all  $k' \neq k$, and
\item $\sch_k \neq \sch_{k'}$ for all $k' \neq k$. \end{enumerate}

Let $D_\sch^\mu:=\{\st \in I | \sch \succ_\st \mu_\st\}$ and, for any nonempty $I' \subseteq I$, $Top_\sch(I')$  be the student $\st \in I'$ who has the highest priority among those in $I'$ at $\succ_\sch$. 

Now we define a particular type of a graph and a cycle on it. This cycle will be used to characterize efficient iBF as well as to define our algorithm.
\begin{definition}\rm
\label{def:FIG}
Given a matching $\mu$, the \textbf{fair improvement graph (FIG)} for $\mu$ is a graph such that, for any $\st \in I$ and $\sch \in S$, 
\begin{enumerate}
\item\label{part:i_to_s}
student $\st \in I$ points to school $\sch \in S$ if $\st= Top_\sch(D_\sch^{\mu})$ and $\st$ is acceptable to $\sch$,  and 
\item school $\sch \in S$ points to student $\st \in I$ if either 
\begin{enumerate}
\item\label{case-standard} $\mu_\st = \sch$, or
\item\label{case-region} $|\mu_\sch|<   q_\sch$ and, [$\st$ is in the same region as $\sch$ and $\mu_\st=\emptyset$] or $\mu_\st$ is in the same region as $\sch$.  \end{enumerate}
\end{enumerate}
A \textbf{fair improvement graph cycle (FIG cycle)} on $\mu$ is a cycle in the FIG for $\mu$.
\end{definition}

Given a matching $\mu$ and a cycle of the form $\mathcal F=(\st_1,\sch_1,\st_2,\sch_2, \dots, \st_m,\sch_m)$, 
call $\mu'$ the \textbf{matching generated by $(\mu,\mathcal F)$} if
\[\mu'_{\st_k}=\sch_k \text{ for each }k \in \{1,\dots,m\}, \text{ and }   \mu'_{j}=\mu_{j} \text{ for all }j\in I \setminus \{\st_1,\dots,\st_m\}.\] 

Let $\tilde \mu$ be an arbitrary iBF  (note that the existence of an iBF is guaranteed as, for instance, the empty matching is an iBF). We define a \textbf{FIG cycles algorithm} starting from $\tilde \mu$ as follows.

\begin{algorithm}[htbp]
\caption*{FIG Cycles Algorithm on $\tilde{\mu}$}
\label{alg:fig_cycles}
\begin{algorithmic}
\State \textbf{Step $0$:} Let $\mu^{0}=\tilde{\mu}$ and  go  to Step 1. 
\State \textbf{Step $l$ $(l \ge 1)$:} 
 If there is no FIG cycle on $ \mu^{l-1}$, terminate the algorithm and output $\mu^{l-1}$. Otherwise, choose a FIG cycle $ \mathcal{F}$ on $  \mu^{l-1} $ arbitrarily and  let $\mu^l$ be the matching generated by $(\mu^{l-1},\mathcal{F})$,
and go to Step $l+1$.
\end{algorithmic}
\end{algorithm}
 Intuitively, the FIG cycles algorithm works as follows. It starts with an arbitrary iBF, such as the empty matching. Then, it runs a FIG cycle, which, when implemented, improves the welfare of the students  involved, while  preserving the fairness and balancedness   and leaving the match of the other students intact. The algorithm keeps implementing such a cycle until there remains no such cycle. \citet{kamakoji-ekkyo} show that the  FIG cycles algorithm starting from any iBF  $\tilde \mu$ runs in polynomial time, and its output is an efficient iBF which weakly Pareto dominates $\tilde \mu$.

We call a mechanism a \textbf{FIG cycles mechanism} if, for any given student preference
profile, it outputs the outcome of a FIG cycles algorithm on some iBF.

%% file: sections/model_daycare_market.tex
\subsection{Relationship to Daycare Matching}
\label{sec:daycare_matching_model}

In this subsection, we relate the basic model outlined above to the daycare matching market in Japan. 

Let $A  \coloneq \{ 0, 1, 2, 3, 4, 5 \}$ denote the set of possible ages. The set of students $I$ is partitioned into subsets $I_0, I_1, \ldots, I_5$, each containing students of the respective age.

Each school has a predetermined capacity for each age  $a \in A$. These capacities are exogenous within the matching process. Importantly, under the current system, schools are not allowed to use vacant capacity from one age to admit students of another.\footnote{While utilization of vacant capacity from a different age group is rarely practiced, it is legally permitted and its possibility has been explored by \citet{kamakoji-fair}.} Consequently, we model this situation by hypothesizing that each school is divided into six distinct schools corresponding to six different ages.

Formally, we partition the set of schools $S$ into subsets $S_0, S_1, \ldots, S_5$, with each subset including hypothetical schools specific to their respective ages.  Each of those hypothetical schools has the capacity that the associated  original school has determined for the corresponding age. For each $a\in A$, each 
student $i\in I_a$ finds all schools not belonging to $S_a$ unacceptable.

\subsubsection{Cross-age Regions and Age-specific Regions}\label{sec:cross_specific_region}

In our counterfactual analysis of the Japanese daycare market, we consider two scenarios, corresponding to two political constraints that seem plausible. In one of those scenarios, for each region, the inflow and outflow of students must be equal to each other \emph{for each age}, while in the other, the balance is required only for the students of all ages combined. In this subsection, we describe how those two scenarios can be mapped to the model we have introduced. 
Let $\tilde{R}$ denote the set of regions, where each region is the union of the set of students and schools in a municipality.  The regions in $\tilde{R}$ are called  \textbf{cross-age regions}. 

In contrast, we also define \textbf{age-specific regions}, where each region is defined as   the union of the set of students of a given age and the set of schools corresponding to that age  in a municipality:
\[
\hat{R} = \Bigl\{ r \cap (I_a \cup S_a) \Big | r \in \tilde{R}, a \in A \Bigr\}.
\]

Regarding balancedness, the age-specific region setting imposes stricter constraints than the cross-age region setting. For example, an exchange in which two students of age zero from region $r_A$ match with schools in region $r_B$, and simultaneously two students of age one from region $r_B$ match with schools in region $r_A$, is admissible under cross-age regions but disallowed under age-specific regions. In daycare matching in Japan, the relevant constraint may be imposed over age-specific regions. This is because per-child cost of care varies significantly depending on the child's age, and thus municipalities concerned about budgetary burden may insist on the balance of inflows and outflows for each age. 
Another canonical case is balancedness imposed at the level of cross-age regions. While this constraint may not exactly balance the budget in the same sense as the balancedness constraint on age-specific regions does, it might be simpler to understand (and thus accept) for policymakers and voters. Moreover, the constraint over cross-age regions is more permissive than the one over age-specific regions, and it may be of interest to quantify welfare improvement caused by a change in constraint from the latter to the former. 
Given those considerations, our counterfactual analysis considers both specifications.

\subsubsection{When Locals are Favored}

In our application to daycare allocation in Tokyo, a  daycare  gives a higher priority to residents of that daycare's region than non-residents. Formally,  in our model,  we say that \textbf{locals are favored} if for each $r \in R$, $\sch \in r$, $\chil \in r$, and $j \not\in  r$, we have $\chil \succ_\sch j$.

\cite{kamakoji-ekkyo} consider a special fair matching, namely the matching that is produced by running the standard  student-proposing Deferred Acceptance mechanism (DA) of \citet{gale/shapley:62}, separately in each region. We call such a matching a \textbf{region-wise student-optimal stable matching}. 
It is straightforward to check that a region-wise student-optimal stable matching is an iBF if locals are favored.

When locals are favored, the outcome of a FIG cycles mechanism starting from a region-wise student-optimal stable matching $\mu$  is an efficient iBF and weakly  Pareto dominates  $\mu$. 
We note that the ``unconstrained'' student-optimal stable matching (i.e., the region-wise student optimal stable matching when $R=\{I\cup S\}$) weakly Pareto dominates the following three matchings:
the region-wise student optimal stable matching, the outcome of the FIG cycles mechanism for the case when $R=\tilde{R}$ (``age-specific regions''), and the analogous matchings when $R=\hat{R}$ (``cross-age regions''). Also, for any outcome of the FIG cycles mechanism for the case when  $R=\tilde R$, there is a  FIG cycles mechanism for the case when  $R=\hat{R}$ whose outcome weakly Pareto dominates it.

%% file: sections/data.tex
\section{Data}\label{sec:data}

Our dataset consists of the following three components: the data from Bunkyo Ward for  the first-round admission process for the April 2022 intake, the  data from Koriyama City for the first-round admission process for the April 2024 intake, and the data from open sources.\footnote{Admission processes of both  Bunkyo Ward and Koriyama City consist of multiple rounds, where the first round  matches most of the children.}
The data from open sources include population data of children under six years old by neighborhood, locations and capacities of daycares within Tokyo City  (outside Bunkyo Ward),  and distances and travel times from each neighborhood to each daycare. This section provides an overview of the dataset, including summary statistics and other relevant details.

\begin{table}[htbp]
\centering
\caption{Summary statistics of children, daycares, and matching outcomes for Bunkyo Ward and Koriyama City}
\input{tables/section5/summary_stats}
\label{tab:5_summary_stats}
\end{table}

\subsection{Data from Bunkyo Ward}
\label{sec:data_bunkyo}

For Bunkyo Ward, we obtained the ROLs submitted to the first-round admission process for the April 2022 intake. The dataset records each child's age, priority score and suborder, residential area (aggregated into seven zones), the names of daycares in their ROLs, and the locations and the capacities of daycares. Under Bunkyo Ward's regulation, children can rank up to five daycares. 
Within each age, applicants are placed in a strict priority order.
Using the submitted ROLs and the priority order, Bunkyo Ward uses  Serial Dictatorship for children at each age to determine the matching. 

Of the 1,692 children who originally applied, 11 withdrew before matching, leaving 1,681 active applications against 2,317 total capacities. The information on the residential areas of children living outside Bunkyo Ward is missing (accounting for 6.7\% of the applicants). All other records are complete.  Summary statistics are shown in  \Cref{tab:5_summary_stats}: in particular, the overall match rate was 81.4\% and the average rank of matched daycares was 1.930.

\subsection{Data from Koriyama City}
\label{sec:data_koriyama}

For Koriyama City, we obtained the ROLs submitted during the first-round admission process for the April 2024 intake. The dataset records each child's age, priority score and suborder, residential address at the level of a geographic subdivision of a municipality, the names of daycares in their ROLs, and the locations and the capacities of daycares.\footnote{Koriyama City has 1,650 such subdivisions, whose average size is 0.459 square kilometers.} Under Koriyama City's regulation, children can rank up to ten daycares.

Withdrawals have already been removed from the received data, leaving 1,415 active applications against a total capacity of 1,417 seats. We note that 2.8\% of applicants live outside Koriyama City, and 8.5\% of records (including applications from outside the city) have missing address information. Summary statistics are shown in \Cref{tab:5_summary_stats}: in particular, the overall match rate was 72.7\%, and the average rank of matched daycares was 1.993. 
Like Bunkyo Ward, Koriyama City also uses Serial Dictatorship to assign children based on their submitted ROLs and the priority order.

\subsection{Data from open sources}
\label{sec:data:open-source}

We collected the population data of children by neighborhood from 2020 Population Census in Japan (\citeyear{kokuseichousa2020}). 
We gathered the official address of every daycare in Tokyo City and Koriyama City from the datasets published by Tokyo Prefecture  and Fukushima Prefecture, which Koriyama City belongs to. We supplemented this with April 2022 vacancy figures for all daycares in Tokyo City (outside Bunkyo Ward) by scraping individual ward websites.\footnote{As we mentioned, the locations and the capacities of daycares in Bunkyo Ward and Koriyama City are obtained from the data set provided by these respective municipaities.} 

To incorporate travel disutility into our utility function, we quantified the distance from each child's home to every daycare in two ways: straight line distance and travel time. The straight line distances were obtained via the Google Maps API by computing the geodesic distance from each residential zone's geographic centroid to each daycare's latitude and longitude. 

Travel times were computed differently for Tokyo City and Koriyama City. For Tokyo City---where car dropoffs and pickups of children are generally prohibited---we estimated travel times using public transportation. Walking times (from home to the nearest station and from the station to the daycare) were retrieved via the Google Maps Routes API, and interstation times via the Ekispert API.\footnote{Ekispert is a major consumer service for searching for travel routes via public transportation in Japan. Since Google does not provide API access for transit time calculations for public transportation in Japan, we used the Ekispert API instead.} Summing these yields the total public transport travel time; the final travel time for each child is then taken as the minimum of this value and the direct walking time, which we also obtained from the Google Maps Routes API.
For Koriyama City, by contrast, we used the Google Maps Routes API to retrieve automobile driving times and did not adjust these times for walking-only trips.\footnote{Although children typically attend nearby daycares on foot rather than by car, the driving time retrieved from Google Maps is generally shorter than the walking time. Therefore, unlike in Tokyo City, we could not define travel time as the minimum of walking and driving times.}

%% file: tables/section5/summary_stats.tex
\small
\begin{talltblr}[
  note{1}={``Total capacity'' refers to the total number of vacant seats available for children of each age across all daycares at the time of matching in our data. It differs from the total of the daycare's official enrollment caps, since some seats are already occupied by children matched in earlier academic years.},
  note{2}={The ``max-ranked applicants'' are those children whose ROL reached the maximum length allowed in the given municipality (five for Bunkyo and ten for Koriyama).},
  note{3}={The ``Average rank of matched daycares'' represents the mean position of the daycare to which each applicant is assigned in their submitted ROL. For unmatched applicants, we assign a rank of ``$(\text{ROL length})+1$.''  
  While this approach may slightly underestimate the rank of the outside option of max-ranked applicants, the impact on the value of the average is negligible given the low unmatch rate among max-ranked applicants.},
  label={none}
]{
  width = 0.9\linewidth,
  colspec = {X[3.5,l] X[1,c] X[1,c]},
  row{1} = {font=\bfseries},
  rows = {valign=m}, 
}
\toprule
                                      & Bunkyo & Koriyama \\ \midrule
\SetCell[c=3]{l} Number of applicants\\
\quad Age 0                            & 515      & 368         \\
\quad Age 1                            & 640    & 559         \\
\quad Age 2                            & 216      & 175         \\
\quad Age 3                            & 196     & 217        \\
\quad Age 4                            & 78   & 63        \\
\quad Age 5                            & 36      & 33         \\ 
\quad Sum                           & 1681     & 1415        \\
\midrule
Number of daycares                  & 127     & 89        \\ 
\SetCell[c=3]{l} Total capacity\TblrNote{1} \\
\quad Age 0                                 & 713      & 512         \\
\quad Age 1                                 & 563      & 421         \\
\quad Age 2                                 & 213      & 134         \\
\quad Age 3                                 & 217      & 209         \\
\quad Age 4                                 & 347      & 81         \\
\quad Age 5                                 & 264      & 60         \\
\quad Sum                                & 2317     & 1417        \\
\midrule
Maximum allowed ROL length & 5 & 10 \\
Average ROL length              & 3.795   & 3.081      \\ 
Share of max-ranked applicants\TblrNote{2}          & 0.572   & 0.028      \\ \midrule 
Match rate                              & 0.814   & 0.727      \\ 
Average rank of matched daycares\TblrNote{3}             & 1.930   & 1.993      \\ 
\bottomrule
\end{talltblr}

%% file: sections/est_results.tex
\section{Empirical Approaches and Results}
\label{sec:est_results}
Using the submitted ROLs, we estimate children's preference parameters.  We model child $\st$'s utility from being matched with daycare $\sch \in \Sch$ as follows:
\begin{align*}
u_{\st \sch} \coloneq U_{\st \sch} + \epsilon_{\st \sch} 
= \underbracket{U(X_{\st \sch} \mid \beta) + \alpha_s}_{\eqcolon U_{\st \sch}} + \ \epsilon_{\st \sch},
\end{align*}
where $U_{\st \sch}$ is the average attractiveness of daycare $\sch$ conditional on the
child-daycare characteristics $X_{\st \sch}$, and $\epsilon_{\st \sch}$ captures unobserved heterogeneity in preferences. 
$U_{\st \sch}$ is decomposed into two components: $U(\cdot \mid \beta)$, a known parametric function with a parameter vector $\beta$, and $\alpha_s$, an average quality of daycare $\sch$. 
Following standard assumptions in the literature, we assume that a preference shock  $\epsilon_\st = (\epsilon_{\st\sch})_{\sch \in \Sch}$  is independent of exogenous variables  $X_\st = (X_{\st \sch})_{\sch \in \Sch}$, i.e., $\epsilon_\st \indep X_\st$, 
and is independently and identically distributed across children and daycares according to the Gumbel distribution.\footnote{The cumulative distribution function (CDF) of the Gumbel distribution is given by $F(x) = \exp\{ -\exp (-x) \}$. Its mean is the Euler constant, $\gamma \simeq 0.577$, and its standard deviation is $\pi / \sqrt{6} \simeq 1.283$.}

We normalize the average utility of the outside option to be $U_{\st\n} = 0$ and set $u_{\st \n} = \epsilon_{\st \n}$. 
We do not assume any particular distribution on $\epsilon_{\st \n}$ at this point, since some strategic assumptions we introduce do not impose any restrictions on the distribution of $\epsilon_{\st \n}$.

\subsection{Strategic Assumptions}
\label{sec:strategic_assumptions}

In both Bunkyo Ward and Koriyama City, children are assigned to daycares via  Serial Dictatorship, with an upper limit constraint of $K$ on the length of the ROLs. In our applications, there are over 80 daycares in each dataset, yet the length of ROLs is limited to 5 in Bunkyo Ward and 10 in Koriyama City. While unconstrained Serial Dictatorship is strategy-proof (children do not benefit from misreporting their preferences), imposing the constraint breaks the property. For instance, applicants may benefit from omitting daycares where they have little chance of admission.

Moreover, a growing body of empirical evidence demonstrates that many applicants do not report their preferences truthfully even when the matching mechanism is strategy-proof (e.g., \citet{rees2018suboptimal}, \citet{fack2019beyond}). This suggests that assuming that all children submit their ROLs truthfully is too demanding in practice.

To address
the issue of misreporting, we impose assumptions on how children report their preferences and use them to identify the underlying preference parameters. In what follows, we introduce several strategic assumptions and estimate the parameters under each of them. We select the best-fitting specification for our counterfactual simulations. 

In this section, we describe the assumptions used to estimate the preference parameters based on observed ROL data. Specifically, we estimate models under four behavioral assumptions: strict truth-telling (STT), weak truth-telling (WTT), optimistic expectation truth-telling (OETT), and undominated strategies and stability (USS). 
Below, we describe the modeling assumptions and the resulting choice probabilities under each of these frameworks.

\subsubsection{Model of Strategic Report}
\label{sec:strategic_assumptions:notation}

We first present a model of strategic reporting. 
Each child $\st$ in the market has a true utility vector $u_\st = (u_{\st\sch})_{\sch \in  S\cup\{\n\}}$ and private information $t_\st$, such as her score, which varies with the estimation model. 

Given $u_\st$ and $t_\st$, child $\st$ submits an ROL, $R_\st = (\sch_1, \sch_2, \ldots, \sch_{\size{R_\st}})$, where $s_r$ denotes $\st$'s $r^{ \text{th}}$ choice and $|R_\st|$ denotes the length of $R_\st$. Since all applicants submitted an ROL with length 1 or greater, we have $1 \leq |R_\st| \leq K$. 
In general, the submitted list $R_\st$ may not coincide with the top $K$ daycares in the descending order of $u_i$. 
We denote by $\sigma_i$ a pure strategy of child $\st$ that assigns an ROL based on her true utility and private information; that is, $\sigma_i (u_\st \mid t_\st) = R_\st$.

When $R_\st=(s_1, s_2, \ldots, s_{\size{R_\st}})$ is given, we sometimes abuse notation to denote $\{s_1, s_2, \ldots, s_{\size{R_\st}}\}$ by $R_\st$, that is, we use the same notation $R_\st$ for the unordered set corresponding to  the original ordered set $R_\st$.

Finally, given a child $\st$'s ROL $R_\st$ and a rank $r$ ($1 \leq r \leq \size{R_\st}$), we define the \emph{lower-contour set} of $\st$ at rank $r$ by $L^r_\st(R_\st) \coloneq S \setminus \{ \sch_1, \ldots, \sch_{r-1} \} $ for $r \geq 2$ and $L^r_\st(R_\st) \coloneq S$ for $r=1$, which is the collection of all daycares except those that $\st$ ranks above $\sch_r$ in $R_\st$.

\subsubsection{Strict Truth-Telling (STT)}
\label{sec:strategic_assumptions:STT}

Our first behavioral assumption is \emph{strict truth-telling (STT)}, under which each child lists all the daycares that $\st$ strictly prefers to the outside option in the descending order of $u_i$, truncated to length $K$. This assumption is widely used in empirical studies on school choice (e.g., \citet{abdulkadiroglu2017welfare}). In practice, STT would be justified if (i) the maximum length $K$ is sufficiently large relative to the number of acceptable daycares for each child and (ii) the cost associated with listing additional daycares is small. 

For STT, we additionally assume that $\epsilon_{i \n}$ is independently drawn from the Gumbel distribution. 
Formally, STT is characterized by the following conditions.

\begin{assumption}[Strict Truth-Telling]
\label{ass:STT}\rm
\quad 
\begin{enumerate}[label=STT \arabic*., wide=1em, leftmargin=5em, font=\itshape, ref=\arabic*]
\item $R_\st$ ranks $\st$'s top $\size{R_\st}$ daycares: for all $r \in \{1,\dots, \size{R_\st}\}$ and $\sch \in L^r_\st(R_\st) \cup \{ \n \}$, we have  $u_{\st\sch_r}\geq u_{\st\sch}$.
\item If $\size{R_\st} < K$,  then the outside option is the most preferred alternative except for the daycares listed in $R_\st$: for all $\sch \in \Sch \setminus R_\st$, $u_{\st\n}> u_{\st\sch}$.
\end{enumerate}
\end{assumption}

Under \Cref{ass:STT}, the probability that child $\st$ submits an ROL $R_\st = (s_1, \ldots, s_{\size{R_\st}})$ given $X_\st$ and $t_\st$ is calculated as 
\begin{align*}
&\prob \Bigl( \sigma_i(u_\st \mid t_\st) = R_\st \mid X_\st , t_\st \Bigr) \nonumber \\
&\hspace{1em} = \left(
\prod_{r=1}^{\size{R_\st}} \frac{ 
    \exp \left(U_{\st \sch_r}\right) 
    }{ 
    \sum_{\sch \in L^r_\st(R_\st) \cup \nset}\exp \left(U_{\st \sch}\right) 
} 
\right) \times \left(
\frac{ 
    1 
    }{ 
    \sum_{\sch \in S\cup\{\n\} \setminus R_\st} \exp \left(U_{\st \sch}\right) 
} 
\right)^{ \ind{\size{R_\st} < K} }.
\end{align*}
Note that under STT, since $\sigma_i$  depends on the private information $t_\st$ only when there are ties in utilities and such events happen with zero probability, choice probabilities are independent of $t_\st$.
Using the choice probability, we estimate the model parameters $\theta = ((\alpha_\sch)_{\sch \in \Sch}, \beta)$ with maximum likelihood estimation. The log-likelihood is given by
\begin{align*}
\log L_{STT}(\theta) = \sum_{\st \in \St} \Biggl[ 
&\left( \sum_{r=1}^{\size{R_\st}} U_{\st \sch_r} \right) 
- \left( \sum_{r=1}^{\size{R_\st}} \log \Biggl( \sum_{\sch \in L^r_\st(R_\st) \cup \nset} \exp\left(U_{\st \sch}\right) \Biggr) \right) \nonumber \\
& + \ind{\size{R_\st} < K} \Biggl\{ 
- \log \Biggl( \sum_{\sch \in S\cup\{\n\} \setminus R_\st} \exp\left(U_{\st \sch}\right) \Biggr) \Biggr\}
\Biggr]. 
\end{align*}

\subsubsection{Weak Truth-Telling (WTT)}
\label{sec:strategic_assumptions:WTT}

Many empirical studies on matching markets and school choice have demonstrated that students rarely rank all available alternatives, even when there is no explicit limitation on the length of ROLs (\citet{abdulkadiroglu2017welfare,he2015gaming,artemov2017strategic}). While one possible explanation is that the number of acceptable daycares is typically small, a more plausible assumption is that compiling an extensive ROL incurs costs, discouraging children from listing every acceptable option (\citet{larrocau2020shortlist}).
An alternative assumption accommodating this situation is the \emph{weak truth-telling (WTT)} proposed by \cite{fack2019beyond}, that each child lists her most preferred daycares up to $K_\st$ ($\leq K$). 
Here, $K_\st$ is heterogeneous among children and assumed to be independent of the preferences of the child. Formally, WTT is characterized by the following assumptions.


\begin{assumption}[Weak Truth-Telling]
\label{ass:WTT}\rm
\quad 
\begin{enumerate}[label=WTT \arabic*., wide=1em, leftmargin=5em, font=\itshape, ref=\arabic*]
\item\label{WTT1} 
$R_\st$ ranks $\st$'s top $\size{R_\st}$ daycares: for all $r\in \{1,\dots,\size{R_\st}\}$ and $\sch \in L^r_\st(R_\st)$,  we have  $u_{\st \sch_r} \geq u_{\st \sch}$.
\item\label{WTT2} The number of daycares listed, $\size{R_\st}$, is independent of the true utilities $(u_{\st\sch})_{\sch \in \Sch}$ for all daycares except the outside option.
\end{enumerate}
\end{assumption}

Here, the assumption WTT~\ref{WTT2} may seem controversial. We do not intend to take any stance on its validity, but list two data generating processes under which it might be justified. 
The first data generation process is as follows. Each child $\st$ first draws true utilities $(u_{\st \sch})_{\sch \in S\cup\{\n\}}$, inclu
ding the utility for the outside option, just as in STT. Independently, the child draws a list length, $K_\st$, from some distribution.
Finally, she submits an ROL of her top $K_\st$ daycares, even if some of those are actually ranked below the outside option in her true utilities. 
The second data generation process is as follows. Each child $\st$ first randomly draws her true utilities $(u_{\st \sch})_{\sch \in \Sch}$ for all daycares except the outside option, and separately draws the length of an ROL they submit, $K_\st$, from some distribution, independently of her utility.
Next, the value of the outside option $u_{\st \n} = \epsilon_\n$ is independently drawn from some distribution in such a way that the number of daycares strictly preferred to the outside option is always no less than $K_\st$. 
Finally, she submits an ROL that lists her top $K_\st$ daycares in order, each of which should be preferred to the outside option by construction.
In contrast, \citet{fack2019beyond} demonstrate that a model in which each additional daycare listed in $R_\st$ imposes a constant marginal cost violates WTT~\ref{WTT2}.

Under \Cref{ass:WTT}, the probability that child $\st$ submits an ROL $R_\st = (s_1, \ldots, s_{K_\st})$ given $X_\st$ and $t_\st$ is calculated as
\begin{align*}
&\prob \Bigl( \sigma_i(u_\st \mid t_\st) = R_\st \Bigm | X_\st , t_\st \Bigr) \nonumber \\ 
&\hspace{1em} = 
\prob \Bigl( \size{\sigma_i(u_\st \mid t_\st)} = K_\st \Bigm| X_\st , t_\st \Bigr) \times 
\prob \Bigl( \sigma_i(u_\st \mid t_\st) = R_\st \Bigm | X_\st , t_\st, \size{\sigma_i(u_\st \mid t_\st)} = K_\st \Bigr) \nonumber \\
&\hspace{1em} = \prob \Bigl( \size{\sigma_i(u_\st \mid t_\st)} = K_\st \Bigm| X_\st, t_\st \Bigr) \times
\prod_{r=1}^{K_\st} \frac{ 
    \exp \left(U_{\st \sch_r}\right) 
    }{ 
    \sum_{\sch \in L^r_\st(R_\st)}\exp \left(U_{\st \sch}\right) 
}. 
\end{align*}
Here, WTT~\ref{WTT2} guarantees that the first factor $\prob \left( \size{\sigma_i(u_\st \mid t_\st)} = K_\st | X_\st , t_\st \right)$ does not depend on true utilities of daycares. 
Thus, to estimate the preference parameters of daycares, it suffices to focus solely on the second factor, using the observed ROL length $K_\st$ in the dataset. The conditional log-likelihood function to maximize is given by 
\begin{align*}
\log L_{WTT}(\theta) = \sum_{\st \in \St} \Biggl[ 
&\left( \sum_{r=1}^{K_\st} U_{\st \sch_r} \right) 
- \left( \sum_{r=1}^{K_\st} \log \Biggl( \sum_{\sch \in L^r_\st(R_\st)} \exp \left (U_{\st \sch} \right) \Biggr) \right) 
\Biggr]. 
\end{align*}

\subsubsection{Optimistic Expectation Truth-Telling (OETT)}
\label{sec:strategic_assumptions:OETT}

Due to the maximum-length constraint on ROLs, children have an incentive to select daycares where they realistically expect to be admitted based on their scores. In fact, in real-life daycare matching markets in Japan, parents often gather information about the cutoff scores (the lowest scores with which children can be matched to each daycare). Local municipalities, politicians, and anonymous bloggers provide past data on these cutoff scores, which helps parents focus on the daycares they can realistically have their children be admitted to.

One approach to formalize the expectation about admission chances in the literature is to obtain students' beliefs about their admission probabilities directly through surveys or by constructing them via bootstrapping from observed ROLs. One would then estimate 
preference parameters under the assumption that each student best responds to these beliefs
(\cite{agarwal2018demand, kapor2020hetero}). This approach, however, becomes  computationally costly as the number of schools grows and the set of potential ROLs expands combinatorially. Some studies such as \cite{calsamiglia2020structual}  
and \cite{son2024dist} avoid computational infeasibility, although doing so necessitates additional assumptions.
In contrast, we take a more direct approach. Building on WTT, we newly develop the \emph{optimistic expectation truth-telling (OETT)} assumption:
On the one hand, each daycare is associated with an exogenously given  ``benchmark cutoff score,''  which for example is based on the information from last year's admission cycle. On the other hand, each child is associated with a ``degree of optimism,'' which quantifies how optimistic she is about the admission to a daycare when comparing her own score and the benchmark cutoff score at the daycare. Each child uses the benchmark cutoff scores and the degree of optimism to form a ``choice set,'' and ranks only those daycares in the choice set.  
Concretely, we assume that each child's choice set includes all daycares whose benchmark cutoff scores minus some optimism degree are at or below her own score. \footnote{One possible alternative specification would be to allow for children's pessimism as well. Our result is unlikely to change significantly, though, as fewer than 2\% of the applicants in either Bunkyo Ward or Koriyama City list only daycares whose cutoff scores are strictly below their own scores.
}
In other words, under OETT, children expect cutoff scores,  optimistically compared to the benchmark cutoff scores, and omit any daycare for which they are ineligible even under their optimistic expectation. 
Finally, each child submits an ROL of their top $K_i$ daycares from this choice set, given an exogenous ROL length $K_i$.

\begin{figure}[tbhp]
  \begin{subfigure}[b]{0.49\textwidth}
    \centering
    \includegraphics[width=\linewidth]{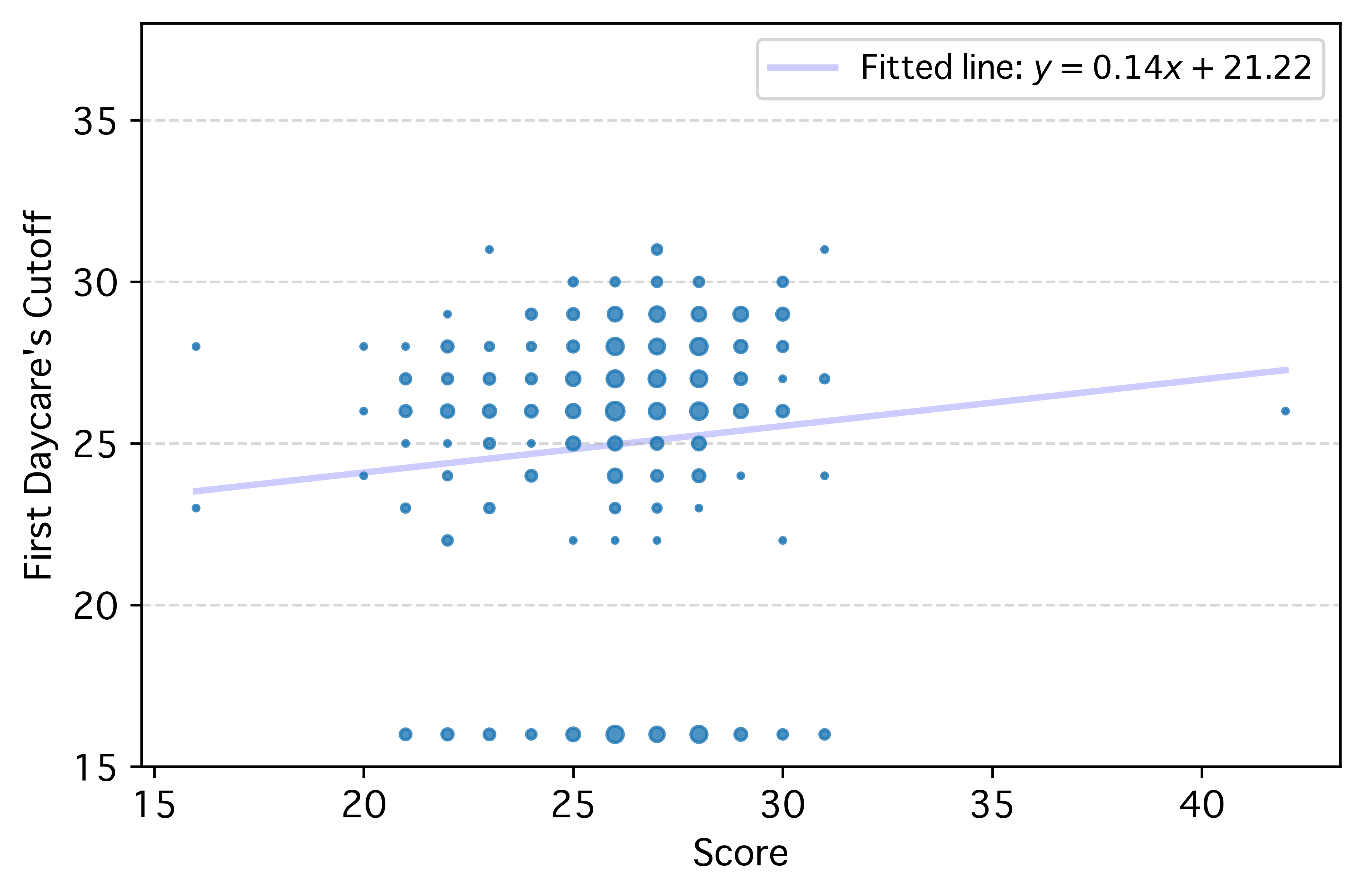}
    \caption{Bunkyo Ward}
    \label{fig:bunkyo_cutoff_scores_in_ROL}
  \end{subfigure}
  \hfill
  \begin{subfigure}[b]{0.49\textwidth}
    \centering
    \includegraphics[width=\linewidth]{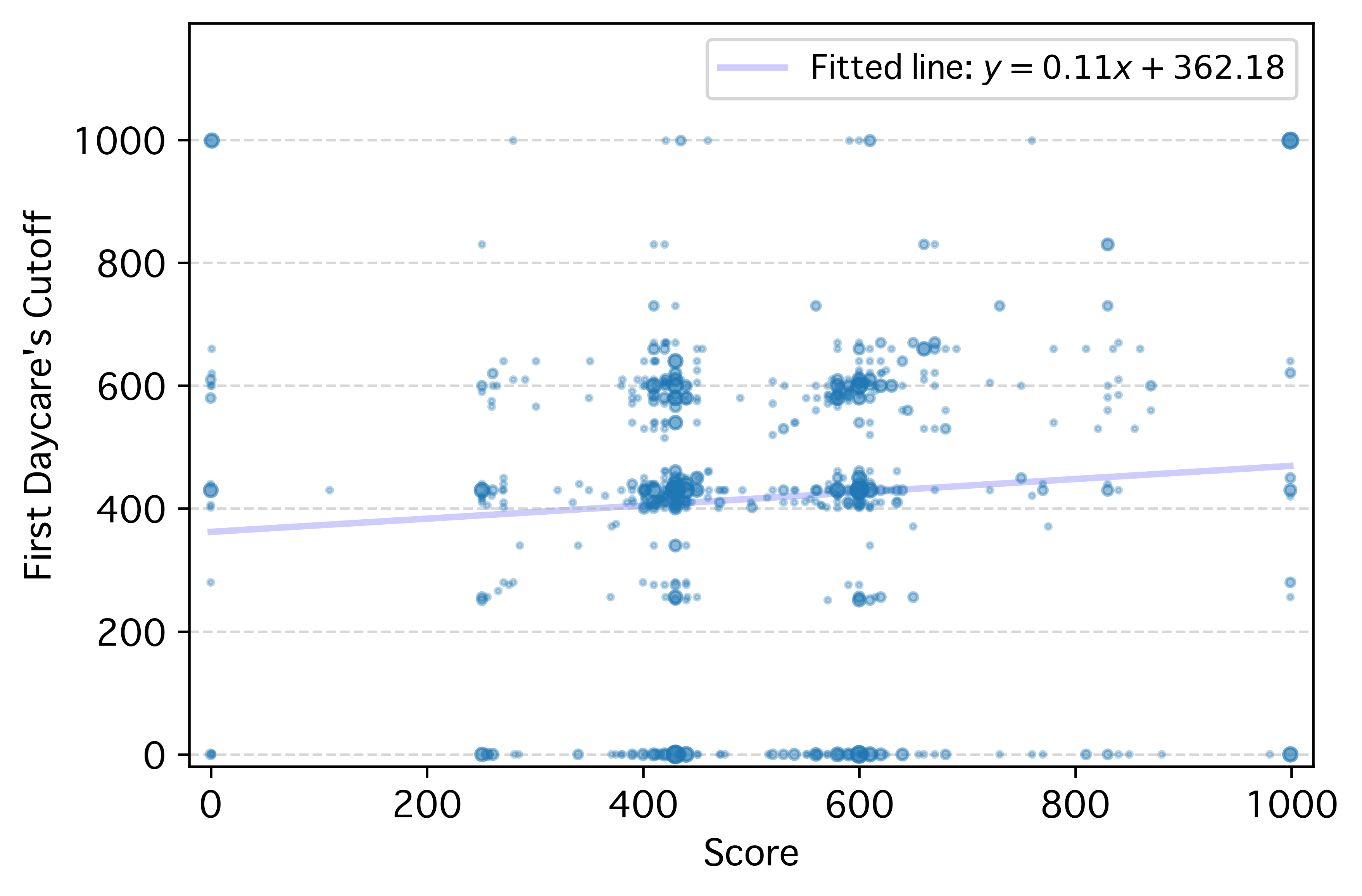}
    \caption{Koriyama City}
    \label{fig:koriyama_cutoff_scores_in_ROL}
  \end{subfigure}
  \caption{
  Scatter plots of each child's score versus the cutoff score of her first-listed daycare for (a) Bunkyo Ward and (b) Koriyama City. Color intensity and size denote the number of children at each coordinate (darker and larger points represent more children), and the solid blue line is the fitted linear regression (the coefficient of the slope is 0.14 and its p-value is 0.019 for Bunkyo Ward; they are 0.11 and 0.006, respectively, for Koriyama City). 
  For a daycare whose capacity is unfilled, the cutoff was set to the minimum observed score in the dataset (16 for Bunkyo Ward; 0 for Koriyama City). In Bunkyo Ward, we also tested all possible cutoff values between the theoretical minimum score (10) and 16 and obtained similar coefficients and p-values. We also tested samples excluding outliers (applicants with scores 16 and 42) and obtained a similar result. 
  }
  \label{fig:cutoff_socres_in_ROL}
\end{figure}

The introduction of both cutoff anticipation and optimism is motivated by our data. 
\Cref{fig:cutoff_socres_in_ROL} plots the relationship between each child's score and the cutoff score of her first-listed daycare.
It illustrates that a child with a higher score is more likely to list a daycare whose admission cutoff is also higher as their first choice. Under WTT, there should be no correlation between one's own score and the cutoff of the first-listed daycare, yet we observe a positive slope. By contrast, under OETT, children tend not to list daycares whose cutoffs are set too high relative to their own scores. Moreover, a non-negligible fraction of children list a daycare whose actual cutoff exceeds their own score, behavior that provides empirical support for incorporating optimism into our model of daycare choice.

Formally, let $e_\st \in \R_+$ denote child $\st$'s score, and let $o_\st \in \R_+$ be child $\st$'s ``degree of optimism.'' 
We also define $\underline{e}_\sch \in \R_+$ to be daycare $\sch$'s benchmark cutoff score, which is an exogenously given number.
Child $\st$'s choice set that consists of ``nearly''  achievable daycares is: 
$$
S(e_\st, o_\st) \coloneq \{ \sch \in \Sch \mid  \underline{e}_\sch - o_\st \leq e_{\st} \}.
$$

In this setting, the private information of child $\st$ is $t_\st = (e_i, o_i)$. The OETT assumption can then be formally characterized by the following conditions.  

\begin{assumption}[Optimistic Expectation Truth-Telling]
\label{ass:OETT}\rm
\quad
\begin{enumerate}[label=OETT \arabic*., wide=1em, leftmargin=5em, font=\itshape, ref=\arabic*]
\item\label{OETT1} $R_i\subseteq S(e_i,o_i)$ holds, and $R_\st$ ranks $\st$'s top $\size{R_\st}$ daycares among the choice set $S(e_\st, o_\st)$: for all $r\in \{ 1,\dots,\size{R_\st} \}$ and $\sch \in L^r_\st(R_\st) \cap S(e_\st, o_\st)$, we have $u_{\st \sch_r} \geq u_{\st \sch}$.

\item\label{OETT2} The score $e_{\st}$, the degree of optimism $o_\st$, and the true utilities $(u_{\st\sch})_{\sch \in \Sch}$ except for the outside option are mutually independent.
\item\label{OETT3} The number of daycares listed, $\size{R_\st}$, is independent of the degree of optimism $o_\st$ and the true utilities $(u_{\st\sch})_{\sch \in \Sch}$ for all daycares except the outside option.
\end{enumerate}
\end{assumption}

Note that $R_i\subseteq S(e_i,o_i)$ in OETT~\ref{OETT1} and the independence of $|R_i|$ and $o_i$ in  OETT~\ref{OETT3} imply $|R_\st| \leq |S(e_\st, 0)|$. This restriction is not binding in our data because there are more than 10 daycares that do not fill their capacities (hence have a cutoff score of $0$) while $|R_i|$ is at most 10 due to the length restriction.
Note also that our assumption allows $|R_\st| $ and $e_\st$ to be correlated. 
For example, we allow a correlation such that a child with a higher score tends to have a shorter ROL length, a pattern that we observe in Koriyama City's data.

Under \Cref{ass:OETT}, the probability that child $\st$ submits ROL $R_\st = (s_1, \ldots, s_{K_\st})$ given $X_\st$ and $t_\st$ is calculated as:
\begin{align*}
&\prob \Bigl( \sigma_i(u_\st \mid t_\st) = R_\st \Bigm | X_\st, t_\st \Bigr) \nonumber \\
&\hspace{1em} = 
\prob \Bigl( \size{\sigma_i(u_\st \mid t_\st)} = K_\st \Bigm| X_\st, t_\st \Bigr) 
\times
\prob \Bigl( \sigma_i(u_\st \mid t_\st) = R_\st \Bigm | X_\st, t_\st, \size{\sigma_i(u_\st \mid t_\st)} = K_\st \Bigr) \nonumber \\
&\hspace{1em} = \prob \Bigl( \size{\sigma_i(u_\st \mid t_\st) } = K_\st \Bigm| X_\st, t_\st \Bigr) 
\times 
\prod_{r=1}^{K_\st} \frac{ 
    \exp \left(U_{\st \sch_r}\right) 
    }{ 
    \sum_{\sch \in L^r_\st(R_\st) \cap S(e_\st, o_\st)}\exp \left(U_{\st \sch}\right)
}.
\end{align*}
Similar to WTT, we consider the conditional log-likelihood maximization using only the second factor given the observed ROL length $K_i$ in the dataset. 
However, because $o_\st$ is unobservable, we must specify its distribution and integrate it out to obtain the marginal likelihood.
Let $F_\psi$ denote the distribution of $o_i$ 
parameterized by 
$\psi$ 
whose support is unbounded above,
 and define $\bar{e}(R_i) \coloneq  \max_{s \in R_i} \underline{e}_s$ as the highest cutoff score among the daycares in child $i$'s ROL $R_i$. Taking into account the constraint that $o_i$ must satisfy $o_i \geq \max \{ \bar{e}(R_i) - e_i, 0 \}$, the conditional log likelihood function to maximize is, 
\begin{align*}
\log L_{OETT}(\theta, \psi) = \sum_{\st \in \St} 
&
\log\left( 
\int
\prod_{r=1}^{K_\st} \frac{ 
    \exp \left(U_{\st \sch_r}\right) 
    }{ 
    \sum_{\sch \in L^r_\st(R_\st) \cap S(e_\st, o_\st)}\exp \left(U_{\st \sch}\right) 
} \ind{o_\st \geq \bar{e}(R_\st) - e_\st}
\ d F_\psi 
\right).
\end{align*}%
In our estimation, we assume that $F_\psi$ is a gamma distribution with shape parameter $\kappa$ and scale parameter $\lambda$.\footnote{
The density function of the gamma distribution with parameter $(\kappa, \lambda)$ is $f(x) = \frac{\lambda^\kappa}{\Gamma(\kappa)} x^{\kappa-1} e^{-\lambda x}$ for $x \geq 0$ and $0$ otherwise, where $\Gamma$ is the gamma function.
} We jointly estimate $\psi = (\kappa, \lambda)$ alongside $\theta$.

\subsubsection{Undominated Strategies and Stability (USS)}
\label{sec:strategic_assumptions:USS}

\cite{fack2019beyond} propose a GMM estimator under the \emph{undominated strategies and stability (USS)} assumption, which is asymptotically weaker than WTT and yet still suffices to identify the preference parameters. 
\footnote{
``Undominated strategies'' requires that submitted ROLs reveal each child's true partial order over daycares (i.e., no swaps of daycares take place); weak truth-telling is a special case of it. ``Stability'' demands that each child should be matched to her most preferred daycare among all ex-post ``reachable'' daycares. \cite{fack2019beyond} show that, while stability need not hold in a finite market even when children are weakly truth-telling, it becomes asymptotically satisfied as the market size grows large in their setting.
}
They conduct Monte Carlo simulations in which ROLs are generated in a Bayes-Nash equilibrium under each of (i) the DA with a maximum-length constraint and (ii) the DA with application costs, where each additional daycare listed after the first choice incurs a constant marginal cost. In their simulations, the GMM estimator under the USS assumption successfully recovers the true preference parameters in both cases, whereas the estimator under WTT fails to recover them.

For USS, as with STT, we assume that $\epsilon_{i \n}$ is independently drawn from the Gumbel distribution. 
We formally state the assumptions of undominated strategies and stability separately.

\begin{assumption}[Undominated Strategies]
\label{ass:Undominated}\rm
\quad 
\begin{enumerate}[label=USS \arabic*., wide=1em, leftmargin=5em, font=\itshape, ref=\arabic*]
\item $R_\st$ does not reverse the order of a true preference relation: for any $r, r' \in \{ 1, \ldots, \size{R_\st} \}$ with $r \leq r'$, we have $u_{\st \sch_r} \geq u_{\st \sch_{r'}}$.
\end{enumerate}
\end{assumption}
Given a matching in the dataset, we say that a daycare $s$ is \textbf{reachable} for child $i$ if either (i) $s$ has a vacant position or (ii) there exists a child who is currently matched with $s$ and has a priority no higher than $i$. We denote the set of daycares that are reachable for $i$ by $S_i$.
\begin{assumption}[Stability]
\label{ass:Stable}\rm
\quad 
\begin{enumerate}[label=USS \arabic*., wide=1em, leftmargin=5em, font=\itshape, ref=\arabic*]
\setcounter{enumi}{1}
\item\label{USS2} Child $\st$ is matched with daycare or the outside option $\bar{\sch}$ that is most preferred among all reachable daycares and the outside option: for all $\sch \in \Sch_\st \cup \{ \n \}$, we have $u_{\st \bar{\sch}} \geq u_{\st \sch}$. 
\item\label{USS3} Conditional on the observables $X_\st$, child $\st$'s error term $\epsilon_i$ and her priority order are independent.
\item\label{USS4} Conditional on the observables $X_\st$, child $\st$'s error term $\epsilon_i$ and her choice set $\Sch_\st$ are independent.\footnote{
We note that 
USS~\ref{USS2} and USS~\ref{USS3} imply USS~\ref{USS4}  in our setting since priority orders are identical across all schools. Nonetheless, we present this condition for better comparison with \citep{fack2019beyond}. They state the condition in a setting with possibly heterogeneous priority orders, in which the implication does not necessarily hold.}
\end{enumerate}
\end{assumption}

\Cref{ass:Undominated} generates moment inequalities reflecting the utility ordering of listed daycares, thus enabling only partial identification.  
Specifically, for child $\st$ whose submitted ROL is $R_\st$, the probability that school $s_1$ is ranked above school $s_2$ can be written as the following. 
\begin{align*}
\prob(\text{$s_1$ is ranked above $s_2$ in $R_\st$} \mid X_\st) &= \prob(u_{\st \sch_1} > u_{\st \sch_2}, \ s_1,s_2 \in R_\st \mid X_\st) \\
&\leq \prob(u_{\st \sch_1} > u_{\st \sch_2} \mid X_\st).
\end{align*} 
The inequality defines a lower bound for the conditional probability of $u_{\st \sch_1}>u_{\st \sch_2}$. Similarly, an upper bound is given by
\begin{align*}
\prob(u_{\st \sch_1} > u_{\st \sch_2} \mid X_\st) 
\leq 1 - \prob(\text{$s_2$ is ranked above $s_1$ in $R_\st$} \mid X_\st).
\end{align*}
These bounds yield the following conditional moment inequalities:
\begin{align*}
\prob(u_{\st \sch_1}>u_{\st \sch_2}\mid X_\st)-E\left[
\mathbf{1}_{\{\text{$s_1$ is ranked above $s_2$ in $R_\st$}\}
}
\mid X_\st\right]&\geq 0, \\
1-E\left[ 
\mathbf{1}_{\{\text{$s_2$ is ranked above $s_1$ in $R_\st$}\}
} \mid X_\st \right]-\prob(u_{\st \sch_1} > u_{\st \sch_2}\mid X_\st)&\geq 0.
\end{align*}
Using the observed ROLs $(R_\st)_{\st \in \St}$, the sample versions of the left-hand sides in the moment inequalities are defined, respectively, as the sample averages of the following values across $i \in I$. 
\begin{align*}
m^{\text{ineq-LB}}_{\st, \sch_1, \sch_2}(\theta) &\coloneq 
  \frac{ \exp\left(U_{\st \sch_1}\right) }{ \exp\left(U_{\st \sch_1}\right) + \exp\left(U_{\st \sch_2}\right) } 
  - 
  \mathbf{1}_{\{\text{$s_1$ is ranked above $s_2$ in $R_\st$}\}} \\
m^{\text{ineq-UB}}_{\st, \sch_1, \sch_2}(\theta) &\coloneq 
  1 
  - 
  \mathbf{1}_{\{\text{$s_2$ is ranked above $s_1$ in $R_\st$}\}}
  - \frac{ \exp\left(U_{\st \sch_1}\right) }{ \exp\left(U_{\st \sch_1}\right) + \exp\left(U_{\st \sch_2}\right) }.
\end{align*} 

In contrast, \Cref{ass:Stable} implies the following conditional probability for child $\st$ to be matched to $\sch \in S\cup\{\n\}$:
\begin{align}
\label{eq:Stable_choice_prob}
\prob(\text{$\st$ is matched to $\sch$} \mid X_\st, \Sch_\st, t_\st) = 
\ind{\sch \in \Sch_\st \cup \{\n\}} \frac{ 
    \exp \left(U_{i \sch}\right) 
    }{ 
    \sum_{\sch' \in \Sch_\st \cup \{ \n \} }\exp \left(U_{\st \sch'}\right) 
} , 
\end{align} 
which allows for point-identification of preference parameters \citep{fack2019beyond}. 
To combine this with the previously stated moment inequalities, we reformulate the maximum likelihood estimation based on \Cref{eq:Stable_choice_prob} into the following moment equalities:
\begin{align*}
m^{\text{eq}, 1}_{\st, \sch}(\theta) &\coloneq 
\mathbf{1}_{\{ \bar{\sch}_\st = \sch \}}
- \ind{\sch \in \Sch_\st} \frac{ 
    \exp \left(U_{\st \sch}\right) 
    }{ 
    \sum_{\sch' \in \Sch_\st \cup \{ \n \}} \exp \left(U_{\st \sch'}\right) 
} \quad \text{for each $\sch \in \Sch$} \\
m^{\text{eq}, 2}_{\st}(\theta) &\coloneq 
\sum_{\sch \in \Sch} 
X_{\st \sch} \left( 
\mathbf{1}_{\{ \bar{\sch}_\st = \sch \} }
- \ind{\sch \in \Sch_\st} \frac{ 
    \exp \left(U_{i \sch}\right) 
    }{ 
    \sum_{\sch' \in \Sch_\st \cup \{ \n \}} \exp \left(U_{\st \sch'}\right) 
} 
\right), 
\end{align*} 
where $\bar{\sch}_\st$ is child $\st$'s matched daycare. 

The objective function for the GMM estimation to be minimized under USS is 
\begin{align*}
\sum_{\sch \in \Sch} 
\left[ \frac{
    \bar{m}^{\text{eq}, 1}_{\sch}(\theta)
}{
    \hat{\sigma}^{\text{eq}, 1}_{\sch}(\theta)
} \right]^{2} + 
\left[ \frac{
    \bar{m}^{\text{eq}, 2}(\theta)
}{
    \hat{\sigma}^{\text{eq}, 2}(\theta)
} \right]^2
 + 
\sum_{\sch_1, \sch_2 \in \Sch} \left(
\left[ \frac{
    \bar{m}^{\text{ineq-LB}}_{\sch_1, \sch_2}(\theta)
}{
    \hat{\sigma}^{\text{ineq-LB}}_{\sch_1, \sch_2}(\theta)
} \right]^{\ 2}_{-} + 
\left[ \frac{
    \bar{m}^{\text{ineq-UB}}_{\sch_1, \sch_2}(\theta)
}{
    \hat{\sigma}^{\text{ineq-UB}}_{\sch_1, \sch_2}(\theta)
} \right]^{\ 2}_{-}
\right), 
\end{align*}
where $[A]_{-} \coloneq \min \{ A, 0 \}$,
and ($\bar{m}_\sch^{\text{eq}, 1}, \bar{m}^{\text{eq}, 2}, \bar{m}_{\sch_1, \sch_2}^{\text{ineq-LB}}, \bar{m}_{\sch_1, \sch_2}^{\text{ineq-UB}}$) and ($\hat{\sigma}_\sch^{\text{eq}, 1}, \hat{\sigma}^{\text{eq}, 2}, \hat{\sigma}_{\sch_1, \sch_2}^{\text{ineq-LB}}, \hat{\sigma}_{\sch_1, \sch_2}^{\text{ineq-UB}}$) are the sample means and  standard deviations of 
($m_{\st, \sch}^{\text{eq}, 1}, m_{\st}^{\text{eq}, 2}, m_{\st, \sch_1, \sch_2}^{\text{ineq-LB}}, m_{\st, \sch_1, \sch_2}^{\text{ineq-UB}}$) computed across all children $\st$.



\subsection{Estimated Parameters}\label{sec:est_params}

We begin by specializing the utility functions as in the following two formulations.
\begin{align*}
\text{\textbf{(Linear)}} \quad u_{\st \sch} & = \alpha_s - \text{$\beta$} d_{\st \sch} + \epsilon_{\st \sch},  \label{eq:util_linear} \\
\text{\textbf{(Log)}} \quad u_{\st \sch} &= \alpha_s - \text{$\beta$} \log d_{\st \sch} + \epsilon_{\st \sch}.
\end{align*}
Here, $d_{\st \sch}$ is our key variable that represents the distance between the place of residence of child $\st$ and the location of daycare $\sch$. 

\input{tables/section6/tab_estimation_reg_bunkyo}
\input{tables/section6/tab_estimation_reg_koriyama}

We use two distance metrics in estimation: straight line distance (km) and travel time (minutes) between the locations. Travel time is measured using public transportation for Bunkyo Ward, while it is measured by driving times for Koriyama City, as explained in \Cref{sec:data:open-source}. 
Other variables are as follows: $\alpha_s$ is a fixed effect for daycare $\sch$ that controls the average quality of each daycare; $\beta$ is a coefficient that measures the disutility of distance; and $\epsilon_{\st \sch}$ is an idiosyncratic error term that follows the Gumbel distribution, independently across both children and daycares. 
Although we treat the daycare slots for each age as a separate ``daycare,'' we nonetheless estimate a single fixed effect $\alpha_\sch$ per facility across all ages, since the small number of applicants at ages 3, 4, and 5 makes age-specific estimates unreliable.
As usual, we impose a location normalization to identify the logit model. Under STT and USS, we set the outside option fixed effect $\alpha_\n$ to zero. 
For WTT and OETT, which exclude the outside option from the alternatives, we instead normalize the fixed effect of one daycare (which we label as $s_1$) to zero.

In OETT, we have in mind a situation in which parents collect and use cutoff information from previous years when forming a choice set, as discussed in the opening paragraph of \Cref{sec:strategic_assumptions:OETT}. However, we do not have cutoff data from the previous years, and thus we use the cutoffs from the year of the data as a proxy for the cutoffs from previous years. Formally, we define daycare $\sch$'s benchmark cutoff score $\underline{e}_\sch$ as the lowest score among the children matched to $\sch$ in the year of the data if it is at full capacity and zero if it is not. 


We estimate 16 models in total, which is a product of the four strategic assumptions (STT, WTT, OETT, USS), two utility formulations (Linear, Log), and two distance metrics (Straight Line Distance, Travel Time). 
We estimate the parameters  $((\alpha_s)_{\sch \in \Sch}, \beta)$ via the maximum likelihood estimation or the generalized method of moments. 
\Cref{tab:reg_bunkyo} reports the estimation results for Bunkyo Ward, while \Cref{tab:reg_koriyama} reports those for Koriyama City. 
Alongside the estimated distance coefficient $\hat\beta$, each table shows the scaled fixed effects $\hat\alpha_\sch / \hat\beta$, which express the relative importance of average daycare quality in distance units. 
Since each model includes a large number of fixed effects, we only provide the mean and the standard deviation.

We first note that all the estimated coefficients in Panel A are significantly different from zero at the 95\% confidence level. Among these, the confidence interval under USS is wider than  those under STT, WTT, and OETT. This is because USS uses the data in a less informative (although  more robust) manner: it incorporates only the matched daycare into the moment equalities, and the other daycares listed in the ROL enter through moment inequalities. In contrast, STT, WTT, and OETT use the full ROL information. \cite{fack2019beyond} make similar points by reporting that the confidence interval under USS is wider than that under WTT.

We also highlight a comparison between WTT and OETT. To begin, we observe that WTT can be seen as a special case of OETT in which the optimism parameter is so large that all daycares are in the choice set of each child. Given this observation, it is of interest to compare and interpret the estimation results from these two models. In Panel B we observe that, across every combination of a distance measure and a utility specification, the standard deviation of the ``school fixed effects divided by coefficients'' under OETT is larger than those under WTT. 
For instance, in \Cref{tab:reg_bunkyo}, the standard deviation for (Straight Line, Linear, OETT) is 0.258, whereas the corresponding value is 0.249 under WTT. Thus, under OETT, variation in average quality of daycares is estimated to matter more (relative to commuting distance) than it does in the other specifications. 
This pattern is consistent with a child who prefers higher-quality daycares in her true preference relation but omits them from her ROLs when admission cutoffs are expected to be high. WTT would misinterpret such omissions as indicating lower utilities for those daycares, which underestimates the effect of the average quality. 

Finally, we note that our specification of the utility function misses variables that may be important in practice, primarily due to data limitation. In particular, parents may find it convenient to have their child matched to a daycare that is close to their workplace, but our specification does not capture such a consideration. As such, our analysis may underestimate the welfare gains from integration of daycare matching markets.

\subsection{Goodness of fit}\label{sec:goodness_of_fit}

Among the estimated models in \Cref{sec:est_params}, we aim to select the one that most accurately reproduces the observed matching patterns and use it for our counterfactual simulations.
To evaluate each model's fit, we simulate the ROLs of the actual children in the dataset using the estimated parameters and the estimation model, run Serial Dictatorship with the observed priority order, and then compare the resulting simulated matching to the actual matching in the data.
\footnote{For each child with a missing address, we randomly sample an address from the empirical distribution of all observed addresses in the dataset.}

We quantify goodness of fit by \emph{area-level matching differences}. Concretely, we partition Bunkyo Ward into the seven residential areas defined in \Cref{sec:data_bunkyo}, and Koriyama City into its three former municipalities defined in \Cref{sec:data_koriyama}. We call them \emph{areas}. For each simulated matching, we count how many children at each age are matched to daycares in each area, compute the absolute difference from the corresponding counts in the actual matching, and finally take the (unweighted) average of these absolute differences over all ages and areas.

Specifically, given the estimated parameters $(\hat\alpha_\sch)_{\sch \in S}$ and $\hat\beta$ (and $\hat\psi = (\hat\kappa, \hat\lambda)$ for OETT), we compute the area-level matching differences by the following procedure:
\begin{enumerate}
\item \textbf{Utility generation (daycares).} For each child $\st$, we generate $\st$'s utility $u_{\st}$ following 
\begin{align*}
\text{\textbf{(Linear Utility)}} \quad u_{\st \sch} &= \hat\alpha_s - \hat\beta d_{\st \sch} + \epsilon_{\st \sch}, \\
\text{\textbf{(Log Utility)}} \quad u_{\st \sch} &= \hat\alpha_s - \hat\beta \log d_{\st \sch} + \epsilon_{\st \sch},
\end{align*}
for each daycare $\sch \in S$, where $\epsilon_{\st \sch}$ is drawn from the Gumbel distribution.

\item \textbf{Utility generation (outside option)}. For STT and USS, we also sample each child's outside option value $u_{\st \n} = \epsilon_{\st \n}$, where  $\epsilon_{\st \n}$ is drawn from the Gumbel distribution.  For WTT and OETT, we do not generate outside option values because 
we do not use them in the ROL construction explained below.

\item \textbf{ROL construction.}  For each child, the daycares and the outside option are sorted according to the descending order of the utilities (with uniform random tie-breaking). Under STT, we then truncate this list right below the $K^{ \text{th}}$ entry or right above the outside option, whichever comes first, to form her ROL. Under WTT, we truncate the list right below the top $K_i$ daycares. For OETT, student $\st$'s ROL starts with the daycare with the highest utility among the set of daycares whose benchmark cutoff score minus $\st$'s optimism  degree  $o_i$  is no greater than her score,
where $o_i$ is drawn from $F_{\hat\psi}$.
The ROL is then truncated right below the top $K_i$ daycares in this set. Under USS, we list all the daycares above the outside option.
\footnote{Without truncation, Serial Dictatorship outputs the unique stable matching (under the applicants' true preferences).}

\item \textbf{Matching via Serial Dictatorship.} Following the ROLs construted in the previous step and the priority order observed in the dataset, we run Serial Dictatorship to determine the simulated matching for children at each age.\footnote{Although we run Serial Dictatorship without any explicit constraint on the ROL length here, note that the length of each submitted ROL is at most $K$ (as imposed in practice) under STT, WTT, and OETT.}

\item \textbf{Goodness of fit computation.} For each age $a$ and area, we first calculate (i) the number of children at age $a$ matched to daycares in that area in the simulated matching and (ii) the number of children at age $a$ matched to daycares in the area in the dataset. Then, we take the average of the absolute differences between (i) and (ii) across all ages and areas, weighted by the population of each age in the dataset.
\end{enumerate}

We performed 100 simulations, and  \Cref{tab:fit_bunkyo} reports the goodness of fit rankings for Bunkyo Ward, while  \Cref{tab:fit_koriyama} does likewise for Koriyama City. In each table, models are sorted in ascending order of area-level matching differences.  We found that the combination of (Travel Time, Log, OETT) is the best fit for Bunkyo Ward and the second best fit for Koriyama City. Given this finding, we use (Travel Time, Log, OETT) in our counterfactual analysis in the next section.\footnote{The best fit for Koriyama City (which is (Travel Time, Linear, OETT)) is in the fifth place for Bunkyo Ward.}

In the estimation procedure under WTT and OETT, we normalize the fixed effect of a reference daycare $\sch_1$ instead of that of the outside option, as discussed in \Cref{sec:est_params}. For our counterfactual simulations, however, we must shift the average quality of daycare $\sch_1$ relative to the average quality of the outside option, which we normalize to zero. 
To achieve this, we choose to estimate the average utility of the outside option using the simulated utilities generated in the previous 100 simulation runs. Specifically, we assume that the utility of the outside option is given by $u^m_{i\n} = \bar{\alpha} + \epsilon^m_{i\n}$, where the error term $\epsilon^m_{i\n}$ in the $m$-th simulation run is drawn independently from the Gumbel distribution, and then estimate $\bar{\alpha}$ by maximum likelihood estimation. 
In each run, if child $i$'s ROL has length $K_i$, we consider the likelihood that 
the outside option is located between the $K_i^\text{ th}$ most preferred daycare and the $(K_i+\ell)^\text{ th}$ most preferred daycare:
\begin{align*}
\prob \left( \tilde{u}^m_{i s_i(K_i)} \geq u^m_{i\n} \geq \tilde{u}^m_{i s_i(K_i + \ell)} \right)
= \prob \left( \tilde{u}^m_{i s_i(K_i)} - \bar\alpha \geq \epsilon^m_{i\n} \geq \tilde{u}^m_{i s_i(K_i + \ell)} - \bar\alpha \right),
\end{align*}
where $\tilde{u}^m_{i s_i(k)}$ denotes the simulated utility of the $k^\text{th}$ most preferred daycare in child $i$'s choice set in the $m$-th run.\footnote{Under WTT, the choice set consists of all the acceptable daycares. For OETT, see \Cref{sec:strategic_assumptions:OETT} for the details on the construction of the choice set.}$^,$\footnote{When $K_i + \ell$ exceeds the size of $\st$'s choice set, we define $\tilde{u}^m_{i s_i(K_i + \ell)} = -\infty$.} 
We maximize the sum of the log-likelihoods over all simulation runs and all children. 
We present the result for $\ell = 5$ here, while we confirmed that the results are qualitatively similar for other values of 
$\ell$.\footnote{We analyzed the cases of $\ell=1, 9$ in addition to $\ell=5$. We note that if $\ell$ is sufficiently large relative to the sizes of the choice sets, the lower bound constraints no longer bind, and the estimate $\hat{\bar{\alpha}}$ diverges to $-\infty$.} After obtaining the estimate $\hat{\bar{\alpha}}$, we shift all daycare fixed effects by subtracting it, i.e., $\hat{\alpha}_\sch \leftarrow \hat{\alpha}_\sch - \hat{\bar{\alpha}}$ for each $s\in S$.

\input{tables/section6/tab_bunkyo_sim_results}
\input{tables/section6/tab_koriyama_sim_results}

%% file: tables/section6/tab_estimation_reg_bunkyo.tex
\begin{table}[bthp]
\centering
\fontsize{8pt}{12pt}\selectfont
\begin{threeparttable}
\renewcommand{\arraystretch}{1.0}
\caption{Estimation Results for Bunkyo Ward}
\label{tab:reg_bunkyo}

\begin{tabular*}{\textwidth}{
    l @{\extracolsep{\fill}} 
    S[table-format=1.3,table-space-text-post={}] @{\hskip 0.0em} 
    S[table-format=1.3,table-space-text-post={}] @{\hskip 0.0em} 
    S[table-format=1.3,table-space-text-post={}] @{\hskip 0.0em} 
    S[table-format=2.3,table-space-text-post={}] @{\hskip 1.0em} 
    S[table-format=1.3,table-space-text-post={}] @{\hskip 0.0em} 
    S[table-format=1.3,table-space-text-post={}] @{\hskip 0.0em} 
    S[table-format=1.3,table-space-text-post={}] @{\hskip 0.0em} 
    S[table-format=1.3,table-space-text-post={}] @{\hskip 1.0em} 
}
\toprule
& \multicolumn{8}{c}{\textbf{Straight Line Distance}} \\ \cmidrule(){2-9}
& \multicolumn{4}{c}{\textbf{Linear Utility}} & \multicolumn{4}{c}{\textbf{Log Utility}} \\
\cmidrule(){2-5} \cmidrule(){6-9}
& \mulcb{STT} & \mulcb{WTT} & \mulcb{OETT} & \mulcb{USS} & \mulcb{STT} & \mulcb{WTT} & \mulcb{OETT} & \mulcb{USS} \\ 
& \mulc{(1)} & \mulc{(2)} & \mulc{(3)} & \mulc{(4)} & \mulc{(5)} & \mulc{(6)} & \mulc{(7)} & \mulc{(8)} \\
\midrule
\multicolumn{9}{l}{\textbf{Panel A: Coefficient $\hat\beta$}} \\
Distance & 3.244 & 3.395 & 3.448 & 3.658 &  &  &  &  \\
& \mulc{\tiny [3.154, 3.333]} & \mulc{\tiny [3.297, 3.492]} & \mulc{\tiny [3.348, 3.547]} & \mulc{\tiny [3.019, 4.325]} &  &  &  &  \\[0.25em]
log(Distance) &  &  &  &  & 3.314 & 3.660 & 3.664 & 3.548 \\
&  &  &  &  & \mulc{\tiny [3.238, 3.389]} & \mulc{\tiny [3.567, 3.753]} & \mulc{\tiny [3.571, 3.757]} & \mulc{\tiny [2.889, 4.215]} \\
\midrule
\multicolumn{9}{l}{\textbf{Panel B: Scaled Daycare Fixed Effects $\hat\alpha_\sch / \hat\beta$}} \\
Mean & -0.252 & 0.014 & -0.092 & -0.489 & -0.188 & 0.015 & -0.072 & -1.027 \\
Std & 0.259 & 0.249 & 0.258 & 0.345 & 0.224 & 0.236 & 0.245 & 0.373 \\
\midrule
\multicolumn{9}{l}{\textbf{Panel C: Distribution Parameters $\hat\psi$}} \\
$\hat\kappa$ &  &  & 2.922 &  &  &  & 2.921 &  \\
$\hat\lambda$ &  &  & 0.530 &  &  &  & 0.516 &  \\
\midrule
Observations & \mulc{1568} & \mulc{1568} & \mulc{1568} & \mulc{1568} & \mulc{1568} & \mulc{1568} & \mulc{1568} & \mulc{1568} \\
Number of Parameters & \mulc{127} & \mulc{126} & \mulc{128} & \mulc{127} & \mulc{127} & \mulc{126} & \mulc{128} & \mulc{127} \\
\midrule \\
\midrule
& \multicolumn{8}{c}{\textbf{Travel Time}} \\ \cmidrule(){2-9}
& \multicolumn{4}{c}{\textbf{Linear Utility}} & \multicolumn{4}{c}{\textbf{Log Utility}} \\
\cmidrule(){2-5} \cmidrule(){6-9}
& \mulcb{STT} & \mulcb{WTT} & \mulcb{OETT} & \mulcb{USS} & \mulcb{STT} & \mulcb{WTT} & \mulcb{OETT} & \mulcb{USS} \\ 
& \mulc{(9)} & \mulc{(10)} & \mulc{(11)} & \mulc{(12)} & \mulc{(13)} & \mulc{(14)} & \mulc{(15)} & \mulc{(16)} \\
\midrule
\multicolumn{9}{l}{\textbf{Panel A: Coefficient $\hat\beta$}} \\
Travel Time & 0.188 & 0.204 & 0.208 & 0.195 &  &  &  &  \\
& \mulc{\tiny [0.183, 0.193]} & \mulc{\tiny [0.198, 0.210]} & \mulc{\tiny [0.202, 0.214]} & \mulc{\tiny [0.165, 0.230]} &  &  &  &  \\[0.25em]
log(Travel Time) &  &  &  &  & 1.948 & 3.933 & 3.945 & 1.888 \\
&  &  &  &  & \mulc{\tiny [1.888, 2.009]} & \mulc{\tiny [3.832, 4.033]} & \mulc{\tiny [3.844, 4.046]} & \mulc{\tiny [1.737, 2.046]}\\
\midrule
\multicolumn{9}{l}{\textbf{Panel B: Scaled Daycare Fixed Effects $\hat\alpha_\sch / \hat\beta$}} \\
Mean & -4.186 & 0.242 & -1.219 & -7.978 & -0.455 & 0.020 & -0.054 & -0.012 \\
Std & 4.537 & 4.213 & 4.340 & 6.276 & 0.439 & 0.224 & 0.232 & 0.597 \\
\midrule
\multicolumn{9}{l}{\textbf{Panel C: Distribution Parameters $\hat\psi$}} \\
$\hat\kappa$ & & & 2.684 &  &  &  & 2.692 &  \\
$\hat\lambda$ & & & 0.453 &  &  &  & 0.454 &  \\
\midrule
Observations & \mulc{1568} & \mulc{1568} & \mulc{1568} & \mulc{1568} & \mulc{1568} & \mulc{1568} & \mulc{1568} & \mulc{1568} \\
Number of Parameters & \mulc{127} & \mulc{126} & \mulc{128} & \mulc{127} & \mulc{127} & \mulc{126} & \mulc{128} & \mulc{127} \\
\bottomrule
\end{tabular*}

\begin{tablenotes}
\renewcommand{\baselinestretch}{0.8}
\small
\item \textit{Notes:} Panel A reports the estimated coefficients, with 95 percent confidence intervals in brackets. Panel B shows the mean and the standard deviation of the estimated fixed effects, each divided by the estimated coefficient. For STT and USS, the fixed effects are normalized by subtracting $\hat\alpha_{\sch_1}$ from every $\hat\alpha_\sch$ so that they are comparable to the WTT and OETT estimates. Panel C reports the estimated parameters of the gamma distribution used in OETT. 
The number of observations equals the count of children included in the estimation (excluding those with missing addresses). 
\end{tablenotes}

\end{threeparttable}
\end{table}

%% file: tables/section6/tab_estimation_reg_koriyama.tex
\begin{table}[bthp]
\centering
\fontsize{8pt}{12pt}\selectfont
\begin{threeparttable}
\renewcommand{\arraystretch}{1.0}
\caption{Estimation Results for Koriyama City}
\label{tab:reg_koriyama}

\begin{tabular*}{\textwidth}{
    l @{\extracolsep{\fill}} 
    S[table-format=-1.3,table-space-text-post={}] @{\hskip 0.0em} 
    S[table-format=-1.3,table-space-text-post={}] @{\hskip 0.0em} 
    S[table-format=-1.3,table-space-text-post={}] @{\hskip 0.0em} 
    S[table-format=-1.3,table-space-text-post={}] @{\hskip 1.0em} 
    S[table-format=-1.3,table-space-text-post={}] @{\hskip 0.0em} 
    S[table-format=-1.3,table-space-text-post={}] @{\hskip 0.0em} 
    S[table-format=-1.3,table-space-text-post={}] @{\hskip 0.0em} 
    S[table-format=-1.3,table-space-text-post={}]
}
\toprule
& \multicolumn{8}{c}{\textbf{Straight Line Distance}} \\ \cmidrule(){2-9}
& \multicolumn{4}{c}{\textbf{Linear Utility}} & \multicolumn{4}{c}{\textbf{Log Utility}} \\
\cmidrule(){2-5} \cmidrule(){6-9}
& \mulcb{STT} & \mulcb{WTT} & \mulcb{OETT} & \mulcb{USS} & \mulcb{STT} & \mulcb{WTT} & \mulcb{OETT} & \mulcb{USS} \\ 
& \mulc{(1)} & \mulc{(2)} & \mulc{(3)} & \mulc{(4)} & \mulc{(5)} & \mulc{(6)} & \mulc{(7)} & \mulc{(8)} \\
\midrule
\multicolumn{9}{l}{\textbf{Panel A: Coefficient $\hat\beta$}} \\
Distance & 0.917 & 1.021 & 1.120 & 1.116 &  &  &  &  \\[0.0em]
& \mulc{\tiny [0.892, 0.943]} & \mulc{\tiny [0.992, 1.050]} & \mulc{\tiny [1.087, 1.152]} & \mulc{\tiny [0.863, 1.436]} &  &  &  &  \\[0.25em]
log(Distance) &  &  &  &  & 1.745 & 1.808 & 1.876 & 1.875 \\
&  &  &  &  & \mulc{\tiny [1.709, 1.781]} & \mulc{\tiny [1.768, 1.848]} & \mulc{\tiny [1.833, 1.919]} & \mulc{\tiny [1.548, 2.331]} \\
\midrule
\multicolumn{9}{l}{\textbf{Panel B: Scaled Daycare Fixed Effects $\hat\alpha_\sch / \hat\beta$}} \\
Mean & 0.102 & -0.187 & -0.278 & -1.244 & -0.112 & -0.097 & -0.171 & -1.020 \\
Std & 1.013 & 0.814 & 1.014 & 1.216 & 0.448 & 0.441 & 0.453 & 0.661 \\
\midrule
\multicolumn{9}{l}{\textbf{Panel C: Distribution Parameters $\hat\psi$}} \\
$\hat\kappa$ & & & 248.050 &  &  &  & 205.532 &  \\
$\hat\lambda$ & & & 0.452 &  &  &  & 0.374 &  \\
\midrule
Observations & \mulc{1295} & \mulc{1295} & \mulc{1295} & \mulc{1295} & \mulc{1295} & \mulc{1295} & \mulc{1295} & \mulc{1295} \\
Number of Parameters & \mulc{89} & \mulc{88} & \mulc{90} & \mulc{89} & \mulc{89} & \mulc{88} & \mulc{90} & \mulc{89} \\
\midrule \\
\midrule
& \multicolumn{8}{c}{\textbf{Travel Time}} \\ \cmidrule(){2-9}
& \multicolumn{4}{c}{\textbf{Linear Utility}} & \multicolumn{4}{c}{\textbf{Log Utility}} \\
\cmidrule(){2-5} \cmidrule(){6-9}
& \mulcb{STT} & \mulcb{WTT} & \mulcb{OETT} & \mulcb{USS} & \mulcb{STT} & \mulcb{WTT} & \mulcb{OETT} & \mulcb{USS} \\ 
& \mulc{(9)} & \mulc{(10)} & \mulc{(11)} & \mulc{(12)} & \mulc{(13)} & \mulc{(14)} & \mulc{(15)} & \mulc{(16)} \\
\midrule
\multicolumn{9}{l}{\textbf{Panel A: Coefficient $\hat\beta$ }} \\
Travel Time & 0.345 & 0.386 & 0.409 & 0.368 &  &  &  &  \\
& \mulc{\tiny [0.336, 0.354]} & \mulc{\tiny [0.376, 0.396]} & \mulc{\tiny [0.398, 0.420]} & \mulc{\tiny [0.308, 0.439]} &  &  &  &  \\[0.25em]
log(Travel Time) &  &  &  &  & 1.920 & 2.417 & 2.492 & 1.945 \\
&  &  &  &  & \mulc{\tiny [1.871, 1.969]} & \mulc{\tiny [2.362, 2.473]} & \mulc{\tiny [2.433, 2.551]} & \mulc{\tiny [1.745, 2.246]} \\
\midrule
\multicolumn{9}{l}{\textbf{Panel B: Scaled Daycare Fixed Effects $\hat\alpha_\sch / \hat\beta$}} \\
Mean & -1.954 & -0.670 & -1.766 & 0.617 & -0.347 & -0.099 & -0.256 & 0.250 \\
Std & 2.197 & 2.013 & 2.238 & 3.027 & 0.430 & 0.325 & 0.333 & 0.600 \\
\midrule
\multicolumn{9}{l}{\textbf{Panel C: Distribution Parameters $\hat\psi$}} \\
$\hat\kappa$ & & & 205.464 &  &  &  & 205.534 &  \\
$\hat\lambda$ & & & 0.375 &  &  &  & 0.374 &  \\
\midrule
Observations & \mulc{1295} & \mulc{1295} & \mulc{1295} & \mulc{1295} & \mulc{1295} & \mulc{1295} & \mulc{1295} & \mulc{1295} \\
Number of Parameters & \mulc{89} & \mulc{88} & \mulc{90} & \mulc{89} & \mulc{89} & \mulc{88} & \mulc{90} & \mulc{89} \\
\bottomrule
\end{tabular*}

\begin{tablenotes}
\renewcommand{\baselinestretch}{0.8}
\small
\item \textit{Notes:} Panel A reports the estimated coefficients, with 95 percent confidence intervals in brackets. Panel B shows the mean and the standard deviation of the estimated fixed effects, each divided by the estimated coefficient. For STT and USS, the fixed effects are normalized by subtracting $\hat\alpha_{\sch_1}$ from every $\hat\alpha_\sch$ so that they are comparable to the WTT and OETT estimates. Panel C reports the estimated parameters of the gamma distribution used in OETT. 
The number of observations equals the count of children included in the estimation (excluding those with missing addresses). 
\end{tablenotes}

\end{threeparttable}
\end{table}

%% file: tables/section6/tab_bunkyo_sim_results.tex
\begin{table}[bthp]
\centering
\scriptsize
\begin{threeparttable}
\renewcommand{\arraystretch}{1.0}
\caption{Goodness of Fit for Bunkyo Ward (Average Across 100 Simulations)}
\label{tab:fit_bunkyo}
\centering
\begin{tabular*}{0.8\textwidth}{
    ccc @{\extracolsep{\fill}} 
    S[table-format=1.3,table-space-text-post={}]
}
\toprule
\mulcb{Distance} & 
\mulcb{Utility} & 
\mulcb{Assumption} & 
\mulcb{\shortstack{Average Area-Level Matching\\Differences (Simulated vs. Actual)}} \\ 
\midrule
Travel Time & Log & OETT & 2.373 \\
Travel Time & Log & WTT & 2.376 \\
Straight Line & Log & OETT & 2.532 \\
Straight Line & Log & WTT & 2.539 \\
Travel Time & Linear & OETT & 2.641 \\
Travel Time & Linear & WTT & 2.689 \\
Straight Line & Linear & OETT & 2.922 \\
Straight Line & Linear & WTT & 2.999 \\
Straight Line & Log & STT & 3.013 \\
Straight Line & Log & USS & 3.035 \\
Straight Line & Linear & USS & 3.128 \\
Travel Time & Linear & USS & 3.139 \\
Travel Time & Linear & STT & 3.627 \\
Straight Line & Linear & STT & 3.789 \\
Travel Time & Log & STT & 5.404 \\
Travel Time & Log & USS & 5.414 \\
\bottomrule
\end{tabular*}
\begin{tablenotes}
\renewcommand{\baselinestretch}{0.8}
\small
\item \textit{Notes:} 
We compare the number of matches for each age and area between the actual data and the simulated data. We run 100 simulations for each combination of distance metric, utility formulation, and strategic assumption.

\end{tablenotes}
\end{threeparttable}
\end{table}

%% file: tables/section6/tab_koriyama_sim_results.tex
\begin{table}[bthp]
\centering
\scriptsize
\begin{threeparttable}
\renewcommand{\arraystretch}{1.0}
\caption{Goodness of Fit for Koriyama City (Average Across 100 Simulations)}
\label{tab:fit_koriyama}
\centering
\begin{tabular*}{0.8\textwidth}{
    ccc @{\extracolsep{\fill}} 
    S[table-format=2.3,table-space-text-post={}]
}
\toprule
\mulcb{Distance} & 
\mulcb{Utility} & 
\mulcb{Assumption} & 
\mulcb{\shortstack{Average Area-Level Matching\\Differences (Simulated vs. Actual)}} \\ 
\midrule
Travel Time & Linear & OETT & 14.911 \\
Travel Time & Log & OETT & 14.977 \\
Straight Line & Log & OETT & 15.009 \\
Straight Line & Log & USS & 15.025 \\
Travel Time & Linear & WTT & 15.069 \\
Travel Time & Log & WTT & 15.079 \\
Straight Line & Log & WTT & 15.124 \\
Straight Line & Log & STT & 15.135 \\
Straight Line & Linear & USS & 15.153 \\
Travel Time & Linear & USS & 15.199 \\
Straight Line & Linear & OETT & 15.202 \\
Travel Time & Linear & STT & 15.283 \\
Travel Time & Log & USS & 15.317 \\
Straight Line & Linear & STT & 15.381 \\
Straight Line & Linear & WTT & 15.471 \\
Travel Time & Log & STT & 15.537 \\
\bottomrule
\end{tabular*}
\begin{tablenotes}
\renewcommand{\baselinestretch}{0.8}
\small
\item \textit{Notes:} 
We compare the number of matches for each age and area between the actual data and the simulated data. We run 100 simulations for each combination of distance metric, utility formulation, and strategic assumption. 
\end{tablenotes}
\end{threeparttable}
\end{table}

%% file: sections/counterfactual.tex
\clearpage

\section{Impact of Integration in Daycare Matching Markets}
\label{sec:simulation}

To assess how market integration influences child welfare, we run counterfactual simulations in two settings that are currently in contrasting situations. 
In Tokyo City, the daycare matching market is fragmented: each of the 23 wards handles its own matching process independently. 
In Koriyama City, the daycare matching market is integrated: as a single municipality, it runs one unified matching process. 
We consider the outcomes in which the Tokyo City market is hypothetically integrated and the Koriyama City market is hypothetically fragmented. 

As discussed in \Cref{sec:cross_specific_region}, we consider two different modes of partial integration, namely cases where balancedness constraints are imposed for cross-age regions and for age-specific regions. 
Specifically, for our simulations, we define the cross-age regions $\tilde{R}$ to be the 23 wards for Tokyo City, and the three former municipalities for Koriyama City. 
The set of age-specific regions $\hat{R}$ is defined accordingly for each city.

As stated in \Cref{sec:goodness_of_fit}, we use (Travel Time, Log, OETT) to  simulate each child's ROL of daycares and apply different mechanisms to compare welfare under different degrees of integration. 

Specifically, we run the following four mechanisms: 
\begin{enumerate-mechanism}
  \item \label{mec:Frag}
  Serial Dictatorship conducted separately within each region in $\tilde{R}$.
  \footnote{
  The outcome of the mechanism is the same if we use $\hat{R}$ instead of $\tilde{R}$ here. The same comment applies to \Cref{mec:DA-all}.
  }$^{,}$
  \footnote{
  In Serial Dictatorship, each child can only be matched to daycares within her own residential region. The same rule applies to Serial Dictatorship in \Cref{mec:FIG-specific,mec:FIG-cross}. 
  Although some children do apply to daycares in Bunkyo Ward and Koriyama City from outside 
  those areas in our data, we maintain this restriction in counterfactual simulations so as to precisely identify the difference between the outcomes of fragmentation and integration. We note that the number of children who live outside of Bunkyo Ward and apply to daycares in Bunkyo Ward is small (6.7\%).
  }
  
  \item\label{mec:FIG-specific} FIG cycles mechanism applied after Serial Dictatorship conducted separately in each region in $\hat{R}$ (age-specific regions). 
  
  \item\label{mec:FIG-cross} FIG cycles mechanism applied after Serial Dictatorship conducted separately in each region in $\tilde{R}$ (cross-age regions).  
  
  \item \label{mec:DA-all} DA conducted across all regions in $\tilde{R}$. 
\end{enumerate-mechanism}

Here, \Cref{mec:Frag} represents complete fragmentation, \Cref{mec:DA-all}  full integration, and \Cref{mec:FIG-specific,mec:FIG-cross} partial integration under different notions of balancedness. 
The difference between \Cref{mec:FIG-specific,mec:FIG-cross} is that  \Cref{mec:FIG-specific} requires balancedness at each age, whereas \Cref{mec:FIG-cross} requires balancedness only in the total number of matches across all ages. 

In our counterfactual simulations, we do not impose any maximum-length constraint on the ROLs and assume that all children submit their preferences truthfully in all mechanisms.
\footnote{
We believe this is not a purely theoretical exercise. Indeed, after discussions with us about increasing ROL lengths, Bunkyo Ward increased the maximum from five (as in our dataset) to ten in 2025.}

\subsection{Preference and Priority Generation}

\subsubsection{Tokyo City}

Since the actual matching data are available only for Bunkyo Ward, we estimate the number of applicants in each of Tokyo's 23 wards as follows. For ward $w$ and age $a \in A = \{ 0,1,2,3,4,5 \}$, we set the number of applicants as follows. 
\begin{align*}
\bigl( \text{Total daycare users in ward $w$} \bigr)
\times
\kappa_a,
\end{align*}
where ``total daycare users in ward $w$'' refers to the number of all children enrolled in any daycare in $w$ as of April 2022, according to the data published by the Tokyo Prefectural Government. The age-specific scaling constant $\kappa_a$ is defined as
\begin{align*}
\kappa_a
&=
\frac{\text{Number of applicants at age $a$ of Bunkyo Ward}}
{\text{Total daycare users in Bunkyo Ward}}.
\end{align*}
Thus, $\kappa_a$ represents the ratio of applicants at age $a$ among all daycare users in Bunkyo Ward, and we assume the same ratio holds true for other wards as well.

For each ward, we sample children's residential locations according to the empirical population distribution for ages 0-4. Because population counts are only available at every geographic-subdivision level, we assume each child resides at the centroid of her respective subdivision when calculating the distance to daycares.

We generate each child's utility in two steps. First, because we lack data on children's preferences over daycares outside Bunkyo Ward, we need to determine each daycare's average quality in Tokyo City. To do so, we draw each daycare's average quality $\tilde\alpha_\sch$ at random from the empirical distribution of the estimated fixed effects $(\hat\alpha_\sch)$ in Bunkyo Ward. 
Then, we generate each child $\st$'s utility for being matched to daycare $\sch$ by
\begin{align*}
u_{is} &= \tilde\alpha_\sch - \hat\beta\,d_{\st \sch} + \epsilon_{\st \sch},
\end{align*}
where $\hat\beta$ is the coefficient estimated from the observed data, $d_{is}$ is the distance between child $\st$ and daycare $s$, and $\epsilon_{is}$ is drawn i.i.d.~from the Gumbel distribution.  
We normalize the average utility of the outside option to zero and set $u_{i \n} = \epsilon_{i \n}$, where $\epsilon_{i\n}$ also follows the Gumbel distribution. 
Each child then constructs their ROL by sorting all daycares in descending order of $u_{\st\sch}$, listing them down to the utility of the outside option $u_{\st \n}$ without a constraint on the ROL length. 
We randomly break ties in utility when they occur, although ties rarely arise in our simulation.

Priority orders are constructed as follows. First, we draw a single ``master'' priority ranking over all children uniformly at random. To obtain each ward's priority list, we reorder that master ranking by moving every child residing in that ward ahead of all non-residents while preserving the relative order within the resident and non-resident groups. This procedure captures the fact that all 23 wards rank children based on criteria that are not identical but broadly similar, and that each ward gives priority to its own residents. In all mechanisms, we use these priorities in which locals are favored.

\subsubsection{Koriyama City}

For Koriyama City, we have access to the dataset of the entire market, so we use the actual number of applicants at each age in our simulations. 
Some children's address information is missing as stated in \Cref{sec:data_koriyama}, and we sample each child's address from the empirical distribution of residential locations in the dataset.

We generate each child $i$'s utility function and then her ROL as in Tokyo City, except that we use the estimated fixed effects directly as the average daycare quality parameters, i.e., $\tilde\alpha_\sch = \hat\alpha_\sch$ for each $\sch$ because, unlike in the Tokyo City case, we have estimated  the fixed effect  for every daycare in Koriyama City.
Priority orders are also constructed in the same manner as  those of Tokyo City, except  that  we use the original master ranking directly in \Cref{mec:DA-all} (in Mechanisms \ref{mec:Frag}, \ref{mec:FIG-specific}, and \ref{mec:FIG-cross}, we use the region-specific orders). 
With this configuration, \Cref{mec:DA-all} mimics Koriyama City's current mechanism, which is Serial Dictatorship with a maximum-length constraint (this constraint  rarely binds in our data, as shown in \Cref{tab:5_summary_stats} in \Cref{sec:data_koriyama}).\footnote{Note that Serial Dictatorship is a special case of DA in which there is a master ranking shared by all schools.\label{fn:SDDA}}  
Since locals are not necessarily favored under this specification of priorities, \Cref{mec:DA-all} does not necessarily Pareto dominate the other mechanisms, in contrast to the Tokyo City case. 
\footnote{Although Pareto dominance does not generally hold, \Cref{sec:counterfactual_result_koriyama} shows that children's utilities improve on average in our simulation.
}

\subsection{Simulation Results}

We conducted 100 simulation runs for each mechanism. Statistics presented in this section are averages across those runs.

\subsubsection{Tokyo City}
\label{sec:counterfactual_result_tokyo}

\Cref{tab:welfare_comp_tokyo} reports the results for Tokyo City. Recall that the average child utility is normalized so that the outside option has zero expected value and unobserved shocks follow the Gumbel distribution whose standard deviation is $\pi / \sqrt{6}$ ($\simeq 1.283$).
The table shows that the full market integration via \Cref{mec:DA-all} raises the average child utility by 39.0\% relative to the fragmented case (from 0.854 to 1.187) and entails the interregional match rate of 21.5\% (where interregional match rate is defined as the share of children who are matched to the daycares outside their home regions). 
By contrast, partial market integration via \Cref{mec:FIG-specific} yields a 15.3\% increase in average child utility (from 0.854 to 0.985) with a 9.1\% interregional match rate, and the one via \Cref{mec:FIG-cross} yields a 16.2\% increase (from 0.854 to 0.993) with a 9.5\% interregional match rate. Thus, partial integration under balancedness captures 39.2\% and 41.5\% of the welfare gains achieved by full integration, respectively.
\Cref{tab:welfare_comp_tokyo} also reports the match rate, which shows that Mechanisms \ref{mec:FIG-specific}, \ref{mec:FIG-cross}, and \ref{mec:DA-all} reduce the unmatch rate by 5.2, 5.5, and 13.1 percentage points, respectively. Therefore,  Mechanisms \ref{mec:FIG-specific} and  \ref{mec:FIG-cross} achieve 40.0\% and 41.9\% of the reduction in the unmatch rate of  \Cref{mec:DA-all}, respectively.
We also calculate the same statistics among children whose closest daycare lies outside their home region. These areas are shown in \Cref{fig:3_voronoi_tokyo} in \Cref{sec:background-tokyo}. \Cref{tab:welfare_comp_tokyo} reports that, among other things, even under the balancedness constraints (\Cref{mec:FIG-specific,mec:FIG-cross}), the average child utility increases by 27.7\% (from 0.696 to 0.889) and 29.3\% (from 0.696 to 0.899), and 14.3\% and 14.9\% of the children are matched to daycares outside their home regions, respectively.

These welfare gains can be interpreted in terms of travel time by dividing them by the coefficient $\hat\beta = 3.945$. The improvements of all children achieved by Mechanisms \ref{mec:FIG-specific}, \ref{mec:FIG-cross}, and \ref{mec:DA-all} relative to \Cref{mec:Frag} correspond to 3.3\%, 3.5\%, and 8.4\% reductions in travel time to daycares, respectively. For the children whose closest daycare lies outside their home region, the corresponding reductions are 4.9\%, 5.2\%, and 12.0\%.

\input{tables/section7/tab_simulation_tokyo}

\begin{figure}[tbhp]
\footnotesize
  \centering
  \begin{subfigure}[b]{0.49\textwidth}
    \centering
    \includegraphics[width=\linewidth]{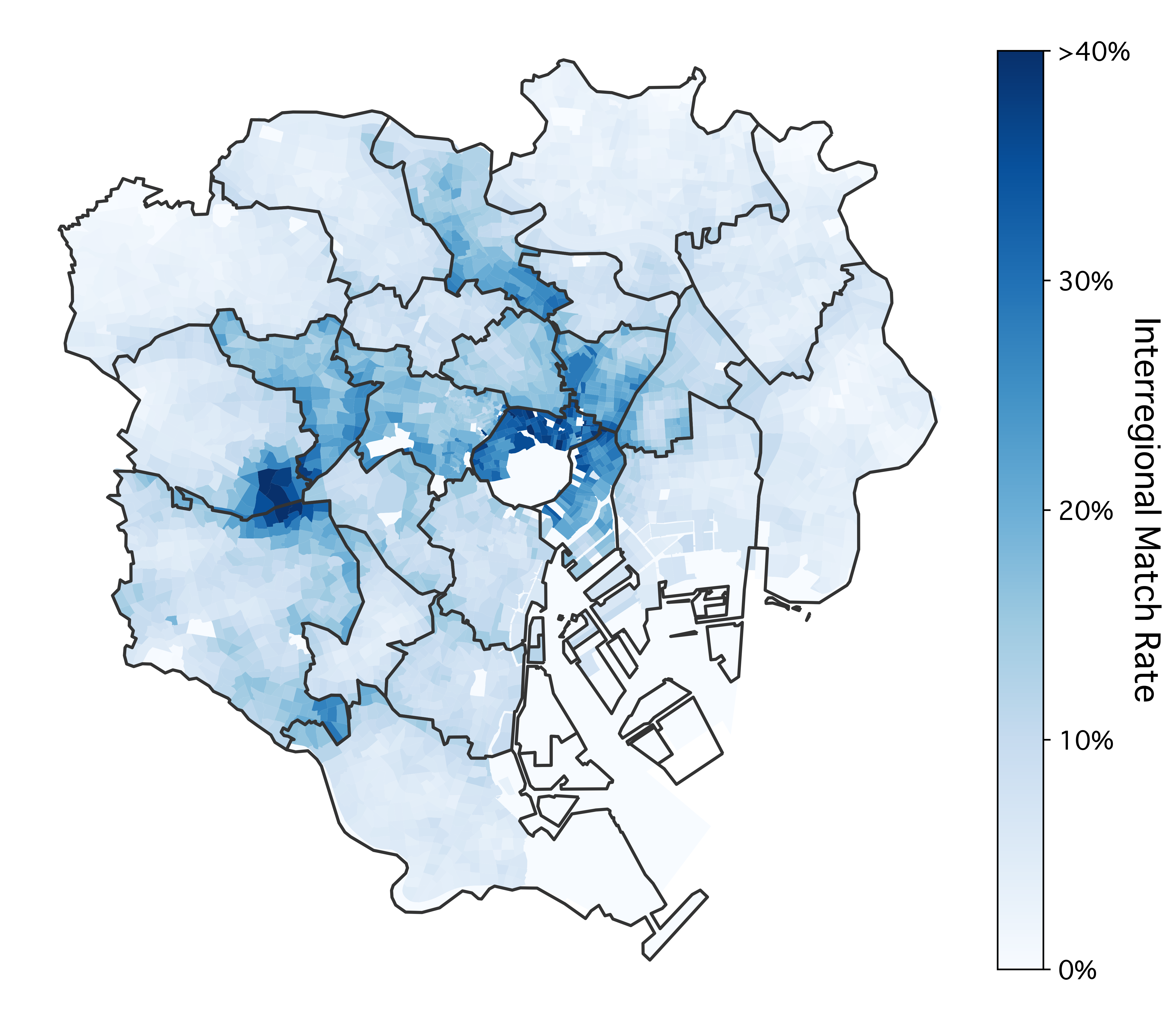}
    \caption{Interregional Match Rates}
    \label{fig:welfare_map_tokyo:inter}
  \end{subfigure}
  \hfill
  \begin{subfigure}[b]{0.49\textwidth}
    \centering
    \includegraphics[width=\linewidth]{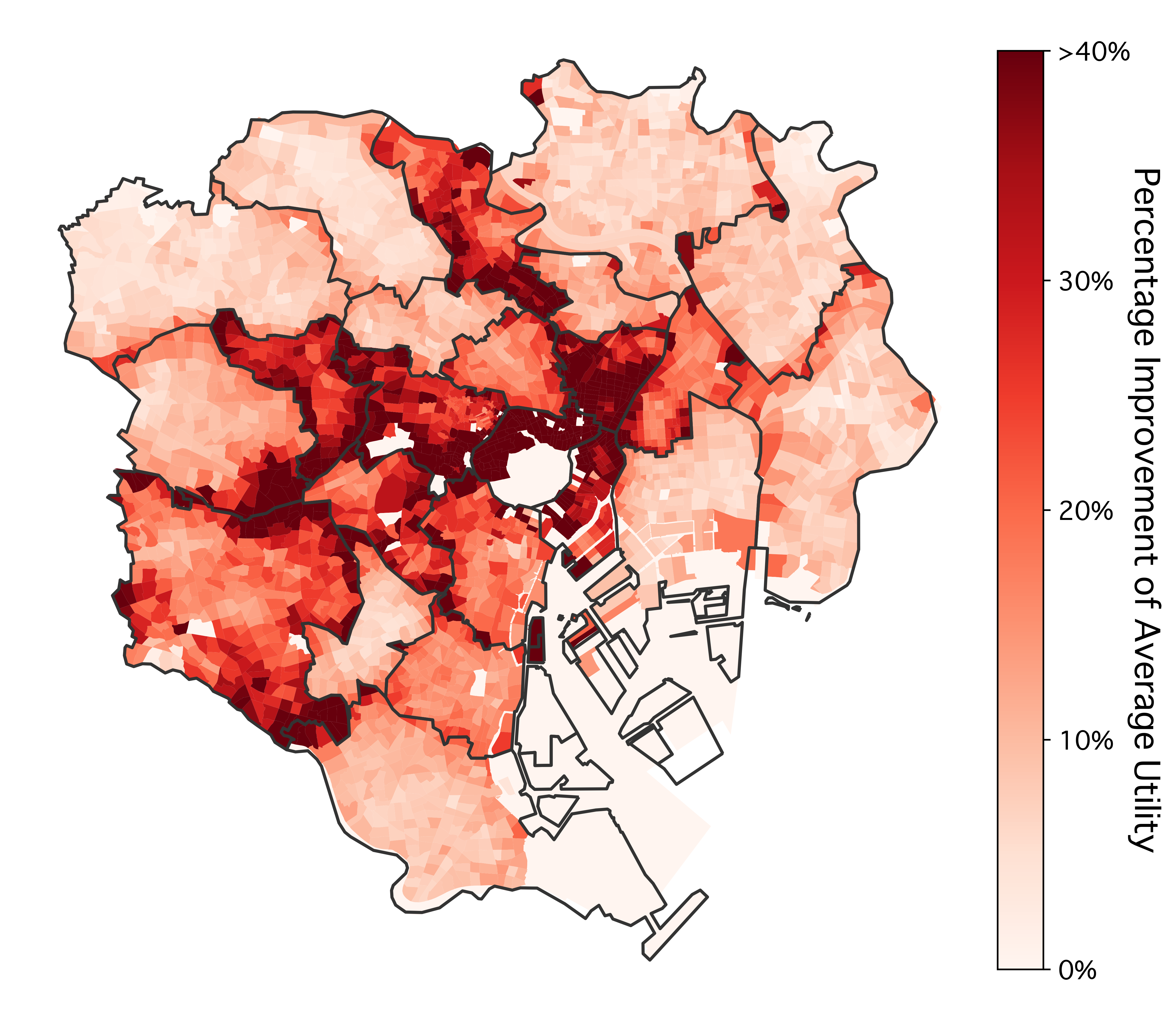}
    \caption{Percentage Improvement of Average Utility}
    \label{fig:welfare_map_tokyo:util}
  \end{subfigure}
  
  \caption{
  \small
  Interregional Match Rates and Average Welfare Improvements of Children in Geographical Subdivisions in Tokyo City under \Cref{mec:FIG-cross}. Panel (a) shows the share of children in each subdivision who are matched to daycares outside their home region, with darker blue areas indicating higher interregional match rates. Panel (b) depicts the percentage increase in average child utility in each subdivision, computed by
    $
    \tfrac{
      \text{Average child utility under Mechanism 3 in the subdivision}
    }{
      \text{Average child utility under Mechanism 1 in the subdivision}
    } - 1.0
    $, with darker red areas indicating larger relative welfare gains.
    Completely white areas denote subdivisions without any children in our simulation (e.g. parks, rivers, or an airport).
  }
  \label{fig:welfare_map_tokyo}
\end{figure}

As illustrated in \Cref{tab:welfare_comp_tokyo}, although the average gains from market integration are modest, the effects are heterogeneous across children. In particular, the welfare gains by the children around the borders are higher than the average over all children. 
\Cref{fig:welfare_map_tokyo} further illustrates the welfare improvements and the interregional match rates across geographic subdivisions under \Cref{mec:FIG-cross}.
Interregional match rates tend to be higher in subdivisions near ward borders. While average utility improvements are also high around the borders, they extend more broadly toward the centers of the wards, compared to the spatial pattern of interregional matches.

\input{tables/section7/tab_simulation_tokyo_welfare}

Additional details of these geographic patterns can be found in  
\Cref{tab:welfare_comp_rel_tokyo} which decomposes the characteristics of children whose matches improve under \Cref{mec:FIG-cross} relative to \Cref{mec:Frag} in \Cref{tab:welfare_comp_tokyo}. 
Although only 4.7\% ($\frac{2,784.7}{59,767}$ from \Cref{tab:welfare_comp_tokyo}) of the children reside in the areas where the closest daycares are outside their home ward, they account for 7.9\% ($\tfrac{5.7}{72.6}$) of interregional matches under \Cref{mec:FIG-cross}. 
By contrast, 6.6\% of the children whose matches improve live in those same areas, which is lower than their 7.9\% share for interregional matches, suggesting that the welfare gains extend more broadly within each ward. 
Similar patterns are also found for \Cref{mec:FIG-specific,mec:DA-all}: compared to the share of children living in border areas among all children, the share is higher among those who are matched to daycares outside their home ward; however, the share  among those with improved matches is lower than this latter share.
One possible explanation for this last observation is that when children match to daycares outside their home region, they leave behind vacant daycare seats; those seats are then filled by other local children, producing additional welfare improvements beyond the interregional matches.

To further clarify this effect, we quantify the ``multiplier effect'' of interregional matches.  \Cref{tab:welfare_comp_rel_tokyo} reveals that 72.6\% of the match improvements are attributed directly to interregional matches, while the remaining 27.4\% arise from slots freed up through these interregional matches, implying an integration multiplier of 1.378 under \Cref{mec:FIG-cross}. The multipliers under \Cref{mec:FIG-specific,mec:DA-all} are 1.368 and 1.311, respectively.

\subsubsection{Koriyama City}
\label{sec:counterfactual_result_koriyama}

\input{tables/section7/tab_simulation_koriyama}

\Cref{tab:welfare_comp_koriyama} reports the results for Koriyama City. 
It shows that the full market integration via \Cref{mec:DA-all}---which corresponds to Koriyama City's current system---raises the average child utility by 12.4\% relative to the fragmented case (from 1.643 to 1.847) and entails the interregional match rate of 22.7\%. By contrast, partial market integration via \Cref{mec:FIG-specific} yields a 6.5\% increase in average child utility (from 1.643 to 1.750) with a 6.8\% interregional match rate, and the one via \Cref{mec:FIG-cross} yields a 7.4\% increase (from 1.643 to 1.765) with a 7.2\% interregional match rate. Thus, partial integration under the balancedness constraint captures 52.2\% and 59.6\% of the welfare gains achieved by full integration, respectively. \Cref{tab:welfare_comp_koriyama} also reports the match rate, which shows that Mechanisms \ref{mec:FIG-specific}, \ref{mec:FIG-cross}, and \ref{mec:DA-all} reduce the unmatch rate by 3.2, 3.7, and 7.0 percentage points, respectively. Therefore, Mechanisms \ref{mec:FIG-specific} and \ref{mec:FIG-cross} achieve 46.3\% and 52.8\% of the reduction in the unmatch rate of \Cref{mec:DA-all}, respectively. 
As we do for the Tokyo data, we calculate the same statistics among children whose closest daycare lies outside their home region as well. 
These areas are shown in \Cref{fig:3_voronoi_koriyama} in \Cref{sec:background-koriyama}. \Cref{tab:welfare_comp_koriyama} reports that, among other things, even under the balancedness constraints (\Cref{mec:FIG-specific,mec:FIG-cross}), 
the average child utility increases by 25.6\% (from 0.898 to 1.128) and 27.7\% (from 0.898 to 1.147), and 
15.8\% and 16.6\% of the children are matched to daycares outside their residential regions, respectively.

These welfare gains can be interpreted in terms of travel time by dividing them by the coefficient $\hat\beta = 2.492$. The improvements of all children achieved by Mechanisms \ref{mec:FIG-specific}, \ref{mec:FIG-cross}, and \ref{mec:DA-all} relative to \Cref{mec:Frag} correspond to 4.3\%, 4.9\%, and 8.2\% reductions in travel time to daycares, respectively. 
For the children whose closest daycare lies outside their home region, the corresponding reductions are 9.2\%, 10.0\%, and 34.3\%.

\begin{figure}[tbhp]
\footnotesize
  \begin{subfigure}[b]{0.49\textwidth}
    \centering
    \includegraphics[width=\linewidth]{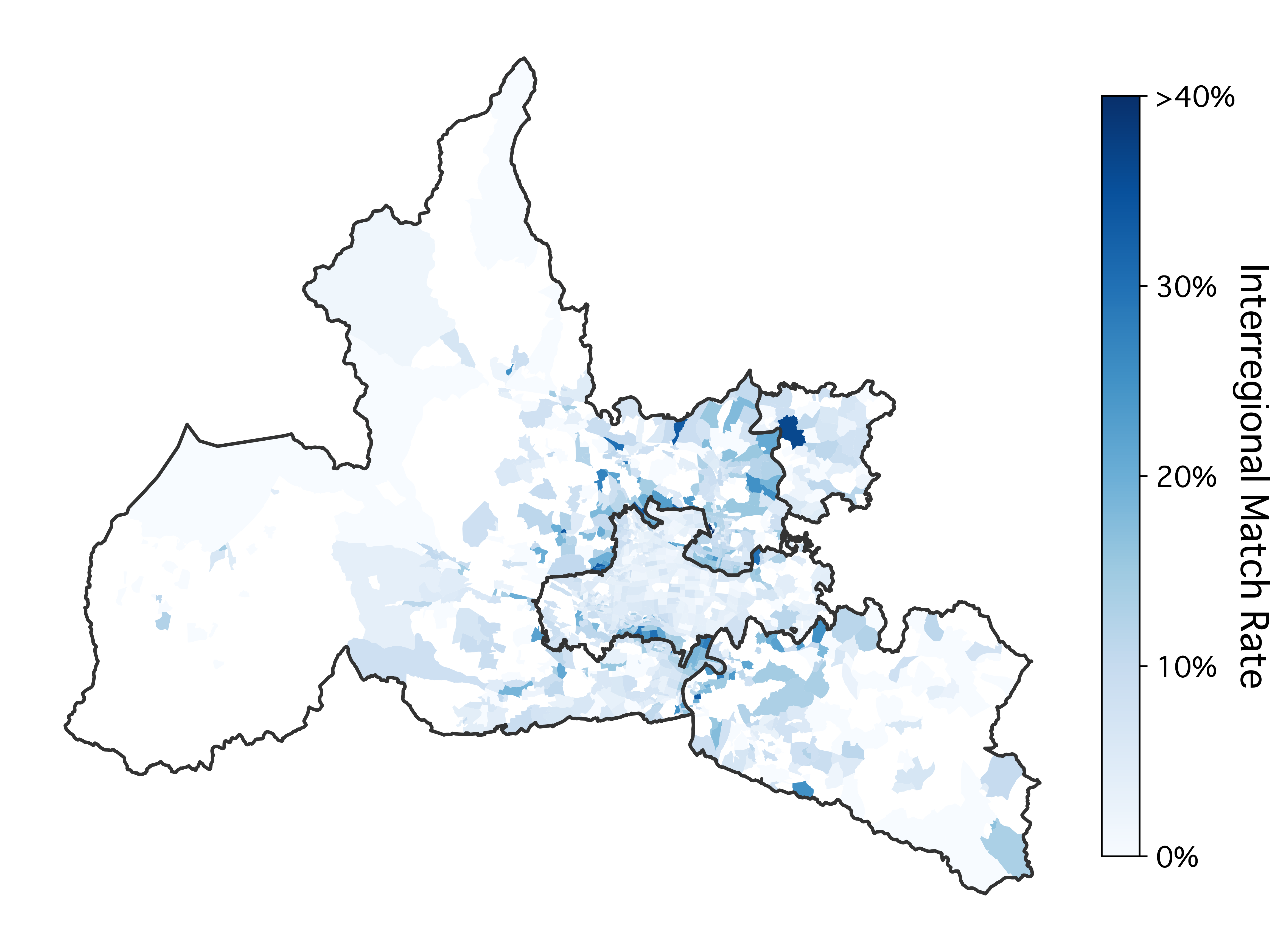}
    \caption{Interregional Match Rates}
    \label{fig:welfare_map_koriyama:inter}
  \end{subfigure}
  \hfill
  \begin{subfigure}[b]{0.49\textwidth}
    \centering
    \includegraphics[width=\linewidth]{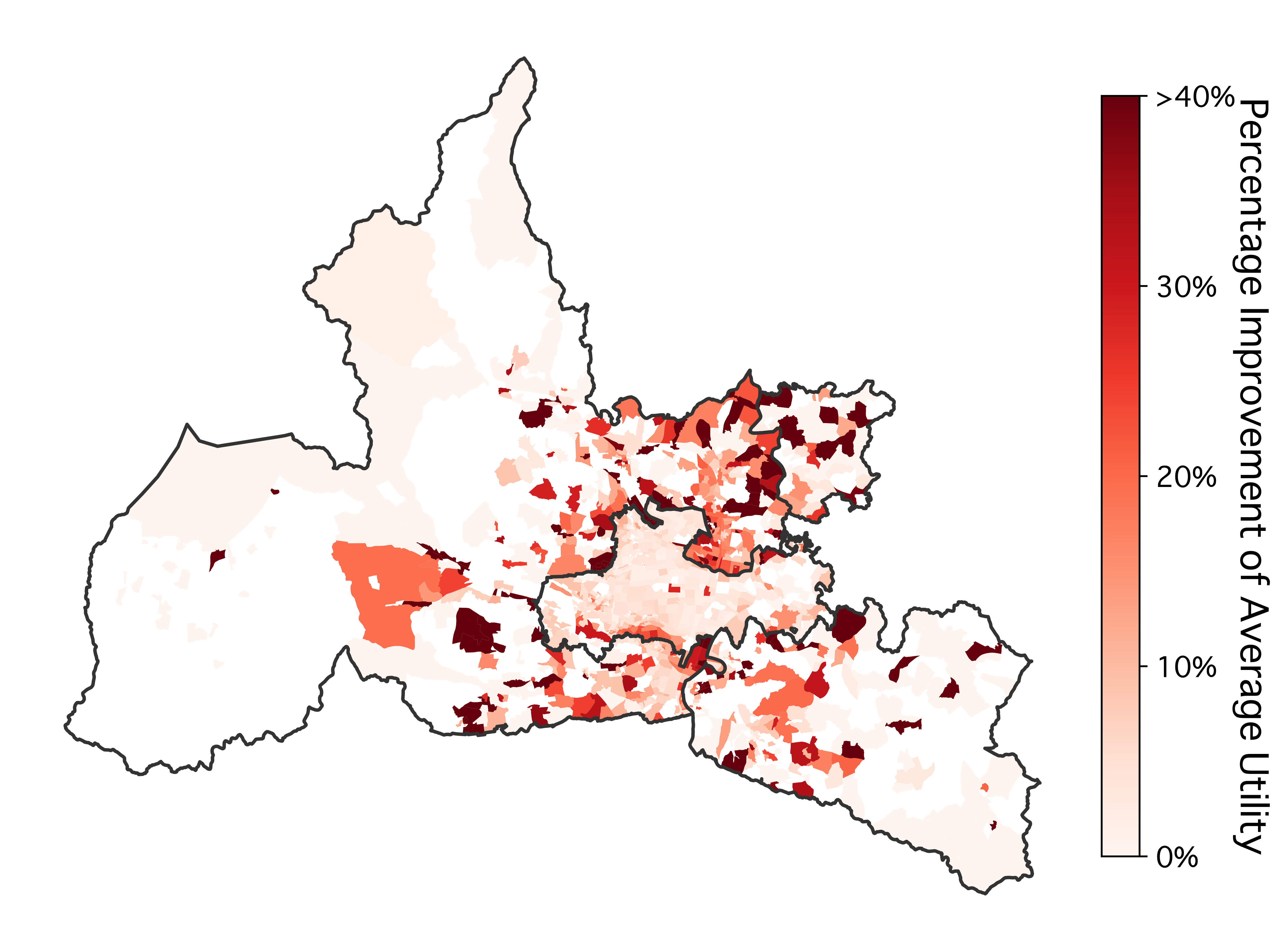}
    \caption{Percentage Improvement of Average Utility}
    \label{fig:welfare_map_koriyama:util}
  \end{subfigure}
  \caption{ 
  \small
  Interregional Match Rates and Average Welfare Improvements of Children in Geographical Subdivisions in Koriyama City under \Cref{mec:FIG-cross}. Panel (a) shows the share of children in each subdivision who are matched to daycares outside their home region, with darker blue areas indicating higher interregional match rates. Panel (b) depicts the percentage increase in average child utility in each subdivision, computed by
    $
    \tfrac{
      \text{Average child utility under Mechanism 3 in each subdivision}
    }{
      \text{Average child utility under Mechanism 1 in each subdivision}
    } - 1.0
    $, with darker red areas indicating larger relative welfare gains. Completely white areas denote subdivisions without any children in our simulation (e.g. parks, rivers, or mountain areas).
  }
  \label{fig:welfare_map_koriyama}
\end{figure}

Heterogeneity in gains from market integration also appears in Koriyama City, as observed in \Cref{tab:welfare_comp_koriyama}. 
\Cref{fig:welfare_map_koriyama} further illustrates the welfare improvements and the interregional match rates across geographic subdivisions under \Cref{mec:FIG-cross}.
Similar to Tokyo City, geographic subdivisions with high interregional match rates are concentrated near regional borders, while average utility improvements extend more broadly toward the centers of regions.

\Cref{tab:welfare_comp_rel_koriyama} decomposes the characteristics of children whose matches improve under \Cref{mec:FIG-cross} relative to \Cref{mec:Frag} in \Cref{tab:welfare_comp_koriyama}. 
Although only 11.5\% ($\frac{149.3}{1,295}$ from \Cref{tab:welfare_comp_koriyama}) of the children reside in the areas where the closest daycares are outside their home region, they account for 36.3\% ($\tfrac{19.5}{53.8}$) of interregional matches under \Cref{mec:FIG-cross}. 
By contrast, 22.1\% of the children whose matches improve live in those same areas, which is lower than their 36.3\% share for interregional matches, suggesting that the welfare gains extend more broadly within each region. A similar pattern is also found for \Cref{mec:FIG-specific}, while the same decomposition cannot be made under \Cref{mec:DA-all} because the matching under that mechanism does not necessarily Pareto dominate the matching under \Cref{mec:Frag} due to the difference in the priority orders used.

We also quantify the ``multiplier effect'' of interregional matches in Koriyama City. \Cref{tab:welfare_comp_rel_koriyama} reveals that 53.8\% of the match improvements are attributed directly to interregional matches, while the remaining 46.2\% arise from slots freed up through these interregional matches, implying an integration multiplier of 1.859 under \Cref{mec:FIG-cross}. The multiplier under \Cref{mec:FIG-specific} is 1.755.

\input{tables/section7/tab_simulation_koriyama_welfare}

We finish this section with several remarks. First, our counterfactual analysis found that integrating matching markets can result in significant welfare gains. Second, we showed that even partial integration results in nontrivial welfare gains, but the magnitude of the gains is meaningfully smaller than that under full integration. Third, we found relatively small difference between the two versions of partial integration. Finally, our analysis revealed that there is considerable heterogeneity in welfare gains---more specifically, children living close to regional borders achieve significantly larger welfare gains than others from both partial and full integration. All in all, our counterfactual simulations confirm our theoretical predictions while also giving additional insights into the magnitude and distribution of welfare gains from different methods of integration.

%% file: tables/section7/tab_simulation_tokyo.tex
\begin{table}[htbp]
\centering
\scriptsize
\caption{Welfare Comparison of Alternative Mechanisms  (Tokyo City)}
\label{tab:welfare_comp_tokyo}
\begin{threeparttable}
\begin{tabular*}{\textwidth}{
    l @{\extracolsep{\fill}} 
    c @{\hskip 0.0em} 
    c @{\hskip 0.0em} 
    c @{\hskip 0.0em} 
    c @{\hskip 0.0em} 
}
\toprule
& \textbf{Fragmented} & \multicolumn{2}{c}{\textbf{Partially Integrated}} & \textbf{Fully Integrated} \\ 
\cmidrule{2-2} \cmidrule{3-4} \cmidrule{5-5}
& \textbf{Mechanism~1} & \textbf{Mechanism~2} & \textbf{Mechanism~3} & \textbf{Mechanism~4} \\
\midrule

\multicolumn{5}{l}{\hspace{-0.5em}\textbf{(a) All Children ($N=59,767$)}} \\
Match rate & 0.537 & 0.590 & 0.592 & 0.668 \\
Interregional match rate & 0.000 & 0.091 & 0.095 & 0.215 \\
Average child match rank & 3.753 & 3.237 & 3.209 & 2.533 \\
Average child utility & 0.854 & 0.985 & 0.993 & 1.187 \\
\shortstack[l]{
Ratio of children whose match becomes  \\
better than under Mechanism~1
} & \mulc{---} & 0.115 & 0.121 & 0.263 \\
\midrule

\multicolumn{5}{l}{
\shortstack[l]{
\hspace{-0.5em}\textbf{(b) Children whose Closest Daycare is}\\
\hspace{-0.5em}\textbf{Outside Their Region ($N=2,784.7$)}
}
} \\
Match rate & 0.469 & 0.555 & 0.558 & 0.666 \\
Interregional match rate & 0.000 & 0.143 & 0.149 & 0.315 \\
Average child match rank & 4.722 & 3.864 & 3.822 & 2.826 \\
Average child utility & 0.696 & 0.889 & 0.899 & 1.169 \\
\shortstack[l]{
Ratio of children whose match becomes  \\
better than under Mechanism~1
} & \mulc{---} & 0.166 & 0.173 & 0.355 \\

\bottomrule
\end{tabular*}
\begin{tablenotes}
\footnotesize
\item \emph{Notes:}  
We present summary statistics from counterfactual simulations of our four mechanisms for (a) all children and (b) children whose closest daycare lies outside their home region 
(highlighted by the orange-shaded areas in \Cref{fig:3_voronoi_tokyo}). 
Note that although we only observe child population at the geographic subdivision level in our data, in some cases within a subdivision, part of it has the closest daycare inside the ward while the rest has the closest daycare outside the ward. Thus, 
if $x$\% of the areas in a given geographic subdivision has the closest daycare outside of its ward, we randomly  assign $x$\% of children in that subdivision in category (b).
 Here, ``Interregional match rate'' is defined as the share of children matched to daycares outside their home region, relative to $N$. 
 In computing the ``Average child match rank,'' any child matched to the outside option is given a rank equal to her ROL length plus one. 
For ``Average child utility,'' we report the values under the normalization that the average quality of the outside option is zero and the error term follows the Gumbel distribution whose standard deviation is $\pi / \sqrt{6}$ ($\simeq 1.283$). 
Given our utility model, dividing differences in welfare by $\hat{\beta} = 3.945$ translates those differences into a percentage reduction in travel time.
\end{tablenotes}
\end{threeparttable}
\end{table}

%% file: tables/section7/tab_simulation_tokyo_welfare.tex
\begin{table}[tbhp]
\centering
\small
\caption{Decomposition of Children whose Matches Improve under \Cref{mec:FIG-cross} (Tokyo City)}
\label{tab:welfare_comp_rel_tokyo}
\begin{threeparttable}
\begin{tabular*}{0.8\textwidth}{
    l @{\extracolsep{\fill}} 
    S[table-format=3.1,table-space-text-post={\%}] 
    S[table-format=3.1,table-space-text-post={\%}] 
    S[table-format=3.1,table-space-text-post={\%}] 
}
\toprule
& \shortstack[c]{
Improvement with \\
Interregional Match
}
& \shortstack[c]{
Improvement without \\
Interregional Match
} & Total \\
\midrule
Border & 5.7\% & 0.9\% & 6.6\% \\ 
Non-Border & 66.8\% & 26.5\% & 93.4\% \\ 
Total & 72.6\% & 27.4\% & 100.0\% \\ 
\bottomrule
\end{tabular*}
\begin{tablenotes}
\footnotesize
\item \emph{Notes:}  
Values are constructed by re-aggregating the data from \Cref{tab:welfare_comp_tokyo}. The total number of children with improved matches is normalized to 100\%. ``Border'' indicates the children whose closest daycare lies outside their home ward. 
Note that any child who is matched with a daycare outside her home region under \Cref{mec:FIG-cross} always obtains a better match than under \Cref{mec:Frag}. 
\end{tablenotes}
\end{threeparttable}
\end{table}

%% file: tables/section7/tab_simulation_koriyama.tex
\begin{table}[bthp]
\centering
\scriptsize
\caption{Welfare Comparison of Alternative Mechanisms (Koriyama City)}
\label{tab:welfare_comp_koriyama}
\begin{threeparttable}
\begin{tabular*}{\textwidth}{
    l @{\extracolsep{\fill}} 
    c @{\hskip 0.0em} 
    c @{\hskip 0.0em} 
    c @{\hskip 0.0em} 
    c @{\hskip 0.0em} 
}
\toprule
& \textbf{Fragmented} & \multicolumn{2}{c}{\textbf{Partially Integrated}} & \textbf{Fully Integrated} \\ 
\cmidrule{2-2} \cmidrule{3-4} \cmidrule{5-5}
& \textbf{Mechanism~1} & \textbf{Mechanism~2} & \textbf{Mechanism~3} & \textbf{Mechanism~4} \\
\midrule

\multicolumn{5}{l}{\hspace{-0.5em}\textbf{(a) All Children ($N=1,295$)}} \\
Match rate & 0.689 & 0.722 & 0.726 & 0.759 \\
Interregional match rate & 0.000 & 0.068 & 0.072 & 0.227 \\
Average child match rank & 3.130 & 2.841 & 2.801 & 2.648 \\
Average child utility & 1.643 & 1.750 & 1.765 & 1.847 \\
\shortstack[l]{
Ratio of children whose match becomes  \\
better than under Mechanism~1
} & \mulc{---} & 0.087 & 0.098 & 0.242 \\
\midrule

\multicolumn{5}{l}{
\shortstack[l]{
\hspace{-0.5em}\textbf{(b) Children whose Closest Daycare is}\\
\hspace{-0.5em}\textbf{Outside Their Region ($N=149.3$)}
}
} \\
Match rate & 0.515 & 0.593 & 0.600 & 0.763 \\
Interregional match rate & 0.000 & 0.158 & 0.166 & 0.503 \\
Average child match rank & 4.118 & 3.428 & 3.373 & 2.021 \\
Average child utility & 0.898 & 1.128 & 1.147 & 1.754 \\
\shortstack[l]{
Ratio of children whose match becomes  \\
better than under Mechanism~1
} & \mulc{---} & 0.173 & 0.188 & 0.519 \\

\bottomrule
\end{tabular*}
\begin{tablenotes}
\footnotesize
\item \emph{Notes:} We present summary statistics from counterfactual simulations of our four mechanisms for (a) all children and (b) children whose closest daycare lies outside their home region 
(highlighted by the orange-shaded areas in \Cref{fig:3_voronoi_koriyama}). 
Note that although we only observe child population at the geographic subdivision level, in some cases within a subdivision, part of it has the closest daycare inside the region while the rest has the closest daycare outside the region. Thus, 
If $x$\% of the areas in a given geographic subdivision has the closest daycare outside of its ward, we assign $x$\% of children in that subdivision in category (b).
 Here, ``Interregional match rate'' is defined as the share of children matched to daycares outside their home region, relative to $N$. 
In computing the ``Average child match rank,'' any child matched to the outside option is given a rank equal to her ROL length plus one. 
For ``Average child utility,'' we report the values under the normalization that the average quality of the outside option is zero and the error term follows the Gumbel distribution whose standard deviation is $\pi / \sqrt{6}$ ($\simeq 1.283$).   
Given our utility model, dividing differences in welfare by $\hat{\beta} = 2.492$ translates those differences into a percentage  reduction in travel time.
\end{tablenotes}
\end{threeparttable}
\end{table}

%% file: tables/section7/tab_simulation_koriyama_welfare.tex
\begin{table}[tbhp]
\centering
\small
\caption{Decomposition of Children whose Matches Improve under \Cref{mec:FIG-cross} (Koriyama City)}
\label{tab:welfare_comp_rel_koriyama}
\begin{threeparttable}
\begin{tabular*}{0.8\textwidth}{
    l @{\extracolsep{\fill}} 
    S[table-format=3.1,table-space-text-post={\%}] 
    S[table-format=3.1,table-space-text-post={\%}] 
    S[table-format=3.1,table-space-text-post={\%}] 
}
\toprule
& \shortstack[c]{
Improvement with \\
Interregional Match
}
& \shortstack[c]{
Improvement without \\
Interregional Match
} & Total \\
\midrule
Border & 19.5\% & 2.6\% & 22.1\% \\ 
Non-Border & 34.3\% & 43.6\% & 77.9\% \\
Total & 53.8\% & 46.2\% & 100.0\% \\ 
\bottomrule
\end{tabular*}
\begin{tablenotes}
\footnotesize
\item \emph{Notes:}  
Values are constructed by re-aggregating the data from \Cref{tab:welfare_comp_koriyama}. The total number of children with improved matches is normalized to 100\%. ``Border'' indicates the children whose closest daycare lies outside their home region. 
Note that any child who is matched with a daycare outside her home region under \Cref{mec:FIG-cross} always obtains a better match than under \Cref{mec:Frag}.
\end{tablenotes}
\end{threeparttable}
\end{table}

%% file: sections/conclusion.tex
\section{Discussion and Conclusion}
\label{sec:conclusion}

\subsection{Summary} This paper 
provided analysis of fragmentation in matching markets. We first documented and discussed a variety of fragmentation cases in practice  
 such as school choice, medical residency matching, and so forth.  We then
examined how integrating fragmented  matching markets can improve outcomes under both \emph{full integration} and \emph{partial integration} scenarios. In the full integration case, there is no constraint on the inflow and outflow of children across regions. In the partial integration case, by contrast, a \textit{balancedness} constraint is imposed, that is, each region's outflow must equal its inflow, preventing any region from gaining or losing children on net. Following  \citet{kamakoji-ekkyo}, we considered a mechanism that finds a Pareto efficient matching among those that respect individual rationality, balancedness, and fairness in the sense of no justified envy among children.  Empirically, using real-world data on daycare allocation, we quantified welfare gains from partial and full integration. In both cases, compared to the status quo of independent regional matching, many children obtain better placements  while no child is made worse off. The counterfactual simulations suggest that allowing interregional child transfers significantly improves average child welfare even under balancedness constraints, but it is less efficient than the outcome under full market integration.

\subsection{Related Literature}
 We review the related literature in some depth to offer our perspective on the current state of knowledge regarding the fragmentation and integration of matching markets. Our main goal with this review is to provide a basis for discussing promising directions for future research and practice in a concluding subsection (\Cref{sec:final}).

 \subsubsection{Fragmentation and Integration in Matching Markets} 
We study fragmentation and integration in matching markets, and we are not the first to study this topic. 
For instance, \citet{ortega2018social} and \citet{klein2024school} explore the welfare impact of consolidating school choice markets under the standard DA. In the context of interdistrict school choice, \citet{hafalir2018interdistrict} develop a model of student assignment across district lines and identify conditions under which the outcome of the unconstrained DA  satisfies a balancedness condition across districts. \citet{kamakoji-integration} investigate when merging local markets benefits every student (i.e., when integration is ``monotonic'' in terms of welfare) under mechanisms that satisfy several desirable properties. In contrast to these studies of \emph{full} integration, \citet{kamakoji-ekkyo} take the balancedness constraint as exogenously given---reflecting political and equity requirements that prevent unrestricted student mobility. They then ask how to efficiently implement \textit{partial} integration, i.e., they try to maximize welfare under a constraint on mobility that reflects the reality shaped by political and equity concerns.\footnote{See also \citet{bloch2020matching} who propose mechanisms for integrating public housing assignment markets under balancedness.} Those approaches are complementary to each other. The present paper further develops both lines of literature by quantifying benefits of both partial and full integration. By doing so, we show that a carefully designed mechanism can achieve some efficiency gains even under partial integration, while the gains from full integration are significantly larger. 

More specifically, \citet{kamakoji-ekkyo} and this paper consider partial integration that is based on a balancedness constraint.
The balancedness constraint was first formalized by  \citet{dur2019two} in a two-sided matching context, inspired by a faculty \emph{tuition exchange} program that requires universities to reciprocally exchange students. Subsequent works have applied similar ideas: \citet{dur2015maintaining} analyze student exchange programs (such as Erasmus in Europe) under bilateral balanced-exchange quotas, and  \citet{dur2024rematching} study college sports transfers with roster balance requirements. In those studies, balancedness is enforced at the level of individual institutions (each school or college must sponsor as many outgoing students as it hosts incoming ones). Our paper departs from this framework by imposing balancedness at an aggregate \textit{regional} level. This generalizes the constraint structure: institution-level balance is a special case (each region containing a single school), whereas our framework allows multiple institutions per region and focuses on net flows across regional boundaries. By addressing balanced exchange in a multi-institution regional setting, our work extends the scope of balanced matching and addresses new practical concerns.\footnote{We note that, however, the specific problems analyzed and especially the desiderata considered by those papers are different from ours, so their analysis and ours are logically unrelated.}

\subsubsection{Estimating Preferences from Revealed Preferences}

In this section, we relate our paper to the growing empirical market design literature. Our study involves various assumptions on the behavior in terms of reporting of ROLs. 
Although assuming truthful reporting is the simplest way to recover applicants' underlying utilities from their ROLs, both experimental and survey evidence document widespread preference misreporting in matching markets. 
Laboratory experiments by 
\citet{chen2006school,pais2008school,calsamiglia2010constrained,klijn2013preference,featherstone2016boston} demonstrate that a non-negligible share of participants misreport their preferences in DA. 
Complementing these experimental results, \citet{hassidim2017mechanism,rees2018suboptimal,larrocau2020shortlist,hassidim2021limits,larroucau2023application} use surveys to confirm that real-world applicants similarly make strategic reports. Especially, \citet{larrocau2020shortlist} provide evidence in Chilean College Admissions that even when an applicant submits an ROL shorter than the permitted maximum, thereby removing any strategic incentive to misreport, they nonetheless tend to omit schools whose admission probability is too small. 
A recent review by \citet{reesjones2023behavioral} comprehensively catalogs these findings, classifies various patterns of misrepresentation, and relates them to the broader behavioral economics literature.

Moreover, strategic reporting of preferences may be especially important in this study because our datasets are elicited under Serial Dictatorship with a maximum length constraint. Such  such a constraint is common in the applications of DA, particularly in school choice.\footnote{As mentioned in footnote \ref{fn:SDDA}, Serial Dictatorship is a special case of DA  in which all schools have identical priorities over students.} For example, selective enrollment high school choice in Chicago \citep{pathak2013school}, secondary school matching in Singapore \citep{teo2001gale}, and high school choice in Turkey \citep{akyol2017preferences} all employ Serial Dictatorship with a maximum length constraint, much like our daycare matching markets in Japan.\footnote{See Table 1 in \citet{fack2019beyond} for a list of real-world DA implementations with limits on the length of submittable ROLs.} 
Although unconstrained Serial Dictatorship is strategy-proof, imposing a maximum length constraint breaks strategy-proofness, inducing applicants to drop daycares with little chances of admission. 

Broadly speaking, there exist two main streams of estimation methods that address strategic reports.\footnote{The survey by \citet{agarwal2020survey} gives a detailed introduction to the literature of estimating preferences from submitted ROLs.} 
The first stream develops fully specified structural models of strategic reporting, as explored by \citet{agarwal2018demand,luflade2019value,calsamiglia2020structual,kapor2020hetero,son2024dist}. They simulate applicants' decision-making environment, such as (potentially incorrect) beliefs or choice sets in the first stage, and then estimate the preference parameters, assuming that they take a best response in the environment. In particular, \cite{kuno2025strategic} uses this approach to estimate the dynamically evolving preferences of children in the Japanese daycare market.



Our paper belongs to the second stream of estimation methods, which exploits an incomplete model of applicants' behavior, as in \citet{he2015gaming,Hwang2015,fack2019beyond}. 
Rather than fully modelling how applicants form beliefs, these approaches impose strategic assumptions, such as truth-telling or stability, to derive either a conditional likelihood or moment (in)equalities that identify preference parameters. 

In particular, our estimation is built on the framework of \citet{fack2019beyond}. We adopt three of their behavioral assumptions they use, STT, WTT, and USS, while we introduce a novel assumption, OETT, under which each child predicts at least with some noises, whether they can enter each daycare.
Our OETT assumption captures the strategic behavior in which each child  drops from their ROL those daycares with little chance of acceptance given  their score and the past cutoffs, which is not captured by the STT or WTT assumption. 
Although USS can accommodate this behavior, it tends to yield higher parameter variances than our OETT approach as it uses moment inequalities.


\subsubsection{Matching with Constraints} In the present paper, we have been  primarily  interested in   the issue of fragmentation and integration of matching markets. However, the present analysis can also be situated in another strand of research, namely the growing literature on \emph{matching with constraints}, because barriers to full integration pose constraints on the design of matching mechanisms.

Matching problems with distributional, fairness, and diversity constraints are quite prevalent in both academic literature and practice. Many matching markets impose restrictions on assignments to ensure equitable outcomes or diversity in the match. In school choice, the  seminal work by  
\citet{abdulkadirouglu2003school} emphasizes  the importance of fairness (eliminating justified envy) through mechanism design  and considers type-specific quotas. Subsequent research incorporates other diversity goals: for example, \citet{hafalir2013effective} and \citet{ehlers2014school} 
impose hard or soft quotas on  students of specific types (such as minority students) within schools.  \citet{echenique/yenmez:12} characterize choice rules for schools that uphold diversity constraints while treating students as substitutes, thus reconciling affirmative action policies with stability of matching.

 Similar issues arise in college admissions and labor market matching as well. \citet{biro/fleiner/irving/manlove:10} consider university admissions with lower and common  quotas, and \citet{goto:16} propose algorithms for matching medical residents to hospitals under regional placement caps aimed at ensuring an equitable geographic distribution of doctors.  \textcolor{blue}{Kamada and Kojima} \citeyearpar{kamakoji-basic,kamakoji-concepts,kamakoji-iff, kamakoji-accommodating}  develop general theoretical frameworks for matching with such distributional constraints, in which they demonstrate that standard stable matchings may not exist, propose stability concepts adapted to the constraints, and show that new algorithms produce a matching satisfying the novel desideratum. 
\citet{kamakoji-basic} conduct counterfactual simulations for data on Japanese medical match, comparing matching mechanisms without the regional cap constraint, the  mechanism under the constraint which is currently used in Japan, and a new and improved mechanism under the constraint.\footnote{Their simulation is based on synthetic data on submitted preferences because the real data were not available, but their synthetic data were generated in such a manner that the data reproduce many summary statistics that are publicly available. We were recently granted access to the real data, 
which we are currently analyzing to conduct a counterfactual analysis in an ongoing project.
}
They find that the ongoing mechanism under constraint  entails  a significantly  greater number of unmatched medical residents  compared to the case without constraints, but their proposed mechanism  reduces the  gap to  about one third. \citet{kamakoji-fair} propose a new mechanism for the environment under a class of constraints called a general upper bound and study the performance of that mechanism using  Japanese daycare matching data. They find that their proposed mechanism can decrease the number of unmatched children by more than a half compared to the present practice.

Our contribution to this literature is to study a new type of constraint—balancedness of inflows and outflows across regional boundaries—and to quantify how much efficiency gain one can achieve with a mechanism designed optimally in the face of the constraint. In particular, we illustrate that even when a strict policy constraint (exact balance of flows) is in place, a significant improvement upon locally constrained outcomes can be made  in a way that remains fair to all participants. 
This finding suggests that it is possible to strike a balance between fairness and welfare optimization under constraints, and our our analysis provides a template for other markets where policymakers seek better outcomes without relaxing constraints.

\subsection{Final Remarks}\label{sec:final} Looking ahead, there are many  promising avenues for future research. First, it would be valuable to investigate other institutional contexts in which the issue of fragmentation and integration is important. For example, one could study interdistrict school choice programs in different countries or allocation of resources like teachers and medical residents across regions, possibly with reciprocity constraints like the balancedness constraint studied here. Each setting may bring its own practical nuances, and empirical analysis in these contexts can quantify potential welfare gains from partial and full integration. For instance, medical schools in Japan have ``regional commitment track.''  Students admitted through this track are required to work in a specific (typically rural) region of the country for several years after graduation. This  track  was introduced to alleviate the problem of doctor shortages  in rural areas of the country, but it makes the market fragmented as well. In our ongoing work, we design a matching mechanism that integrates the market for the students in the regional commitment track with the rest of the market, which improves the doctor welfare, while still addressing the problem of doctor shortage.

From both theoretical and practical perspectives, it may be important to consider generalizations of the balancedness constraint. Although it is arguably a realistic constraint in various applications (see our discussion in the Introduction), policymakers in practice might allow some degree of imbalance, such as permitting a limited net outflow or inflow for each region within a certain bound, to accommodate fluctuations in demand or capacity. Extending our approach to handle such ``soft'' balance constraints or more complex multi-region exchange arrangements poses a challenging yet important problem. Such generalizations would likely require new algorithmic ideas (for instance, introducing additional types of cycles or trades) to ensure feasibility and efficiency. Exploring these extensions, along with more empirical evaluation, will help to broaden the impact of partial integration in market design practice. At the same time, even though one might think that exact balance is restrictive, it is noteworthy that our FIG cycles mechanism captures a significant portion of potential welfare gains in our counterfactual analysis.

In our present context, further empirical investigation may be interesting. For instance, our empirical analysis does not capture parents' benefit from being able to have their child matched to a daycare that is close to their workplaces  due to data limitation. Thus,  welfare gains from integration of daycare matching markets in practice may be larger than the one suggested by our present analysis. A more refined empirical analysis could also take into account other variables such as daycare space, the number of teachers, and other attributes.  Such analysis would be useful for policymakers.

Finally, there remain many practical issues unexplored in this paper that nevertheless affect the  effectiveness and even the  feasibility  of integration.   In particular,  logistical considerations are  important in practice. For instance, transportation of students is a major concern for parents of children in daycares and schools. Some municipalities provide transportation such as school buses, but it is often very costly for them (see e.g., \citet{angrist2022still}). In other municipalities, parents need to take their children by themselves, and it may deter households  from utilizing interdistrict school choice schemes.\footnote{ \url{https://www.hopskipdrive.com/blog/navigating-the-school-commute-parent-perspectives/}} Whether, and if so how,  transportation should be facilitated is well beyond the scope of this paper, but it may be an especially important question in considering geographical integration of matching markets.
Political and policymaking complication is another major issue to tackle. Our analysis of partial integration addressed a particular form of political concern, namely
districts' fear of losing students---and the 
associated funding---or being overwhelmed by outside demand. But policymakers have other concerns. One concern we heard from 
officials is privacy, given that an integrated matching mechanism would presume that information about applicants from different  
municipalities can be shared. 
Another concern is that it is costly and difficult to even communicate effectively between politicians and officials in multiple municipalities and coordinate their policies. Those are well outside the scope of the present study, but they may prove crucial for effective market design and, in our view, warrant further investigations.

%% file: ref.bib
@string{amm = {{A}merican {M}athematical {M}onthly}}

@string{ema = {{E}conometrica}}

@article{hafalir2013effective,
	Author = {Hafalir, Isa E and Yenmez, M Bumin and Yildirim, Muhammed A},
	Journal = {Theoretical Economics},
	Number = {2},
	Pages = {325--363},
	Publisher = {Wiley Online Library},
	Title = {Effective affirmative action in school choice},
	Volume = {8},
	Year = {2013}}

@techreport{abdulkadiroglu2017minimizing,
  title={Minimizing Justified Envy in School Choice: The Design of New Orleans' OneApp},
  author={Atila Abdulkad\.{i}ro\u{g}lu and Yeon-Koo Che and Pathak, Parag A and Roth, Alvin E and Tercieux, Olivier},
  year={2017},
  institution={working paper, National Bureau of Economic Research}
}

@article{dur2015maintaining,
  title={Maintaining Diversity in Student Exchange},
  author={Dur, Umut and Kesten, Onur and {\"U}nver, M Utku},
  journal={Available at SSRN 4064921},
  year={2015}
}

@article{dur2024rematching,
  title={Rematching with Contracts under Labor Mobility Restrictions: Theory and Application},
  author={Dur, Umut and Hammond, Robert G and {\"U}nver, M Utku},
  journal={Available at SSRN 4781912},
  year={2024}
}

@article{ehlers2014school,
	Author = {Ehlers, Lars and Hafalir, Isa E and Yenmez, M Bumin and Yildirim, Muhammed A},
	Journal = {Journal of Economic Theory},
	Publisher = {Elsevier},
	Title = {School choice with controlled choice constraints: Hard bounds versus soft bounds},
	Year = {2014}}

@article{hafalir2018interdistrict,
  title={Interdistrict school choice: A theory of student assignment},
  author={Hafalir, Isa E and Kojima, Fuhito and Yenmez, M Bumin},
  journal={Journal of Economic Theory},
  volume={201},
  pages={105441},
  year={2022},
  publisher={Elsevier}
}

@article{bloch2020matching,
  title={Matching through institutions},
  author={Bloch, Francis and Cantala, David and Gibaja, Dami{\'a}n},
  journal={Games and Economic Behavior},
  volume={121},
  pages={204--231},
  year={2020},
  publisher={Elsevier}
}

@unpublished{kamakoji-fair,
	Author = {Yuichiro Kamada and Fuhito Kojima},
	Note = {Forthcoming, \textit{The Review of Economic Studies}},
	Title = {Fair Matching under Constraints: Theory and Applications},
	Year = {2023}}

@article{goto:16,
	Author = {Masahiro Goto and Fuhito Kojima and Ryoji Kurata and Akihisa Tamura
and Makoto Yokoo},
	Journal = {\textit{American Economic Journal: Microeconomics}, forthcoming},
	Title = {Designing Matching Mechanisms under General Distributional Constraints},
	Year = {2017}}

@article{kamakoji-accommodating,
  title={Accommodating various policy goals in matching with constraints},
  author={Yuichiro Kamada and Fuhito Kojima},
  journal={The Japanese Economic Review},
  volume={71},
  pages={101--133},
  year={2020},
  publisher={Springer}
}

@article{kamakoji-basic,
	Author = {Yuichiro Kamada and Fuhito Kojima},
	Journal = {American Economic Review},
	 volume={105},
	Number = {1},
	Pages = {67--99},
	Title = {Efficient Matching under Distributional Constraints: Theory and Applications},
	Year = {2015}}

@article{kamakoji-concepts,
	Author = {Yuichiro Kamada and Fuhito Kojima},
	Journal = {Journal of Economic Theory},
	Pages = {107--142},
	Publisher = {Elsevier},
	Title = {Stability concepts in matching with distributional constraints},
	Volume = {168},
	Year = {2017}}

@article{kamakoji-iff,
	Author = {Yuichiro Kamada and Fuhito Kojima},
	Journal = {Theoretical Economics},
	Pages = {761--793},
		Volume = {13},
	Title = {Stability and Strategy-Proofness for Matching with Constraints: A Necessary and Sufficient Condition},
	Year = {2018}}

@article{kamakoji-ekkyo,
  title={Efficient iBF: Balanced Integration of Fragmented Matching Markets for Welfare Improvement},
  author={Yuichiro Kamada and Fuhito Kojima},
  year={2025},
  journal={Working Paper}
}

@article{kamakoji-integration,
  title={Choice or Competition: Does Integration Benefit Everyone?},
  author={Yuichiro Kamada and Fuhito Kojima},
  year={2025},
  journal={forthcoming, \textit{Theoretical Economics}}
}

@techreport{angrist2022still,
  title={Still worth the trip? school busing effects in boston and new york},
  author={Angrist, Joshua and Gray-Lobe, Guthrie and Idoux, Clemence M and Pathak, Parag A},
  year={2022},
  institution={National Bureau of Economic Research}
}

@article{echenique/yenmez:12,
	Author = {Echenique, Federico and Yenmez, Bumin},
	Journal = {American Economic Review},
	Number = {8},
	Pages = {2679--2694},
	Title = {How to control controlled school choice},
	Volume = {105},
	Year = {2015}}

@article{dur2019two,
  title={Two-sided matching via balanced exchange},
  author={Dur, Umut Mert and {\"U}nver, M Utku},
  journal={Journal of Political Economy},
  volume={127},
  number={3},
  pages={1156--1177},
  year={2019},
  publisher={The University of Chicago Press Chicago, IL}
}

@article{gale/shapley:62,
	Author = {David Gale and Lloyd S. Shapley},
	Journal = AMM,
	Pages = {9-15},
	Title = {College Admissions and the Stability of Marriage},
	Volume = 69,
	Year = 1962}

@article{pathak2013school,
  title={School admissions reform in Chicago and England: Comparing mechanisms by their vulnerability to manipulation},
  author={Pathak, Parag A and S{\"o}nmez, Tayfun},
  journal={American Economic Review},
  volume={103},
  number={1},
  pages={80--106},
  year={2013}
}

@article{roth:02,
	Author = {Alvin E Roth},
	Journal = EMA,
	Pages = {1341-1378},
	Title = {The Economist as Engineer: Game Theory, Experimentation, and Computation as Tools for Design Economics, Fisher-Schultz Lecture},
	Volume = 70,
	Year = {2002}}

@article{biro/fleiner/irving/manlove:10,
	Author = {Pet\'{e}r Bir\'o and Tam\'as Fleiner and Robert W Irving and David F Manlove},
	Journal = {Theoretical Computer Science},
	Pages = {3136-3153},
	Title = {The College Admissions problem with lower and common quotas},
	Volume = 411,
	Year = 2010}

@article{rees2018suboptimal,
  title={Suboptimal behavior in strategy-proof mechanisms: Evidence from the residency match},
  author={Rees-Jones, Alex},
  journal={Games and Economic Behavior},
  volume={108},
  pages={317--330},
  year={2018},
  publisher={Elsevier}
}

@techreport{artemov2017strategic,
  title={Strategic `Mistakes': Implications for Market Design Research},
  author={Artemov, Georgy and Che, Yeon-Koo and He, Yinghua},
  year={2019},
  institution={mimeo}
}

@article{fack2019beyond,
  title={Beyond Truth-Telling: Preference Estimation with Centralized School Choice and College Admissions},
  author={Fack, Gabrielle and Grenet, Julien and He, Yinghua},
  journal={American Economic Review},
  volume={109},
  number={4},
  pages={1486--1529},
  year={2019}
}

@article{agarwal2019market,
  title={Market failure in kidney exchange},
  author={Agarwal, Nikhil and Ashlagi, Itai and Azevedo, Eduardo and Featherstone, Clayton R and Karaduman, {\"O}mer},
  journal={American Economic Review},
  volume={109},
  number={11},
  pages={4026--70},
  year={2019}
}

@techreport{robinson2019gets,
  title={Who gets placed where and why? An empirical framework for foster care placement},
  author={Robinson-Cort\'es, Alejandro},
  year={2021},
  institution={Working Paper}
}

@article{kapor2020hetero,
Author = {Kapor, Adam J and Neilson, Christopher A and Zimmerman, Seth D},
Title = {Heterogeneous Beliefs and School Choice Mechanisms},
Journal = {American Economic Review},
Volume = {110},
Number = {5},
Year = {2020},
Month = {May},
Pages = {1274–1315},
DOI = {10.1257/aer.20170129},
URL = {https://www.aeaweb.org/articles?id=10.1257/aer.20170129}
}

@article{son2024dist,
Author = {Lee, Jaewon and Son, Suk Joon},
Title = {Distributional Impacts of Centralized School Choice},
Journal = {Working Paper},
Year = {2024},
URL = {https://drive.google.com/file/d/1cRRiB7_Sha22Dhx_5wIlQ0Xdj6plVFAe/view}
}

@article{massenzio2023navigating,
  title={Navigating the ophthalmology \& urology match with a significant other},
  author={Massenzio, Samantha S and Uhler, Tara A and Massenzio, Erik M and Sun, Emily and Srikumaran, Divya and Clifton, Marisa M and Green, Laura K and Sun, Grace and Wang, Jiangxia and Woreta, Fasika A},
  journal={Journal of surgical education},
  volume={80},
  number={1},
  pages={135--142},
  year={2023},
  publisher={Elsevier}
}

@article{ashlagi2014free,
  title={Free riding and participation in large scale, multi-hospital kidney exchange},
  author={Ashlagi, Itai and Roth, Alvin E},
  journal={Theoretical Economics},
  volume={9},
  number={3},
  pages={817--863},
  year={2014},
  publisher={Wiley Online Library}
}

@article{ortega2018social,
  title={Social integration in two-sided matching markets},
  author={Ortega, Josu{\'e}},
  journal={Journal of Mathematical Economics},
  volume={78},
  pages={119--126},
  year={2018},
  publisher={Elsevier}
}

@article{klein2024school,
  title={School choice with independent versus consolidated districts},
  author={Klein, Thilo and Aue, Robert and Ortega, Josu{\'e}},
  journal={Games and Economic Behavior},
  volume={147},
  pages={170--205},
  year={2024},
  publisher={Elsevier}
}

@article{almarzooq2021single,
  title={The single match: reflections on the National Resident Matching Program’s sustained partnership with learners},
  author={Almarzooq, Zaid I and Lillemoe, Heather A and White-Manigault, Ebony and Wickham, Thomas and Curtin, Laurie S},
  journal={Academic Medicine},
  volume={96},
  number={8},
  pages={1116--1119},
  year={2021},
  publisher={LWW}
}

@article{abdulkadirouglu2003school,
  title={School choice: A mechanism design approach},
  author={Atila Abdulkad\.{i}ro\u{g}lu and S{\"o}nmez, Tayfun},
  journal={American Economic Review},
  volume={93},
  number={3},
  pages={729--747},
  year={2003},
  publisher={American Economic Association}
}

@article{roth1984evolution,
  title={The evolution of the labor market for medical interns and residents: a case study in game theory},
  author={Roth, Alvin E},
  journal={Journal of political Economy},
  volume={92},
  number={6},
  pages={991--1016},
  year={1984},
  publisher={The University of Chicago Press}
}

@article{roth2004kidney,
  title={Kidney exchange},
  author={Roth, Alvin E and S{\"o}nmez, Tayfun and {\"U}nver, M Utku},
  journal={The Quarterly Journal of Economics},
  volume={119},
  number={2},
  pages={457--488},
  year={2004},
  publisher={MIT Press}
}

@article{agarwal2018demand,
    author = {Agarwal, Nikhil and Somaini, Paulo},
    title = {Demand Analysis Using Strategic Reports: An Application to a School Choice Mechanism},
    journal = {Econometrica},
    volume = {86},
    number = {2},
    pages = {391-444},
    doi = {https://doi.org/10.3982/ECTA13615},
    url = {https://onlinelibrary.wiley.com/doi/abs/10.3982/ECTA13615},
    eprint = {https://onlinelibrary.wiley.com/doi/pdf/10.3982/ECTA13615},
    year = {2018}
}

@article{agarwal2020survey,
   author = "Agarwal, Nikhil and Somaini, Paulo",
   title = "Revealed Preference Analysis of School Choice Models", 
   journal= "Annual Review of Economics",
   year = "2020",
   volume = "12",
   number = "Volume 12, 2020",
   pages = "471-501",
   doi = "https://doi.org/10.1146/annurev-economics-082019-112339",
   url = "https://www.annualreviews.org/content/journals/10.1146/annurev-economics-082019-112339",
   publisher = "Annual Reviews",
   issn = "1941-1391",
   type = "Journal Article",
  }

@article{larrocau2020shortlist,
    author = {Larroucau, Tom\'{a}s and Rios, Ignacio},
    year = {2020},
    month = {09},
    pages = {},
    title = {Do "Short-List" Students Report Truthfully? Strategic Behavior in the Chilean College Admissions Problem},
    doi = {10.13140/RG.2.2.10082.79045}
}

@article{calsamiglia2010constrained,
    Author = {Calsamiglia, Caterina and Haeringer, Guillaume and Klijn, Flip},
    Title = {Constrained School Choice: An Experimental Study},
    Journal = {American Economic Review},
    Volume = {100},
    Number = {4},
    Year = {2010},
    Month = {September},
    Pages = {1860–74},
    DOI = {10.1257/aer.100.4.1860},
    URL = {https://www.aeaweb.org/articles?id=10.1257/aer.100.4.1860}
}

@article{chen2006school,
    title = {School choice: an experimental study},
    journal = {Journal of Economic Theory},
    volume = {127},
    number = {1},
    pages = {202-231},
    year = {2006},
    issn = {0022-0531},
    doi = {https://doi.org/10.1016/j.jet.2004.10.006},
    url = {https://www.sciencedirect.com/science/article/pii/S0022053104002418},
    author = {Yan Chen and Tayfun S\"{o}nmez},
}

@article{pais2008school,
    title = {School choice and information: An experimental study on matching mechanisms},
    journal = {Games and Economic Behavior},
    volume = {64},
    number = {1},
    pages = {303-328},
    year = {2008},
    issn = {0899-8256},
    doi = {https://doi.org/10.1016/j.geb.2008.01.008},
    url = {https://www.sciencedirect.com/science/article/pii/S0899825608000286},
    author = {Pais, Joana and Pint\'{e}r, \'{A}gnes},
}

@article{
    klijn2013preference, 
    title={Preference intensities and risk aversion in school choice: a laboratory experiment}, 
    volume={16}, 
    DOI={10.1007/s10683-012-9329-5}, 
    number={1}, 
    journal={Experimental Economics}, 
    author={Klijn, Flip and Pais, Joana and Vorsatz, Marc}, 
    year={2013}, 
    pages={1–22}
}

@article{featherstone2016boston,
    title = {Boston versus deferred acceptance in an interim setting: An experimental investigation},
    journal = {Games and Economic Behavior},
    volume = {100},
    pages = {353-375},
    year = {2016},
    issn = {0899-8256},
    doi = {https://doi.org/10.1016/j.geb.2016.10.005},
    url = {https://www.sciencedirect.com/science/article/pii/S0899825616301208},
    author = {Clayton R Featherstone and Muriel Niederle},
}

@article{hassidim2017mechanism,
Author = {Hassidim, Avinatan and Marciano, D\'{e}borah and Romm, Assaf and Shorrer, Ran I.},
Title = {The Mechanism Is Truthful, Why Aren't You?},
Journal = {American Economic Review},
Volume = {107},
Number = {5},
Year = {2017},
Month = {May},
Pages = {220–24},
DOI = {10.1257/aer.p20171027},
URL = {https://www.aeaweb.org/articles?id=10.1257/aer.p20171027}
}

@article{hassidim2021limits,
author = {Hassidim, Avinatan and Romm, Assaf and Shorrer, Ran I.},
title = {The Limits of Incentives in Economic Matching Procedures},
journal = {Management Science},
volume = {67},
number = {2},
pages = {951-963},
year = {2021},
doi = {10.1287/mnsc.2020.3591},
URL = {https://doi.org/10.1287/mnsc.2020.3591}
}

@article{he2015gaming,
  title={Gaming the Boston school choice mechanism in Beijing},
  author={He, Yinghua},
  year={2017},
  publisher={TSE Working Paper}
}

@article{larroucau2023application,
    author = {Fabre, Ana\"{i}s and Larroucau, Tom\'{a}s and Martinez, Manuel and Neilson, Christopher and Rios, Ignacio},
    title = {Application Mistakes and Information Frictions in College Admissions},
    journal = {Working Paper},
    year = {2023}, 
    URL = {https://tlarroucau.github.io/Mistakes_College.pdf}
}

@article{reesjones2023behavioral,
    author = {Rees-Jones, Alex and Shorrer, Ran},
    title = {Behavioral Economics in Education Market Design: A Forward-Looking Review},
    journal = {Journal of Political Economy Microeconomics},
    volume = {1},
    number = {3},
    pages = {557-613},
    year = {2023},
    doi = {10.1086/725054},
    URL = {https://doi.org/10.1086/725054}
}

@article{teo2001gale,
    author = {Teo, Chung-Piaw and Sethuraman, Jay and Tan, Wee-Peng},
    title = {Gale-Shapley Stable Marriage Problem Revisited: Strategic Issues and Applications},
    journal = {Management Science},
    volume = {47},
    number = {9},
    pages = {1252-1267},
    year = {2001},
    doi = {10.1287/mnsc.47.9.1252.9784},
    URL = {https://doi.org/10.1287/mnsc.47.9.1252.9784}
}

@article{akyol2017preferences,
    title = {Preferences, selection, and value added: A structural approach},
    journal = {European Economic Review},
    volume = {91},
    pages = {89-117},
    year = {2017},
    issn = {0014-2921},
    doi = {https://doi.org/10.1016/j.euroecorev.2016.09.009},
    url = {https://www.sciencedirect.com/science/article/pii/S0014292116301489},
    author = {Pelin Akyol and Kala Krishna}
}

@article{calsamiglia2020structual,
author = {Calsamiglia, Caterina and Fu, Chao and G\"{u}ell, Maia},
title = {Structural Estimation of a Model of School Choices: The Boston Mechanism versus Its Alternatives},
journal = {Journal of Political Economy},
volume = {128},
number = {2},
pages = {642-680},
year = {2020},
doi = {10.1086/704573},
URL = {https://doi.org/10.1086/704573},
}

@article{ambagtsheer2020global,
  title={Global Kidney Exchange: opportunity or exploitation? An ELPAT/ESOT appraisal},
  author={Ambagtsheer, Frederike and Haase-Kromwijk, Bernadette and Dor, Frank JMF and Moorlock, Greg and Citterio, Franco and Berney, Thierry and Massey, Emma K},
  journal={Transplant International},
  volume={33},
  number={9},
  pages={989--998},
  year={2020},
  publisher={Wiley Online Library}
}

@article{ergin2020efficient,
  title={Efficient and incentive-compatible liver exchange},
  author={Ergin, Haluk and S{\"o}nmez, Tayfun and {\"U}nver, M Utku},
  journal={Econometrica},
  volume={88},
  number={3},
  pages={965--1005},
  year={2020},
  publisher={Wiley Online Library}
}

@article{ergin2017dual,
  title={Dual-donor organ exchange},
  author={Ergin, Haluk and S{\"o}nmez, Tayfun and {\"U}nver, M Utku},
  journal={Econometrica},
  volume={85},
  number={5},
  pages={1645--1671},
  year={2017},
  publisher={Wiley Online Library}
}

@article{dickerson2017multi,
  title={Multi-organ exchange},
  author={Dickerson, John P and Sandholm, Tuomas},
  journal={Journal of Artificial Intelligence Research},
  volume={60},
  pages={639--679},
  year={2017}
}

@article{watanabe2022multiorgan,
  title={Multiorgan Exchange: A Liver-Kidney Paired Exchange Perspective},
  author={Watanabe, Moyuru},
  journal={Available at SSRN 4287517},
  year={2022}
}

@article{anand2025Multi,
  title={Multi-Modal Paired Exchange},
  author={Anand Siththaranjan},
journal={Mimeo},
Note = {https://anands29.github.io/assets/pdf/multi-modal.pdf},
  year={2025}
}

@misc{kokuseichousa2020,
  title = {2020 Population Census of Japan},
  author = {{Statistics Bureau, Ministry of Internal Affairs, and Communications}},
  url = {{https://www.stat.go.jp/data/kokusei/2020/kekka.html}},
  year = {2022}
}

@book{Hwang2015,
  title={A robust redesign of High School Match},
  publisher={PhD Thesis, The University of Chicago},
  author={Hwang, Il Myoung},
  year={2015}
}

@article{abdulkadiroglu2017welfare,
    Author = {Abdulkadiroğlu, Atila and Agarwal, Nikhil and Pathak, Parag A.},
    Title = {The Welfare Effects of Coordinated Assignment: Evidence from the New York City High School Match},
    Journal = {American Economic Review},
    Volume = {107},
    Number = {12},
    Year = {2017},
    Month = {December},
    Pages = {3635–89},
    DOI = {10.1257/aer.20151425},
    URL = {https://www.aeaweb.org/articles?id=10.1257/aer.20151425}
}

@misc{kuno2025strategic,
      title={Strategic Waiting in Centralized Matching: Daycare Assignment}, 
      author={Kan Kuno},
      year={2025},
      eprint={2311.07920},
      archivePrefix={arXiv},
      primaryClass={econ.GN},
      url={https://arxiv.org/abs/2311.07920}, 
}

@article{luflade2019value,
      title={The value of information in centralized school choice systems}, 
      author={Margaux Luflade},
      year={2019},
      journal={Mimeo},
      url={https://drive.google.com/file/d/1IKxiuzuCrIvQuTuIO3cj9GH_sDvRmLoo/view}, 
}
